\documentclass[letterpaper]{article} % DO NOT CHANGE THIS
\usepackage{aaai23}  % DO NOT CHANGE THIS
\usepackage{times}  % DO NOT CHANGE THIS
\usepackage{helvet}  % DO NOT CHANGE THIS
\usepackage{courier}  % DO NOT CHANGE THIS
\usepackage[hyphens]{url}  % DO NOT CHANGE THIS
\usepackage{graphicx} % DO NOT CHANGE THIS
\usepackage{natbib}  % DO NOT CHANGE THIS AND DO NOT ADD ANY OPTIONS TO IT
\usepackage{caption} % DO NOT CHANGE THIS AND DO NOT ADD ANY OPTIONS TO IT
\usepackage{algorithm}
\usepackage{algorithmic}

\usepackage{newfloat}
\usepackage{listings}
\DeclareCaptionStyle{ruled}{labelfont=normalfont,labelsep=colon,strut=off} % DO NOT CHANGE THIS
\floatstyle{ruled}
\newfloat{listing}{tb}{lst}{}
\floatname{listing}{Listing}
\usepackage{tikz}
\usetikzlibrary{positioning}
\usetikzlibrary{cd,shapes,decorations,arrows,calc,arrows.meta,fit,positioning}
\tikzset{
    -Latex,auto,node distance =.5 cm and .5 cm,%semithick,
    state/.style ={rectangle, draw, minimum width = 0.4 cm},
    point/.style = {circle, draw, inner sep=0.04cm,fill,node contents={}},
    bidirected/.style={Latex-Latex,dashed},
    pago/.style={o-Latex},
    pagt/.style={o-o},
    el/.style = {inner sep=2pt, align=left, sloped}
}

\usepackage{stackrel, amssymb}
\usepackage{mathtools}
\usepackage{amsthm}
\usepackage{bm}
\usepackage{bbm}
\usepackage{thmtools}
\usepackage{thm-restate}
\usepackage{xhfill}

\declaretheorem{lemma}

\declaretheorem{remark}

\theoremstyle{definition}
\declaretheorem{definition}

\usepackage{apptools}
\usepackage{tabularx}
\usepackage{booktabs}

\AtAppendix{\counterwithin{lem}{section}}

\usepackage[labelformat=simple]{subcaption}

\def\*#1{\mathbf{#1}}
\makeatletter
\newcommand*{\indep}{%
  \mathbin{%
    \mathpalette{\@indep}{}%
  }%
}
\newcommand*{\nindep}{%
  \mathbin{%                   % The final symbol is a binary math operator
    \mathpalette{\@indep}{\not}% \mathpalette helps for the adaptation
  }%
}
\newcommand*{\@indep}[2]{%
  \sbox0{$#1\perp\m@th$}%        box 0 contains \perp symbol
  \sbox2{$#1=$}%                 box 2 for the height of =
  \sbox4{$#1\vcenter{}$}%        box 4 for the height of the math axis
  \rlap{\copy0}%                 first \perp
  \dimen@=\dimexpr\ht2-\ht4-.2pt\relax
  \kern\dimen@
  {#2}%
  \kern\dimen@
  \copy0 %                       second \perp
} 
\makeatother

\title{Causal Effect Identification in Cluster DAGs}
\author{
      Tara V. Anand\equalcontrib\textsuperscript{\rm 1},       %\textls[-80]{\small{\texttt{tara.v.anand@columbia.edu}}} \\
      Adèle H. Ribeiro\equalcontrib\textsuperscript{\rm 2},
      Jin Tian\textsuperscript{\rm 3},
      Elias Bareinboim\textsuperscript{\rm 2}
}
\affiliations{
    \textsuperscript{\rm 1}Department of Biomedical Informatics, Columbia University\\
    \textsuperscript{\rm 2}Department of Computer Science, Columbia University\\
    \textsuperscript{\rm 3}Department of Computer Science, Iowa State University\\
    \small{\texttt{tara.v.anand@columbia.edu}, \texttt{adele@cs.columbia.edu}, \texttt{jtian@iastate.edu}, \texttt{eb@cs.columbia.edu}}
}

\begin{document}

\maketitle

\begin{abstract}
Reasoning about the effect of interventions and counterfactuals is a fundamental task found throughout the data sciences.
A collection of principles, algorithms, and tools has been developed for performing such tasks in the last decades \citep{pearl:2k}.  One of the pervasive requirements found throughout this literature is the articulation of assumptions, which commonly appear in the form of causal diagrams. 
Despite the power of this approach, there are significant settings where the knowledge necessary to specify a causal diagram over all variables is not  available, particularly in complex, high-dimensional domains. In this paper, we introduce  a new graphical modeling tool called \textit{cluster DAGs} (for short, C-DAGs) that allows for the partial specification of relationships among variables based on limited prior knowledge, alleviating the stringent requirement of specifying a full causal diagram. A C-DAG specifies relationships between clusters of variables, while the relationships between the variables within a cluster are left unspecified, and can be seen as a graphical representation of an equivalence class of causal diagrams that share the relationships among the clusters. We develop the foundations and  machinery for valid inferences over  C-DAGs about the clusters of variables 
at each layer of Pearl's Causal Hierarchy \citep{pearl:mackenzie2018, bar:etal2020} - $\mathcal{L}_1$ (probabilistic), $\mathcal{L}_2$ (interventional), and $\mathcal{L}_3$ (counterfactual).  
In particular, we prove the soundness and completeness of d-separation for probabilistic inference in C-DAGs. 
Further, we demonstrate  the validity of Pearl's do-calculus rules over C-DAGs and  show that the standard ID identification algorithm is sound and complete to systematically compute causal effects from observational data given a C-DAG. Finally, we show that C-DAGs are valid for performing counterfactual inferences about clusters of variables.
\end{abstract}

\section{Introduction}

% 1. usefulness of causality 
% 2. id problem and ample literature , do-calc and id algorithms 
% 3. challenging for its applicability 
% 4. 

One of the central tasks found in data-driven disciplines is to infer the effect of a treatment $X$ on an outcome $Y$, which is formally written as the interventional distribution $P(Y | do(X=x))$, from observational (non-experimental) data collected from the phenomenon under investigation. These relations are considered essential in the construction of explanations and for making decisions about interventions that have never been implemented before  \citep{pearl:2k,spirtes:etal00,bareinboim:pea16,PetersJanzingSchoelkopf17}.%,pearl:mackenzie2018}.  

Standard tools necessary for identifying the aforementioned  do-distribution, such as d-separation, do-calculus \cite{pearl:95a}, and the ID-algorithm \cite{tian:pea02-general-id,shpitser:pea06a,huang:val06-complete, lee:etal19} take as input a combination of an observational distribution and a qualitative description of the underlying causal system, often articulated in the form of a causal diagram \citep{pearl:2k}. However,  specifying a causal diagram requires knowledge about the causal relationships among all pairs of observed variables, which is not always available in many real-world applications. This is especially  true and acute in complex, high-dimensional settings, which curtails the applicability of causal inference theory and tools.
 In the context of medicine, for example, electronic health records include data on lab tests, %conditions observed, 
drugs, demographic information, and other clinical attributes, but medical knowledge is not yet advanced enough to lead to the construction of causal diagrams over all of these variables, limiting the use of the graphical approach to inferring causality \cite{kleinberg_review_2011}. In many cases, however, contextual or temporal information about variables is available, which may partially inform how these variables are situated in a causal diagram relative to other key variables. For instance, a data scientist may know that covariates
%variables like age ($A$), blood pressure ($B$), comorbidities ($C$), and medication history ($D$) 
occur temporally before a drug is prescribed or an outcome occurs. They may even suspect that some pre-treatment variables are causes of the treatment and the outcome variables. However, they may be uncertain about the relationships among each pair of covariates, or it may be burdensome to explicitly define them. Given that a misspecified causal diagram may lead to wrong causal conclusions, this issue raises the question of whether a coarser representation of the causal diagram, where no commitment is made to the relationship between certain variables, would still be sufficient to determine the causal effect of interest. 

% which we will defined as a cluster. %, by what we will define as a cluster. 

In this paper, our goal is to develop a framework for %identification of causal effects 
causal inferences
in partially understood domains such as the medical domain discussed above. 
% First, we introduce a novel graphical model that allows the specification of the causal relationships over clusters of variables, while the relationships among the variables within the clusters are left unspecified. 
%\hl{Change next discussion}
We will focus on formalizing the problem of causal effect identification considering that the data scientist does not have prior knowledge to fully specify a causal diagram %or causal relationships 
over all pairs of variables.
%, but does have prior knowledge to % specify the causal relationships over clusters or groupings of variables. partially specify causal relationships between variables in different clusters or groups. 
First, we formally define and characterize %a novel causal graphical model 
a novel class of graphs
called \textit{cluster DAGs} (or C-DAG, for short), which will allow for encoding of partially understood causal relationships between variables in different abstracted clusters, representing a group of variables among which causal relationships are not %fully 
understood or specified. 
%Within a cluster, any relationships between variables are possible, so a C-DAG represents an equivalence class of causal diagrams compatible with the knowledge available.
%Such a model can be considered an equivalence class of graphical models, allowing for ambiguity in the specification of the relationships within clusters. 
Then, we develop the foundations and machinery for valid probabilistic and causal inferences, akin to Pearl’s d-separation and do-calculus for when such a coarser graphical representation of the system is provided based on the limited prior knowledge available. 
In particular, we follow Pearl's Causal Hierarchy \citep{pearl:mackenzie2018, bar:etal2020} and develop the machinery for inferences in C-DAGS at all three inferential layers -- $\mathcal{L}_1$ (associational), $\mathcal{L}_2$ (interventional), and $\mathcal{L}_3$ (counterfactual).
%\xadd{Then, following Pearl's causal hierarchy, we show that a C-DAG is optimal for inferences over clusters (macro-variables) at all three inferential layers -- $\mathcal{L}_1$ (associational), $\mathcal{L}_2$ (interventional), and $\mathcal{L}_3$ (counterfactual). Specifically, we show that a C-DAG is a (probabilistic) Bayesian Network (BN) over macro-variables and d-separation is sound and complete for extracting conditional independence relations over such macro-variables. Moreover, we show that a C-DAG is a Causal Bayesian Network (CBN) over macro-variables and causal tools such as do-calculus and ID algorithm are sound and complete for causal reasoning over such macro-variables. Finally, we show that a C-DAG is induced by a Structural Causal Model (SCM) over macro-variables which is equivalent to the underlying SCM over the original variables on statements about the macro-variables. Therefore, tools for counterfactual reasoning are also sound and complete in C-DAGs (?)} 
%, akin to Pearl’s d-separation and do-calculus for when such a coarser graphical representation of the system is provided based on the limited prior knowledge available.
%\footnote{Investigating different structure learning approaches for the full DAG is not part of the scope of this paper.}
The results  are fundamental first steps in terms of semantics and graphical conditions to perform probabilistic, interventional, and counterfactual inferences over clusters of variables. %The new proposed semantics provide the basis for the development of more practical and relaxed causal inferences and learning tools.
Specifically, we outline our technical contributions below. %in the following.
%\xcomment{Write the contributions emphasizing better what is done, maybe adding references to the Theorem.}
%\input{figs/1_fig1_compact}
%\vspace{-0.5em}
%\hl{change the list of contributions}
\begin{enumerate}
\setlength\itemsep{0em}
    \item We introduce a new graphical modelling tool called \emph{cluster DAGs} (or C-DAGs) over %a set of
    macro-variables representing clusters of variables where the relationships among the variables inside the clusters are left unspecified \textit{(Definition \ref{def:cdag})}. %We show that interventional distributions admit a convenient factorization following the coarser structure of C-DAGs. This gives a C-DAG the semantics of an equivalence class of all compatible underlying causal diagrams \textit{(Theorem \ref{thm:markovrelative})}.
    Semantically, a C-DAG represents an equivalence class of all %compatible
    underlying graphs over the original variables that share the relationships among the clusters.
   % \item We show that 
  
  \item We show that a C-DAG is a (probabilistic) Bayesian Network (BN) over macro-variables and Pearl's d-separation is sound and complete for extracting conditional independencies over macro-variables if the underlying graph over the original variables is a BN (\textit{Theorems \ref{thm:dsep_cdag_dag} and  \ref{thm:cdag_bn}}).
    
%    \item We prove the soundness and completeness of Pearl's d-separation and do-calculus rules extended to the coarse graphical representation of  C-DAGs \textit{(Theorems \ref{thm:dsep_cdag_dag}, \ref{thm:docalc-cdags} and \ref{thm:do-calc_completenes})}, despite C-DAGs’ abstracted representation of possibly numerous underlying causal diagrams.
    
  \item We show that a C-DAG is a Causal Bayesian Network (CBN) over macro-variables and Pearl's do-calculus is sound and complete for causal inferences about macro-variables in C-DAGs if the underlying graph over the original variables is a CBN (\textit{Theorems \ref{thm:docalc-cdags}, \ref{thm:do-calc_completenes}, and \ref{thm:markovrelative}}).  The results can be used to show that the ID-algorithm is sound and complete to systematically infer causal effects from the  observational distribution and partial domain knowledge encoded as a C-DAG (\textit{Theorem \ref{thm:ids}}). 
%    \item %Given the developed language and foundations of C-DAGs, we prove that the ID-algorithm is sound and complete to systematically infer causal effects from the %combination of an 
    %observational distribution and partial domain knowledge encoded as a C-DAG \textit{(Theorem \ref{thm:ids})}. 
% \item    We prove that interventional distributions admit a convenient factorization following the C-DAG's structure \textit{(Theorem \ref{thm:markovrelative})}. This factorization can be used to show that the ID-algorithm is sound and complete to systematically infer causal effects from the  observational distribution and partial domain knowledge encoded as a C-DAG \textit{(Theorem \ref{thm:ids})}.
\item We show that, assuming the underlying graph $G$ is induced by an SCM $\mathcal{M}$, then 
there exists an SCM $\mathcal{M}_{\*C}$ over macro-variables $\*C$ such that its induced causal diagram is $G_{\*C}$ and it is equivalent to $\mathcal{M}$ on statements about the macro-variables (\textit{Theorem~\ref{thm:ctf_cdags}}). Therefore, the CTFID algorithm \citep{correa21nestedctf} for the identification of nested counterfactuals from an arbitrary combination of observational and experimental distributions can be extended to the C-DAGs.

%and,a C-DAG is induced by a Structural Causal Model (SCM) over macro-variables which is equivalent to the underlying SCM over the original variables on statements about the macro-variables. Therefore, tools for counterfactual reasoning are also sound and complete in C-DAGs
    \end{enumerate}

%\input{figs/3_med-example}

%     \item We introduce a new class of graphs called cluster DAGs (or C-DAGs) over a set of clusters of variables where the relationships amongst the variables inside the clusters are left unspecified \textit{(Definition \ref{def:cdag})}.
%     \item We show that despite C-DAGs' abstracted representation of %numerous possible 
%     possibly numerous 
%     underlying paths, separation rules can be developed to account for the clusters. We prove the soundness and completeness of the d-separation rules extended to C-DAGs \textit{(Theorem \ref{thm:dsep_cdag_dag})}. 
%     \item We prove that the inference rules known as Pearl's do-calculus are sound and complete for the coarse representation of C-DAGs \textit{(Theorems \ref{thm:docalc-cdags} and \ref{thm:do-calc_completenes})}.
%     \item  We prove that interventional distributions admit a convenient factorization following the C-DAG's structure \textit{(Theorem \ref{thm:markovrelative})}. This factorization can be used to show that the ID-algorithm is sound and complete to systematically infer causal effects from the combination of an observational distribution and partial domain knowledge encoded as a C-DAG \textit{(Theorem \ref{thm:ids})}. 

%\vspace{-0.2in}
\subsection{Related work}

%\xadd{There are other attempts in the literature to address the difficulty of specifying a full causal diagram.  the work by Beckers and Halpern (AAAI 19) and other recent work on causal abstraction. }
Since a group of variables may constitute a semantically meaningful entity, causal models over abstracted clusters of variables have attracted increasing attention for the development of more interpretable tools \cite{scholkopf2021toward,Shen2018psdd}.
\cite{parviainen16groupbn} studied the problem of, given a DAG, under what assumptions a DAG over macro-variables can represent
the same conditional independence relations between the macro-variables. 
%\cite{Shen2018psdd} mentioned the use of Bayesian network (BN) over clusters of variables.
Recent developments on causal abstraction  %, however, 
have focused on the distinct problem of investigating %functions that map 
mappings of a cluster of (micro-)variables to a single (macro-)variable, while preserving some causal properties \citep{chalupka15visual, chalupka16multi, rubensteinWBMJG17, halpern19abstraction}. %and, importantly, assuming a structural causal model defined on the level on which the variables were measured, is given. 
The result is a new structural causal model defined on a higher level of abstraction, but with causal properties similar to those in the low-level model.\footnote{In \cite{halpern19abstraction}'s notation, we investigate the case of a \textit{constructive $\tau$-abstraction} where the mapping $\tau$ only groups the low-level variables into high-level (cluster) variables.} Other related works include chain graphs \citep{lauritzen:ric02} and ancestral causal graphs \citep{zhang:08} developed  to represent collections of causal diagrams equivalent under  certain properties. By contrast, our work proposes a new graphical representation of a class of compatible causal diagrams, representing limited causal knowledge when the full structural causal model is unknown.

Causal discovery algorithms %from observational data 
can be an alternative for when prior knowledge is insufficient to fully delineate a causal diagram \citep{pearl:2k, spirtes:etal00,PetersJanzingSchoelkopf17}. %\cite{zhang:08}.
%There are extensive literature on learning causal graphs from observational data. 
%\xst{However, causal structure learning remains an open, challenging problem in high-dimensional settings, particularly when unobserved/hidden variables are allowed, as is the case considered in this work.} %\xst{which is the setting we are considering (dashed bidirected edges in the causal diagram represent unobserved confounders)}. 
However, in general, it is impossible to fully recover the causal diagram based solely on observational data, without making strong assumptions about the underlying causal model, including causal sufficiency (all variables have been measured), the form of the functions (e.g., linearity, additive noise), and the distributions of the error terms (e.g. Gaussian, non-Gaussian, etc) \citep{glymour2019review}. Then, there are cases where a meaningful causal diagram cannot be learned and prior knowledge is necessary for its construction. Our work focuses on establishing a language and corresponding machinery to encode partial knowledge and infer causal effects over clusters, alleviating some challenges in causal modeling in high-dimensional settings.

\section{Preliminaries}
% SCM
% Causal diagram definition
% Causal diagram and kinship nomenclature

%We introduce in this section the necessary concepts and notation used throughout the paper.

\textbf{Notation.} A single variable is denoted by a (non-boldface) uppercase letter $X$ and its realized value by a small letter $x$. 
A boldfaced uppercase letter $\*X$ denotes a set (or a cluster) of variables. We use kinship relations, defined along the full edges in the graph, ignoring bidirected edges. We denote by $Pa(\*X)_G$, $An(\*X)_G$, and $De(\*X)_G$, the sets of %union of $\*X$ with its 
parents, ancestors, and descendants in $G$, respectively. A vertex $V$ is said to be \emph{active} on a path relative to
%a set 
$\*Z$ if 1) $V$ is a collider and $V$ or any of its descendants are in $\*Z$ or 2) $V$ is a non-collider and is not in $\*Z$. A path $p$ is said to be \emph{active} given (or conditioned on) $\*Z$ if every vertex on $p$ is active relative to $\*Z$. Otherwise, $p$ is said to be \emph{inactive}. Given a graph $G$, $\*X$ and $\*Y$ are d-separated by $\*Z$ if every path between $\*X$ and $\*Y$ is inactive given $\*Z$. We denote this d-separation by $(\*X \indep \*Y \mid \*Z)_{G}$. 
The mutilated graph $G_{\overline{\*X}\underline{\*Z}}$ is the result of
removing from $G$ edges %\xst{coming into} 
with an arrowhead into $\*X$ (e.g., $A \rightarrow \*X$, $A \leftrightarrow \*X$), %\xst{variables in $\*X$} 
and edges with a tail from 
%\xst{going out of variables in}
$\*Z$ (e.g., $A \leftarrow \*Z$).

%$G_\*C$ denotes the subgraph of $G$ over $\*C$.

%For $\*X, \*Y,\*Z$, we denote the conditional independence of $\*X$ and $\*Y$ conditioned on $\*Z$ by $\*X \indep \*Y \mid \*Z$. 
%we denote the d-separation between $\*X$ and $\*Y$ given $\*Z$ by $(\*X \indep \*Y \mid \*Z)_{G}$. \xadd{We call a path $p$ \emph{inactive} if and only if $(\*X \indep \*Y \mid \*Z)_{G}$ and 1) $p$ contains a chain of nodes $A  \rightarrow B \rightarrow C$ or a fork $A \leftarrow B \rightarrow C$ such that $B \in \*Z$ or 2) $p$ contains a collider $A \rightarrow B \leftarrow C$ such that $(B \cup De(B)) \cap \*Z = \emptyset$}. Otherwise, we call the path $p$ active.
% Jin
%\hl{Jin: need to review key concepts under d-separation, e.g. active/inactive path. Done!}

\textbf{Structural Causal Models (SCMs)}
%We use the language of Structural Causal Models (SCMs) \citep[pp.~204-207]{pearl:2k} as our semantical framework. % to represent causal systems.
Formally, an SCM $\mathcal{M}$ is a 4-tuple $\langle \*U, \*V, \mathcal{F}, P(\*U)\rangle$, where $\*{U}$ is a set of exogenous (latent) variables and $\*{V}$ is a set of endogenous (measured) variables. $\mathcal{F}$ is a collection of functions $\{f_i\}_{i=1}^{|\*{V}|}$ such that each endogenous variable $V_i\in\*{V}$ is %determined by 
a function $f_i\in\mathcal{F}$ %, where $f_i$ is a mapping from the respective domain 
of $\*{U}_i\cup Pa(V_i)$, % to $V_i$,
where $\*{U}_i\subseteq\*{U}$ and $Pa(V_i)\subseteq\*{V}\setminus V_i$. The uncertainty is encoded through a probability distribution over the exogenous variables, $P(\*{U})$.
Each SCM $\mathcal{M}$ induces a directed acyclic graph (DAG) with bidirected edges -- or an acyclic directed mixed graph (ADMG) -- $G(\*V, \*E)$, known as a \emph{causal diagram}, that encodes the structural relations among $\*{V}\cup\*{U}$, where every $V_i \in \*V$ is a vertex, there is a directed edge $(V_j \rightarrow V_i)$ for every $V_i \in \*V$ and $V_j \in Pa(V_i)$, and there is a dashed bidirected edge $(V_j \dashleftarrow \!\!\!\!\!\!\!\!\! \dashrightarrow V_i)$ for every pair $V_i, V_j \in \*V$ such that $\*U_i \cap \*U_j \neq \emptyset$ ($V_i$ and $V_j$ have a common exogenous parent). % We constrain our results to causal systems that are acyclic. 
%Following standard practice, we omit the exogenous variables in the causal diagrams. 
%Within the structural semantics, 
Performing an intervention $\*{X}\!\!=\!\! \*{x}$ is represented through the do-operator, \textit{do}($\*{X}\!=\!\*{x}$), which represents the operation of fixing a set $\*X$ to a constant $\*{x}$, and % regardless of their original mechanisms. Such an intervention 
induces a submodel $\mathcal{M}_\*{x}$, which is $\mathcal{M}$ with $f_X$ replaced to $x$ for every $X \in \*X$. The post-interventional distribution induced by $\mathcal{M}_\*{x}$ is denoted by $P(\*v \setminus \*x |do(\*x))$.\\ %or simply $P_\*x(\*v)$. %For further details, we refer readers to \citep{pearl:2k}.
For any subset $\*Y \subseteq \*V$, the \emph{potential response} $\*Y_{\*x}(\*u)$ is defined as the solution of $\*Y$ in the submodel $M_{\*x}$ given $\*U = \*u$. %Drawing values of exogenous variables $\*U$ following the distribution 
$P(\*U)$ then induces a \emph{counterfactual variable} $\*Y_{\*x}$.

\textbf{Pearl's Causal Hierarchy (PCH) / The Ladder of Causation } \citep{pearl:mackenzie2018, bar:etal2020} is a formal framework that divides inferential tasks into three different layers, namely, 1) associational, 2) interventional, and 3) counterfactual (see Table\ref{pearl:hierarchy}). %shows the typical activity, model and question in each layer. Any association or predictive analysis is placed in the first layer of the PCH. Effects of interventions are placed on the second layer, while counterfactuals are placed on the third layer. 
An important result formalized under the rubric of the Causal Hierarchy Theorem (CHT) \citep[Thm. 1]{bar:etal2020} states that inferences at any layer of the PCH almost never can be obtained by using solely information from lower layers.

\begin{table}[t]
{  \scriptsize
  \centering
  \begin{tabular}{@{}p{0.3cm}p{1.5cm}p{0.9cm}p{0.9cm}p{3.1cm}@{}} \hline %\toprule
   & \textbf{Level \newline (Symbol)}  &\textbf{Typical\newline Activity} & \textbf{Typical Model} & \textbf{Typical\newline Question}   \\ \hline %\midrule
   $\mathcal{L}_1$ & Associational \newline $P(y|x)$ & Seeing  & BN & What is? \newline How would seeing $X$ change my belief in $Y$?  \\ \midrule
   $\mathcal{L}_2$ & Interventional \newline $P(y|do(x),c)$ & Doing & CBN & What if? \newline What if I do $X$?  \\ \midrule
   $\mathcal{L}_3$ & Counterfactual \newline $P(y_x|x',y')$ & Imagining & SCM & Why?  \newline  What if I had acted differently?   \\ \hline %\bottomrule
  \end{tabular}}
  \caption{The Ladder of Causation / Pearl's Causal Hierarchy %Pearl's Ladder of Causation
  %\citep{pearl2018book,bar:etal2020}.
  }
  \label{pearl:hierarchy}
  \vspace{-1\intextsep}	
\end{table}

\section{C-DAGs: Definition and Properties}
\label{sec:cdags}

%\section{C-DAGs: Definition and Properties}

% SCM $\mathcal{M} = \langle \*{V}, \*{U}, \mathcal{F}, P(\*{U})\rangle$ and the corresponding causal diagram $G(\*{V}, \*{E})$. Given a partition $\*C=\{\*{C}_1, \ldots, \*{C}_k\}$ of $\*V$, we do not have knowledge about the relations within each cluster

Standard causal inference tools typically require assumptions articulated through causal diagrams. We investigate the situations where the knowledge necessary to specify the underlying causal diagram $G(\*{V}, \*{E})$ over the individual variables in $\*V$ may not be available. % but we are still interested in performing probabilistic, causal, and counterfactual inferences that are valid in $G$. 
%\xadd{We propose that variables among which the causal relationships are unknown (or uncertain) are grouped as a cluster. In this way, the set of individual variables $\*V$ will be partitioned into a set of of variables $\*{C}_1, \ldots, \*{C}_k$ and only relationships among clusters are needed to be specified. Then, we address the following technical problem: }
 In particular, we assume that variables are grouped into a set of clusters of variables $\*{C}_1, \ldots, \*{C}_k$ that form a partition of $\*V$ (note that a variable may be grouped in a cluster by itself) such that we do not have knowledge about the relationships amongst the variables inside the clusters $\*C_i$ but we have some knowledge about the relationships between variables in different groups. We are interested in performing probabilistic and causal inferences about these clusters of variables; one may consider each cluster as defining a \emph{macro-variable} and our aim is to reason about these macro-variables. 

%\subsection{Formal Statement of the Problem}

Formally, we address the following technical problem: 

%\hl{causal diagram of DAG?}
\vspace{.3em}
\textbf{Problem Statement:} Consider a causal diagram $G^*$ over $\*V$ and a set of clusters of variables $\*C = \{\*C_1, \ldots, \*C_k\}$ forming a partition of $\*V$. We aim to perform probabilistic, interventional, or counterfactual inferences about the macro-variables. Can we construct a causal diagram $G^*_{\*C}$ over the macro-variables in $\*C$ such that inferences by applying standard tools (d-separation, do-calculus, ID algorithm) on $G^*_{\*C}$ are valid \emph{in the  sense they lead to the same conclusions as inferred on $G^*$}?

%we can reason about the macro-variables by directly applying d-separation, do-calculus, ID algorithm on $G_C$? The inferences on $G_C$ must be sound in the sense that the inferences on $G_C$ will lead to the same conclusions as inferred on $G^*$. 

%We are interested in performing probabilistic and causal inferences about the macro-variables $C_1, \ldots, C_k$. Can we construct a DAG/ADMG $G_C$ over macro-variables such that we can reason about the macro-variables by directly applying d-separation, do-calculus, ID algorithm on $G_C$? The inferences on $G_C$ must be sound in the sense that the inferences on $G_C$ will lead to the same conclusions as inferred on $G^*$. 

%\subsection{C-DAGs: Definition and Properties}

To this end, 
%we formally introduce 
we propose a graphical object called \textit{cluster DAGs} (or C-DAGs) to capture our partial knowledge about the underlying causal diagram over individual variables. %\xst{, which is a coarser representation of a causal diagram:} \hl{can we say that is a coarser representation of a causal diagram at this moment?}
\begin{definition}[\textbf{Cluster DAG or C-DAG}]
\label{def:cdag}
%\hl{For inferences in Layer 1, $G$ doesn't need to be a causal diagram. Could we say that $G$ is just a DAG? Then in Layer 2 we discuss the particular case in which $G$ is a causal diagram?}
Given  %causal diagram 
 an ADMG $G(\*{V}, \*{E})$ (a DAG with bidirected edges)  and a partition $\*C=\{\*{C}_1, \ldots, \*{C}_k\}$ of $\*V$, construct a graph $G_{\*{C}}(\*{C}, \*{E}_\*{C})$ over $\*C$ with a set of edges $\*E_\*C$ defined as follows:
\begin{enumerate}
    \itemsep0em 
    \item  An edge $\*{C}_i \rightarrow \*{C}_j$ is in $\*E_\*C$ if exists some $V_i \in \*{C}_i$ and $V_j \in  \*{C}_j$ such that $V_i\in Pa(V_j)$ in $G$;
    \item  A dashed bidirected edge $\*{C}_i \dashleftarrow \!\!\!\!\!\!\!\!\! \dashrightarrow \*{C}_j$ is in $\*E_\*C$ if exists some $V_i \in \*{C}_i$ and $V_j \in \*{C}_j$ such that %there is a bidirected edge 
    $V_i \dashleftarrow \!\!\!\!\!\!\!\!\! \dashrightarrow V_j$ in $G$. %$V_i$ and $V_j$ have a common exogenous parent.
\end{enumerate}
If $G_{\*{C}}(\*{C}, \*{E}_\*{C})$ contains no cycles, then we say that $\*C$ is an \emph{admissible partition} of $\*V$. We then call $G_{\*{C}}$  a \emph{cluster DAG}, or \emph{C-DAG}, compatible with $G$. 
\end{definition} 
Throughout the paper, we will use the same symbols (e.g. $\*{C}_i$) to represent both a cluster node in a C-DAG $G_{\*{C}}$ and the set of variables contained in the cluster. %\xst{in a compatible causal diagram $G$.} 

\begin{figure}[t]
\begin{subfigure}[b]{0.24\linewidth}
\centering
        \begin{tikzpicture}[scale = .5]
			\node(X) at (0,0) [label = below:$X$, point, blue];
			\node(D) at (0, 2)[label=above:$D$, point]; 
			\node(B) at (.8, 1.1)[label=above:$B$, point]; 
			\node(S) at (1.5, 0)[label=below:$S$, point]; 
			\node(C) at (1.5, 2)[label=above:$C$, point]; 
			\node(A) at (3, 2)[label=above:$A$, point]; 
			\node(Y) at (3, 0)[label = below:$Y$, point, red];
		
			\path (D) edge (X);
			\path (X) edge (S);
			\path (S) edge (Y);
			\path[bidirected] (X) edge[bend right =0] (B); 
		    \path (B) edge (C); 
		    \path (C) edge (Y); 
		    \path[bidirected] (C) edge[bend left = 30] (Y);
		    \path[bidirected] (D) edge[bend left = 30] (C); 
		    \path (A) edge (Y); 
		    \path (A) edge (C); 
        \end{tikzpicture}
        \caption*{$(a)$}
\end{subfigure}
\begin{subfigure}[b]{0.24\linewidth}
\centering
        \begin{tikzpicture}[scale = .5]
            \node(X) at (0,0) [label = below:$X$, point, blue];
% 			\node(D) at (0, 2)[label=left:$D$, point]; 
% 			\node(B) at (.8, .9)[label=below:$B$, point]; 
			\node(S) at (1.5, 0)[label=below:$S$, point]; 
% 			\node[state](SB) at (1.5, 0) {$\{B, S\}$};
% 			\node(C) at (1.5, 2)[label=above left:$C$, point]; 
% 			\node(A) at (3, 2)[label=above:$A$, point]; 
		  %  \node[state](CA) at (1.5, 2) {$\{A, C\}$};
		    \node[state](ABCD) at (1.5, 2) {$\*Z$};
			\node(Y) at (3, 0)[label = below:$Y$, point, red];
		
			\path (ABCD) edge (X);
			\path (X) edge (S);
			\path (S) edge (Y);
			\path[bidirected] (X) edge[bend left =30] (ABCD); 
		  %  \path (S) edge (ABCD); 
		    \path (ABCD) edge (Y); 
		    \path[bidirected] (ABCD) edge[bend left = 30] (Y);
		  %  \path[bidirected] (D) edge[bend left = 30] (CA); 
		  %  \path (ABCD) edge (Y); 
		  %  \path (A) edge (C); 
        \end{tikzpicture}
        \caption*{$(b)$}
\end{subfigure}
\begin{subfigure}[b]{0.24\linewidth}
\centering
        \begin{tikzpicture}[scale = .5]
			\node(X) at (0,0) [label = below:$X$, point, blue];
			\node(D) at (0, 2)[label=above:$D$, point]; 
% 			\node(B) at (.8, .9)[label=below:$B$, point]; 
% 			\node(S) at (1.5, 0)[label=below:$S$, point]; 
			\node[state](SB) at (1.5, 0) {$\*W$};
% 			\node(C) at (1.5, 2)[label=above left:$C$, point]; 
% 			\node(A) at (3, 2)[label=above:$A$, point]; 
		    \node[state](CA) at (1.5, 2) {$\*Z$};
			\node(Y) at (3, 0)[label = below:$Y$, point, red];
		
			\path (D) edge (X);
			\path (X) edge (SB);
			\path (SB) edge (Y);
			\path[bidirected] (X) edge[bend right =30] (SB); 
		    \path (SB) edge (CA); 
		    \path (CA) edge (Y); 
		    \path[bidirected] (CA) edge[bend left = 30] (Y);
		    \path[bidirected] (D) edge[bend left = 30] (CA); 
		    \path (CA) edge (Y); 
		  %  \path (A) edge (C); 
        \end{tikzpicture}
        \caption*{$(c)$}
\end{subfigure}
\begin{subfigure}[b]{0.24\linewidth}
\centering
        \begin{tikzpicture}[scale = .5]
		    \node(X) at (0,0) [label = below:$X$, point, blue];
% 			\node(D) at (0, 2)[label=left:$D$, point]; 
% 			\node(B) at (.8, .9)[label=below:$B$, point]; 
			\node[state](S) at (1.5, 0){$\*W$}; 
% 			\node[state](SB) at (1.5, 0) {$\{B, S\}$};
% 			\node(C) at (1.5, 2)[label=above left:$C$, point]; 
% 			\node(A) at (3, 2)[label=above:$A$, point]; 
		  %  \node[state](CA) at (1.5, 2) {$\{A, C\}$};
		    \node[state](ABCD) at (1.5, 2) {$\*Z$};
			\node(Y) at (3, 0)[label = below:$Y$, point, red];
		
			\path (ABCD) edge (X);
			\path (X) edge (S);
			\path (S) edge (Y);
			\path(S) edge (ABCD);
			\path[bidirected] (X) edge[bend left =30] (ABCD); 
		  %  \path (S) edge (ABCD); 
		    \path (ABCD) edge (Y); 
		    \path[bidirected] (ABCD) edge[bend left = 30] (Y);
		    \path[bidirected] (X) edge[bend right = 30] (S); 
		  %  \path (ABCD) edge (Y); 
		  %  \path (A) edge (C); 
        \end{tikzpicture}
        \caption*{$(d)$}
\end{subfigure}
		\caption{$(a)$: a possible ADMG over %possible causal diagram over
		%of representing the effect of taking lisinopril on having a stroke where 
		lisinopril ($X$), stroke ($Y$), age ($A$), 
		blood pressure ($B$), comorbidities ($C$),  medication history ($D$), and sleep quality ($S$). $(b)$: a C-DAG of $(a)$ with $\*Z=\{A, B, C, D\}$. %where $P(y \mid do(x))$ is identifiable by front-door adjustment. 
		$(c)$: a C-DAG of $(a)$ with $\*W = \{S, B\}$, $\*Z = \{A, C\}$. %where $P(y \mid do(x))$ is not identifiable.
		%(in practice, it would likely not make sense to group the pre-treatment variable of blood pressure with the post-treatment variable of sleep quality).
		% $(d)$: an invalid C-DAG of $(a)$ with $\*W = \{S, B\}, \*Z = \{A, C, D\}$; note %that $\{X, Y, \*W, \*Z\}$ is an inadmissible partition as 
		% a cycle is created among $(X, \*W, \*Z)$.
  $(d)$: an invalid C-DAG of $(a)$, as the partition $\{X, Y, \*W, \*Z\}$, with $\*W = \{S, B\}, \*Z = \{A, C, D\}$ is inadmissible due to the cycle among $(X, \*W, \*Z)$.}
		 \label{fig:med-example}
\vspace{-1\intextsep}		 
\end{figure}
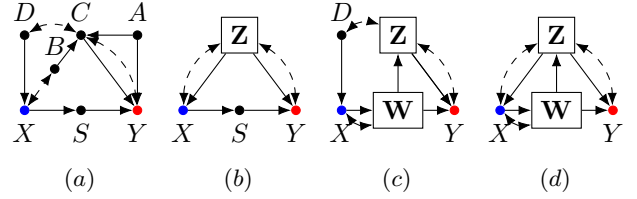

\textbf{Remark 1.} The definition of C-DAGs does not allow for cycles in order to utilize standard graphical modeling tools that work only in DAGs. An inadmissible partition of $\*V$ means that the partial knowledge available for constructing $G_{\*{C}}$ is not enough for drawing conclusions  using the tools developed in this paper.

\begin{figure}[t]
%\begin{wrapfigure}{r}{0.35\textwidth}
%\vspace{-2\intextsep}
\begin{minipage}[t]{0.32\linewidth}
\begin{subfigure}[b]{\linewidth}
\centering	
\begin{tikzpicture}[scale = .6]
			\node(X) at (0,0) [label = below:X, point, blue];
			\node(Y) at (2, 0)[label = below:Y, point, red];
			\node[state](Z) at (1, 2) {$\*Z$}; 
			\path (X) edge (Y);
			\path (Z) edge (X);
			\path (Z) edge (Y);
		\end{tikzpicture} 
		\caption*{$G_{\*C_1}$}
\end{subfigure}
\end{minipage}
\begin{minipage}[t]{.32\linewidth}
\begin{subfigure}[b]{\linewidth}
\centering	
\begin{tikzpicture}[scale = .6]
			\node(X) at (0,0) [label = below:$X$, point];
			\node(Y) at (2, 0)[label = below:$Y$, point];
			\node(Z1) at (0, 2)[label=left:{$Z_1$}, point]; 
			\node(Z3) at (2, 2)[label=right:{$Z_3$}, point]; 
			\node(Z2) at (1, 1) {$Z_2$};     %[label=below:{$Z_2$}, point]; 
			\path (X) edge (Y);
			\path (Z1) edge (Z2);
			\path (Z1) edge (X);
			\path (Z3) edge (Z2);
			\path (Z3) edge (Y);
			%\draw[gray, fill=gray!30, opacity =.5] (-.1,1.7) rectangle (2.1, 2.3);
	\end{tikzpicture}
		\caption*{$(a)$}
		%\vspace{0.7em}
\end{subfigure}
\end{minipage}
\begin{minipage}[t]{.32\linewidth}
\begin{subfigure}[b]{\linewidth}
\centering	
\begin{tikzpicture}[scale = .6]
			\node(X) at (0,0) [label = below:$X$, point];
			\node(Y) at (2, 0)[label = below:$Y$, point];
			\node(Z1) at (0, 2)[label=left:{$Z_1$}, point]; 
			\node(Z3) at (2, 2)[label=right:{$Z_3$}, point]; 
			\node(Z2) at (1, 1) {$Z_2$};     %[label=below:{$Z_2$}, point]; 
			\path (X) edge (Y);
			\path (Z1) edge (Z2);
			\path (Z2) edge (Y);
			\path (Z2) edge (X);
			\path (Z1) edge (X);
			\path (Z3) edge (Z2);
			\path (Z3) edge (Y);
			%\draw[gray, fill=gray!30, opacity =.5] (-.1,1.7) rectangle (2.1, 2.3);
	\end{tikzpicture}
		\caption*{$(b)$}
\end{subfigure}
\end{minipage}

\begin{minipage}[t]{.32\linewidth}
\begin{subfigure}[b]{\linewidth}
\centering	
		\begin{tikzpicture}[scale = .6]
			\node(X) at (0,0) [label = below:X, point, blue];
			\node(Y) at (2, 0)[label = below:Y, point, red];
			\node[state](Z) at (1, 2) {$\*Z$}; 
			\path (X) edge (Y);
			%\path (Z) edge (X);
			\path (Z) edge (Y);
			\path[bidirected] (Z) edge[bend right = 60] (X);
			\path[bidirected] (Z) edge[bend left = 60] (Y);
		\end{tikzpicture} 
		\caption*{$G_{\*C_2}$}
\end{subfigure}
\end{minipage}
\begin{minipage}[t]{.32\linewidth}
\begin{subfigure}[b]{\linewidth}
\centering	
\begin{tikzpicture}[scale = .6]
			\node(X) at (0,0) [label = below:$X$, point];
			\node(Y) at (2, 0)[label = below:$Y$, point];
			\node(Z1) at (0, 2)[label=left:{$Z_1$}, point]; 
			\node(Z3) at (2, 2)[label=right:{$Z_3$}, point]; 
			\node(Z2) at (1, 1) {$Z_2$};     %[label=below:{$Z_2$}, point]; 
			\path (X) edge (Y);
			\path (Z1) edge (Z2);
			\path (Z3) edge (Y);
			\path (Z3) edge (Z2);
			%\path (Z3) edge (Y);
			\path[bidirected] (Z1) edge[bend left =20] (Z3);
			\path[bidirected] (Z3) edge[bend left =60] (Y);
			\path[bidirected] (Z1) edge[bend right =60] (X);
			%\draw[gray, fill=gray!30, opacity =.5] (-.1,1.7) rectangle (2.1, 2.3);
	\end{tikzpicture}
		\caption*{$(c)$}
		%\vspace{0.7em}
\end{subfigure}
\end{minipage}	
\begin{minipage}[t]{.32\linewidth}
\begin{subfigure}[b]{\linewidth}
\centering	
\begin{tikzpicture}[scale = .6]
			\node(X) at (0,0) [label = below:$X$, point];
			\node(Y) at (2, 0)[label = below:$Y$, point];
			\node(Z1) at (0, 2)[label=left:{$Z_1$}, point]; 
			\node(Z3) at (2, 2)[label=right:{$Z_3$}, point]; 
			\node(Z2) at (1, 1) {$Z_2$};     %[label=below:{$Z_2$}, point]; 
			\path (X) edge (Y);
			\path (Z1) edge (Z2);
			\path (Z2) edge (Y);
			\path (Z3) edge (Z2);
			%\path (Z3) edge (Y);
			\path[bidirected] (Z1) edge[bend left =20] (Z3);
			\path[bidirected] (Z3) edge[bend left =60] (Y);
			\path[bidirected] (Z1) edge[bend right =60] (X);
			%\draw[gray, fill=gray!30, opacity =.5] (-.1,1.7) rectangle (2.1, 2.3);
	\end{tikzpicture}
		\captionsetup{labelformat=empty}
		\caption*{$(d)$}
\end{subfigure}
\end{minipage}
        \caption{$G_{\*C_1}$ is the C-DAG for diagrams (a) and (b) and $G_{\*C_2}$ is the C-DAG for diagrams (c) and (d), where $\*{Z} = \{Z_1, Z_2, Z_3\}$. $P(y | do(x))$ is identifiable in $G_{\*C_1}$ by backdoor adjustment over $\*Z$ and is not identifiable in $G_{\*C_2}$.} 
	\label{fig:C-EC}
 \vspace{-1\intextsep}	
\end{figure}
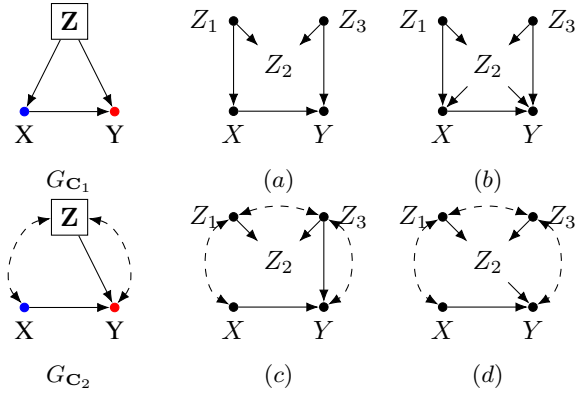 
%\vspace{-0.5\intextsep}
%\end{wrapfigure} 

\textbf{Remark 2.} Although a C-DAG is defined in terms of an underlying graph %causal diagram %\hl{or DAG?} 
$G$, \textit{in practice, one will construct a C-DAG when complete knowledge of the graph $G$ is unavailable. }
As an example of this construction, consider %the diagram in
the model of the effect of lisinopril ($X$) on the outcome of having a stroke ($Y$) in Fig.~\ref{fig:med-example}(a). If not all the relationships specified in Fig.~\ref{fig:med-example}(a) are known, a data scientist cannot construct a full causal diagram, but may still have enough knowledge to create a C-DAG. For instance, they may have partial knowledge that the covariates occur temporally before lisinopril is prescribed, or that a stroke occurs and the suspicion that some of the pre-treatment variables are causes of $X$ and $Y$. 
Specifically, they can create the cluster $\*Z = \{A, B, C, D\}$ with all the covariates, and then construct a C-DAG with edges $\*Z \rightarrow X$ and $\*Z \rightarrow Y$. Further, the data scientist may also suspect that some of the variables in $\*Z$ are confounded with $X$ and others with $Y$, an uncertainty that is encoded in the C-DAG through the bidirected edges $\*Z \dashleftarrow \!\!\!\!\!\!\!\!\! \dashrightarrow X$ and $\*Z \dashleftarrow \!\!\!\!\!\!\!\!\! \dashrightarrow Y$. 
With the additional knowledge that sleep quality ($S$) %is measured after the treatment is applied (i.e., post-treatment), and 
acts as a mediator between the treatment and outcome, the C-DAG in Fig.~\ref{fig:med-example}(b) can be constructed. 
Note that this C-DAG is consistent with the true underlying causal diagram in Fig.~\ref{fig:med-example}(a), but was constructed without knowing this diagram and using much less knowledge than what is encoded in it. %\xst{Partial knowledge of this sort couldn't be formally articulated and used up to this point.} 
Alternatively, if clusters $\*W = \{S, B\}$ and $\*Z = \{A, C\}$ are created, then the C-DAG shown in Fig.~\ref{fig:med-example}(c) would be constructed. Note that both (a) and (b) are considered valid C-DAGs because no cycles are created. Finally, if a clustering with $\*W = \{S, B\}$ and $\*Z = \{A, C, D\}$ is created,  this would lead to the C-DAG shown in Fig.~\ref{fig:med-example}(d), which is invalid. The reason is that a cycle $X \rightarrow \*W \rightarrow \*Z \rightarrow X$ is created due to the connections $X \rightarrow S$, $B \rightarrow C$, and $D \rightarrow X$ in the diagram (a).

%Note that if we consider partition  $\{\{X\}, \{Y\}, \*Z, \*W\}$, where $\*W = \{S, B\}$ and $\*Z = \{A, C, D\}$, this would lead to the graph shown in Fig.~\ref{fig:med-example}(d). The edge $\{X\} \rightarrow \*{W}$ is present since $S \in \*W$ is a function of (i.e., listens to)  $X$; the edge $\*W \rightarrow \*Z$ is present because $C \in \*Z$ listens to $B \in \*W$; the edge $\*Z \rightarrow X$ is present since $X$ listens to $D \in \*Z$. These edges together create a cycle, therefore $G_{\*C''}$ is not a valid C-DAG.

\begin{figure}
\centering
%\vspace{-2\intextsep}
\includegraphics[width=0.74\linewidth]{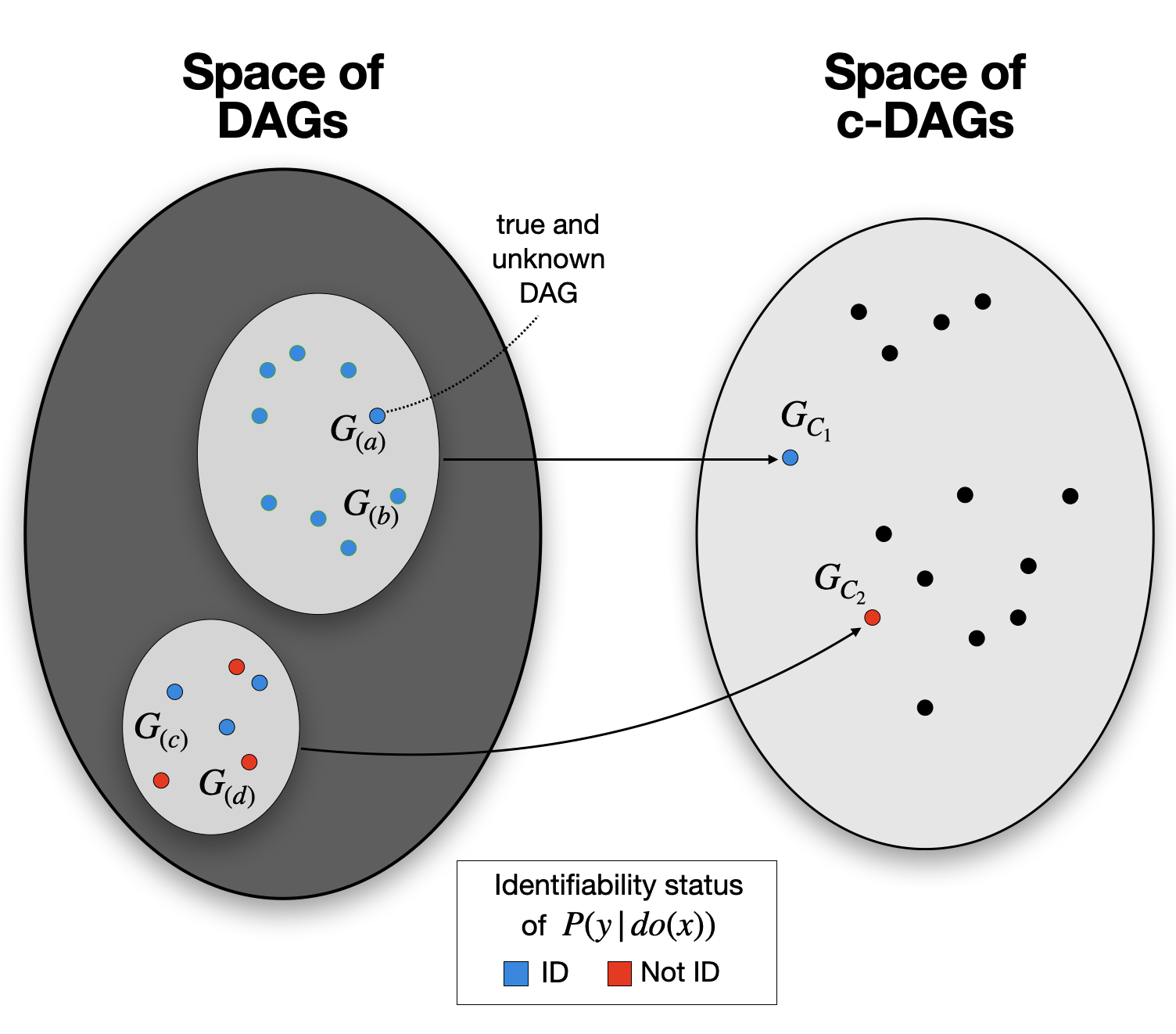}
\caption{Identifying $P(y | do(x))$ in a C-DAG  means identifying such an effect for the entire equivalence class. In $G_{\*C_1}$, the effect is identifiable (blue) because it is identifiable in $G_{(a)}$, $G_{(b)}$, and all the other causal diagrams represented. In $G_{\*C_2}$, the same effect is non-identifiable (red), as the encoded partial knowledge is compatible with some causal diagrams in which the effect is not identifiable (e.g., $G_{(d)}$).}
\vspace{-1\intextsep}
\label{fig:clusterEC}
\end{figure}

\textbf{Remark 3.} It is important to note that a C-DAG $G_{\*{C}}$ as defined in Def.~\ref{def:cdag} is merely a graph over clusters of nodes ${\*C_1, ..., \*C_k }$, and does not have a priori the semantics and properties  of a BN or CBN over macro-variables $\*C_i$. It's not clear, for example, whether the cluster nodes $\*C_i$ satisfy the Markov properties w.r.t. the graph $G_{\*{C}}$. Rather, a C-DAG can be seen as a graphical representation of an equivalence class (EC, for short) of graphs that share the relationships among the clusters while allowing for any possible relationships among the variables within each cluster. 
For instance, in Fig.~\ref{fig:C-EC}, the diagrams (a) and (b) can be represented by C-DAG $G_{\*{C}_1}$ (top) and can, therefore, be thought of as being members of an EC represented by $G_{\*{C}_1}$. The same can be concluded for diagrams  (c) and (d), both represented by C-DAG $G_{\*{C}_2}$. 
The graphical representation of this ECs are shown in Fig.~\ref{fig:clusterEC}, where on the left we have the space of all possible ADMGs, and on the right the space of C-DAGs. 
% we identify an effect in a C-DAG (e.g., the C-DAG $G_{\*C_1}$ in Fig.~\ref{fig:C-EC}) if it is identifiable in all compatible causal diagrams (i.e., diagrams $(a)$, $(b)$ and all other diagrams compatible with $G_{\*C_1}$).

%The semantics of a C-DAG must account for all these possible underlying models.
 %For instance, the two causal diagrams $G_1$ in Fig.~\ref{fig:C-EC}(a) and $G_2$ in (b) can be represented by the C-DAG $G_{\*{C}}$ in (c), and can therefore be thought of as being members of the equivalence class represented by $G_{\*{C}}$.

%\begin{wrapfigure}{r}{0.5\textwidth}
%\end{wrapfigure}

Given the semantics of a C-DAG as an equivalence class of ADMGs, what valid inferences can one perform about the cluster variables given a C-DAG $G_{\*{C}}$? %Can these inferences be performed directly, without enumerating all the possible underlying ADMGs? 
What properties of C-DAGs are shared by all compatible ADMGs?

In principle, definite conclusions from C-DAGs can only be drawn  about properties shared among all EC's members. Going back to Fig.~\ref{fig:clusterEC}, we identify an effect in a C-DAG (e.g., $G_{\*C_1}$ in Fig.~\ref{fig:C-EC}) whenever this effect is identifiable in all members of the EC; e.g., causal diagrams $(a)$, $(b)$, and all other diagrams compatible with $G_{\*C_1}$.
Note that in this particular EC, all dots are marked with blue, which means that the effect is identifiable in each one of them. 
On the other hand, if there exists one diagram in the EC where the effect is not identifiable, this effect will not be identifiable in the corresponding C-DAG. For instance, the effect is not identifiable in the C-DAG $G_{\*C_2}$ due to diagram $(d)$ in Fig.~\ref{fig:C-EC}.

%\xst{Interestingly enough, it's certainly possible that the effect is identifiable in the true, underlying causal diagram, but identifiability will not be warranted if there exists another member of the EC such that this effect is not identifiable. In fact, this suggests a trade-off of expressiveness versus identifiability power. On the one hand, C-DAGs are more expressive than causal diagrams allowing just partial specification of knowledge to be articulated, while causal diagrams are more brittle, requiring full knowledge among all pairs of variables. On the other hand, the identification status of any query will be more likely to be positive in the case of causal diagrams than in C-DAGs, since they are much stricter. The modeling task requires one to find a balance between the amount of knowledge put in the model (C-DAGs) in a way such that identification may be achieved. }

Once the semantics of C-DAGs is well-understood, we turn our attention to computational issues. One naive approach to causal inference with cluster variables, e.g. identifying $Q = P(\*C_i|do(\*C_k))$, goes as follows --  first enumerate all causal diagrams compatible with $G_{\*{C}}$; then, evaluate the identifiability of $Q$ in each diagram; finally, output $P(\*C_i|do(\*C_k))$  if all the diagrams entail the same answer, otherwise output ``non-identifiable''. However, in practice, this approach is intractable in high-dimensional settings -- given a cluster $\*C_i$ of size $m$, the number of possible causal diagrams over the variables in $\*C_i$ is super-exponential in $m$. Can valid inferences be performed about cluster variables using C-DAGs directly, without going through exhaustive enumeration? What properties of C-DAGs are shared by all the compatible causal diagrams? The next sections present theoretical results to address these questions.

Finally, note that not all properties of C-DAGs are shared across all compatible diagrams. To illustrate, consider the case of \textit{backdoor paths}, i.e., paths between $X$ and $Y$ with an arrowhead into $X$, in Fig.~\ref{fig:C-EC}. 
The path $X \dashleftarrow \!\!\!\!\!\!\!\!\! \dashrightarrow \*Z \rightarrow Y$ in $G_{\*{C}_2}$ is active when not conditioning on $\*Z$. However, the corresponding backdoor paths in diagram  (c) are all inactive.
%the \textit{backdoor path} $X \leftarrow \*Z \rightarrow Y$, i.e., a path between $X$ and $Y$ with an arrowhead into $X$, is active in the C-DAG $G_{\*{C}}$ in Fig.~\ref{fig:C-EC}(c).  All such paths are active in $G_2$ in (b), but the only backdoor path, $X \leftarrow Z_1 \rightarrow Z_2 \leftarrow Z_3 \rightarrow Y$, in $G_1$ in (a) is inactive. 
Therefore, a d-connection in a C-DAG does not necessarily correspond to a d-connection in all diagrams in the EC.

\section{C-DAGs for $\mathcal{L}_1$-Inferences}%: d-Separation}
%\section{D-Separation and do-Calculus in C-DAGs}
\label{sec:dsep}

%\hl{Move the next part to the section on Layer 1 inferences... }
%On the other hand, we show in the next section a pleasant and surprising result (Theorem \ref{thm:dsep_cdag_dag}) that d-separations in a C-DAG %as defined next in Definition \ref{def:dsepcdag} 
%do hold in all compatible causal diagrams. 
%This powerful result is indeed critical to deriving causal inference rules and algorithms that are applicable to \textit{all} the causal diagrams compatible with a given C-DAG (Theorems \ref{thm:docalc-cdags}-\ref{thm:ids}) regardless of the unknown relationships within each cluster.
%\hl{Provide a better introduction for this section explaining what are BN and that dseps implies c.i.}

In this section, we study probabilistic inference with C-DAGs - $\mathcal{L}_1$ inferences. We assume the underlying graph $G$ over $\*V$ is a Bayesian Network (BN) \emph{with no causal interpretation}.\footnote{For a more detailed discussion on the tension between layers $\mathcal{L}_1$ and $\mathcal{L}_2$, please refer to  \cite[Sec.~1.4.1]{bar:etal2020}.  } We aim to perform probabilistic inferences about macro-variables with $G_{\*C}$ that are valid in $G$ regardless of the unknown relationships within each cluster.

First, we extend d-separation \cite{pearl:88a}, a fundamental tool in probabilistic reasoning in BNs, to C-DAGs.  % In particular, we show that d-separations in a C-DAG hold in all compatible DAGs. 
As noticed earlier, a d-connecting path in a C-DAG does not necessarily imply that the corresponding paths in a compatible ADMG $G$ are connecting. Such paths can be either active or inactive. However, d-separated paths in a C-DAG correspond to only d-separated paths  in all compatible ADMGs.\footnote{In Appendix \ref{app:pathanalysis}, we investigate in detail how path analysis is extended to C-DAGs. We note that d-separation in ADMGs has also been called m-separation \citep{richardson2003admg}.}
These observations together, lead to the following definition where the symbol $\ast$ represents either an arrow head or tail:

\begin{definition}[\textbf{d-Separation in C-DAGs}] 
\label{def:dsepcdag}
A path $p$ in a C-DAG $G_\*C$ is said to be d-separated (or blocked) by a set of clusters $\mathbf{Z} \subset \*C$ if and only if $p$ contains a triplet
\begin{enumerate}
    \itemsep0em 
    \item 
    $\*C_i\ast\!\! - \!\!\ast \*C_m  \rightarrow \*C_j$
    %$\*{X} \ast\!\! - \!\!\ast \*{Z} \rightarrow \*{Y}$ 
    such that the non-collider cluster $\*C_m$ is in $\*Z$, or 
    \item $\*C_i \ast\!\!\!\rightarrow \*C_m \leftarrow\!\!\!\ast \*C_j$ such that the collider cluster $\*C_m$ and its descendants are not in $\*Z$.
\end{enumerate}
A set of clusters $\*Z$ is said to d-separate two sets of clusters $\*X, \*Y \subset \*C$, denoted by $(\*X \indep \*Y \mid \*Z)_{G_{\*C}}$, if and only if $\*Z$ blocks every path from a cluster in $\*X$ to a cluster in $\*Y$.
\end{definition}

%As highlighted previously, %the $\mathcal{L}_1$ optimality
%Thm.~\ref{thm:cdag_bn} does not guarantee that a dependence implied by a C-DAG $G_\mathbf{C}$ is also implied by any compatible DAG $G$. In other words, 
%a d-connecting path in a C-DAG not necessarily implies that the corresponding paths in a compatible DAG are connecting. Such paths can be either active or inactive. However, it guarantees that any conditional independence implied by a C-DAG $G_\mathbf{C}$ is also implied by any compatible DAG $G$. In other words, d-separated paths in a C-DAG correspond to only d-separated paths in all compatible DAGs.\footnote{In Appendix \ref{app:pathanalysis}, we investigate in detail how path analysis is extended to C-DAGs.}

%Specifically, we show in 

We show in the following proposition that  the d-separation rules are  sound and complete in C-DAGs in the following sense: 
whenever a d-separation holds in a C-DAG, %according to these rules,
it holds for all ADMGs compatible with it; on the other hand, if a d-separation does not hold in a C-DAG, then there exists at least one ADMG compatible with it for which the same d-separation statement does not hold. 

\begin{restatable}{theorem}{dsepcdags}(\textbf{Soundness and completeness of d-separation in C-DAGs}).
\label{thm:dsep_cdag_dag}
%Consider 
In a C-DAG $G_{\*{C}}$, let $\*{X}, \*{Z}, \*{Y} \subset \*{C}$. If $\*{X}$ and $\*{Y}$ are d-separated by $\*{Z}$ in $G_{\*{C}}$, then, in any ADMG $G$ compatible with $G_{\*{C}}$,
%letting $\*{X}$, $\*{Y}$, $\*{Z}$ be the corresponding variables contained in $\*{X}, \*{Z}, \*{Y}$, then 
$\*{X}$ and $\*{Y}$ are d-separated by $\*{Z}$ in $G$, i.e., % that is, 
\begin{align}
(\*X \indep \*Y \mid \*Z)_{G_{\*C}} \implies (\*X \indep \*Y \mid \*Z)_{G}.  
\end{align} 
%$\*{X} = \bigcup \*{X_C}$ and $\*{Y} = \bigcup \*{Y_C} $ are d-separated by $\*{Z} = \bigcup \*{Z_C}$ where $\*X, \*Y, \*Z \subset \*V$ in  $G$.
If $\*{X}$ and $\*{Y}$ are not d-separated by $\*{Z}$ in $G_{\*{C}}$, then, there exists an ADMG $G$ compatible with $G_{\*{C}}$ where $\*{X}$ and $\*{Y}$ are not d-separated by $\*{Z}$ in $G$. 
\end{restatable}
Theorem \ref{thm:dsep_cdag_dag} implies that $G_\mathbf{C}$ does not imply any conditional independence  that is not implied by the underlying $G$. %(i.e., no additional assumption is made at the macro level) 

ADMGs are commonly used to represent BNs with latent variables (which may imply Verma constraints on $P(v)$ not captured by independence relationships \citep{tian:pea02testable-implications}) %. An ADMG $G$ is a BN if 
where the observational distribution $P(\*{v})$ factorizes according to the ADMG $G$ as follows
%\footnote{One subtlety is  ADMGs may imply Verma constraints on $P(v)$ not captured by independence relations \citep{tian:pea02testable-implications}.}
\begin{align}
P(\*v) = \sum_{\*u} P(\*u) \prod_{k:V_k \in \*V } P(v_k | pa_{v_k}, \*u_k),
\label{eq:g-factor}
\end{align}
where $Pa_{V_k}$ are the endogenous parents of $V_k$ in $G$ and $\*U_k \subseteq \*U$ are the latent parents of $V_k$.
%\footnote{One important subtlety is that ADMGs may imply Verma constraints on $P(v)$ not captured by independence relationships \citep{tian:pea02testable-implications}.}
 %\footnote{One important subtlety is that  d-separation itself does not automatically hold in DAGs with latent variables, or ADMGs.} 
 We show next  that the observational distribution $P(\*{v})$ factorizes according to the graphical structure of the C-DAG $G_\mathbf{C}$ as well. % can be treated as a BN with latent variables over macro-variables $\*C$.  

\begin{restatable}{theorem}{cdagsbn}(\textbf{C-DAG as BN}).
% \begin{theorem}[C-DAG as BN]
\label{thm:cdag_bn}
%If $G(\*V,\*E)$ is a BN over $\*V$, then the C-DAG $G_\mathbf{C}(\*C, \*E_{\*C})$ constructed w.r.t. $G$ as Def.~\ref{def:cdag} is a BN over $\*C$. Formally, 
%Let $G_\mathbf{C}(\*C, \*E_{\*C})$ be a C-DAG constructed w.r.t. $G(\*V,\*E)$ as Def.~\ref{def:cdag}. 
Let $G_\*{C}$  be a C-DAG compatible with an ADMG $G$. If the observational distribution $P(\*{v})$ factorizes according to $G$ by Eq.~(\ref{eq:g-factor}),  
%\begin{align}
%P(\*v) = \sum_{\*u} P(\*u) \prod_{k:V_k \in \*V } P(v_k | pa_{v_k}, \*u_k),
%\label{eq:g-factor}
%\end{align}
%where $Pa_{V_k}$ are the endogenous parents of $V_k$ in $G$ and $\*U_k \subseteq \*U$ are the latent parents of $V_k$, 
then %for any admissible partition $\mathbf{C}$ of $\mathbf{V}$, 
the observational distribution $P(\mathbf{v}) = P(\mathbf{c})$ factorizes according to $G_\mathbf{C}$, i.e.,
\begin{align}
P(\*c) = \sum_{\*u} P(\*u) \prod_{k:\*C_k \in \*C } P(\*c_k | pa_{\*C_k}, \*u'_k), \label{eq:g-factorc}
\end{align}
where $Pa_{\*C_k}$ are the parents of the cluster $\*C_k$, %in $G_\*{C}$ 
and $\*U'_k \subseteq  \*U$ such that, for any $i, j$, $\*U'_i \cap \*U'_j \neq \emptyset$ if and only if there is a bidirected edge ($\*C_i \dashleftarrow \!\!\!\!\!\!\!\!\! \dashrightarrow \*C_j$) between $\*C_i$ and $\*C_j$ in $G_\*{C}$. 

\end{restatable}
%In Eq.~(\ref{eq:g-factorc}) and ,  $\*{X}, \*{C}, \*{C}_k, Pa_{\*C_k}$ are %sets of cluster nodes in $G_\*{C}$  while in the distribution $P(.)$ statements, $\*{X}, \*{Y}, \*{Z}, \*{W}$ represent the sets of variables contained in the corresponding sets of clusters. 
Thm.~\ref{thm:cdag_bn} implies that 
if the underlying ADMG $G$ represents a BN with latent variables over $\*V$, then the C-DAG $G_\mathbf{C}$ represents a BN over micro-variables $\*C$.
%The aforementioned d-separation criterion allows us to decide from a C-DAG $G_\*C$ whether two sets of clusters $\*X$ and $\*Y$ are d-separated given a third set $\*Z$ in \emph{any} compatible causal diagram $G$. This will be essential to establishing conditions for causal identifiability in C-DAGs.

\section{C-DAGs for $\mathcal{L}_2$-Inferences} %: do-Calculus and Causal Effect Identification}
\label{sec:do-calculus}

%\hl{Provide an introduction for this section explaining that causal inferences can be done from the causal diagram. This requires now a causal semantics, which will follow from the construction of $G_C$ from a causal diagram $G$}.

We study now interventional ($\mathcal{L}_2$) reasoning with C-DAGs. We assume the underlying graph $G$ over $\*V$ is a CBN. Our goal is to perform causal reasoning about macro-variables  with $G_{\*C}$ that are guaranteed to be valid in each $G$ of the underlying EC. We focus on extending to C-DAGs Pearl's celebrated \textit{do-calculus} \citep{pearl:95a} and the ID-algorithm \citep{tian:02thesis, shpitser:pea06a,huang:val06-complete}. % for identifying causal effects. 

%We aim to perform causal inferences with $G_{\*C}$. Armed with the understanding coming from the d-separation rules in C-DAGs, 
%First we proof the following $\mathcal{L}_2$-optimality result:

%\begin{proposition}[\textbf{$\mathcal{L}_2$-optimality of C-DAGs}]
%\label{prop:optL2}
%For any admissible partition $\mathbf{C}$ of $\mathbf{V}$ and any $\*X \subseteq \*C$, 
%the interventional distribution $P(\*v\setminus \*x | do(\*x)) = P(\*c\setminus \*x | do(\*x))$ factorizes according to 
%a causal C-DAG $G_\mathbf{C}$ constructed following Def.\ref{def:cdag} w.r.t. a causal diagram $G$. 
%\hl{add here truncated factorization}
%Moreover, $G_\mathbf{C}$ does not imply any causal or confounding relationship among clusters in $\*C$ that is not implied by $G$. %(i.e., no additional assumption is made at the macro level) 
%\end{proposition}

%%%%%%%%%%%%
%, known as the truncated factorization product, i.e.:%. 
% Let $G_\*{C}$ %(\*{C}, \*{E}_\*{C})$ 
%be a C-DAG compatible with a causal diagram $G$ associated with an SCM $\mathcal{M}=\langle \*U, \*V, \mathcal{F}, P(\*U)\rangle$. For any $\*X \subseteq \*C$, the following holds
% then representing the equivalence class of all compatible CBNs over the variables $\*V$.  %where a cluster with $N$ variables is treated as an $N$-dimensional random variable. 

\subsection*{Do-Calculus in C-DAGs}
\label{sec:docalc-cdags}
Do-calculus is a fundamental tool in causal inference from causal diagrams and has been used extensively for solving a variety of identification tasks.  We show that if the underlying ADMG $G$ is a CBN on which do-calculus rules hold, then do-calculus rules are valid in the corresponding C-DAG $G_C$.  
We first present a key lemma for proving the soundness of do-calculus in C-DAGs that the mutilation operations in a C-DAG to create $G_{\*C_{\underline{\*{X}}}}$ and $G_{\*C_{\overline{\*{X}}}}$ carry over to all compatible underlying ADMGs. This result is shown in the following:

\begin{restatable}{lemma}{mutilation}
\label{lem:graph_mutilation} 
%If a C-DAG $G_\*C$ is compatible with a causal diagram $G$, then, for $\*X_\*C, \*Z_\*C \in \*C$, $G_{\*{C}_{\overline{\*{X_\*C}} \underline{\*{Z_\*C}}}}$ is compatible with $G_{\overline{\*{X}}\underline{\*{Z}}}$, where $\*X = \bigcup \*X_\*C$ and $\*Z = \bigcup \*Z_\*C$.
If a C-DAG $G_\*C$ is compatible with an ADMG $G$, then, for $\*X, \*Z \subset \*C$, the mutilated C-DAG $G_{\*{C}_{\overline{\*{X}} \underline{\*{Z}}}}$ is compatible with the mutilated ADMG $G_{\overline{\*{X}}\underline{\*{Z}}}$. % where $\*X, \*Z$ are the set of variables in the corresponding set of clusters. 
\end{restatable}

%One of the 
%A fundamental tool used in causal inference is Pearl's celebrated \textit{do-calculus} \cite{pearl:95a}. The do-calculus has been used extensively for solving a variety of causal effect identification tasks. 
 %we show next that the do-calculus rules are also sound in C-DAGs.
%Pearl's do-calculus is a fundamental causal inference tool and has been widely used for solving a variety of causal effect identification tasks. Armed with the d-separation rules in C-DAGs, we show in Theorem \ref{thm:docalc-cdags} that  the  do-calculus rules in causal diagrams carry over to the C-DAGs.
%Given these d-separation rules, we will show in Theorem \ref{thm:docalc-cdags} that we can use the existing do-calculus rules for determining causal effects in cluster causal diagrams. Possibly active paths should be considered active when trying to establish a d-separation necessary to apply the rules of do-calculus. 

The soundness of do-calculus in C-DAGs as stated next follows from %Propositions \ref{prop:cdags_connections} and \ref{prop:connection_directed_Gc},
Theorem~\ref{thm:dsep_cdag_dag} and Lemma~\ref{lem:graph_mutilation}. 
%\newpage
\begin{restatable}{theorem}{docalc}(\textbf{Do-calculus in causal C-DAGs}).
\label{thm:docalc-cdags}
Let $G_\*{C}$ %(\*{C}, \*{E}_\*{C})$ 
be a C-DAG compatible with an ADMG $G$. If $G$ is a CBN encoding interventional distributions $P(\cdot|do(\cdot))$, then  %associated with an SCM $\mathcal{M}$. %and let $P$ be the probability distribution induced by $\mathcal{M}$. 
for any disjoint subsets of clusters $\*{X}, \*{Y}, \*{Z}, \*{W} \subseteq \*{C}$, the following three rules hold: 

$$
\begin{aligned}
    &\textbf{Rule 1:} \ P(\*y|do(\*x), \*z, \*w) = P(\*y | do(\*x), \*w) \\
    & 
     \qquad \qquad \qquad
     \text{if } (\*{Y} \indep \*{Z} | \*{X}, \*{W})_{G_{\*{C}_{\overline{\*X}}}}  \\ 
\end{aligned}
$$

$$
\begin{aligned}
    & \textbf{Rule 2:} \ P(\*{y} | do(\*{x}), do(\*{z}), \*{w}) = P(\*{y} | do(\*{x}), \*{z}, \*{w}) \\
     & 
     \qquad \qquad \qquad
     \text{if } (\*{Y} \indep \*{Z} | \*{X}, \*{W})_{G_{\*{C}_{\overline{\*{X}} \underline{\*{Z}}}}} \\
    & \textbf{Rule 3:} \ P(\*y | do(\*x), do(\*z), \*w) = P(\*y | do(\*x), \*w) \\
    & 
    \qquad \qquad \qquad
    \text{if } (\*Y \perp \! \!\! \perp \*Z | \*X, \*W)_{G_{\*{C}_{\overline{\*X}\overline{\*Z(\*W)}}}}\\
\end{aligned}
$$
where $G_{\*{C}_{\overline{\*X}\underline{\*Z}}}$ is obtained from $G_{\*{C}}$ by removing the edges %incoming to $\*X$ and outgoing from $\*Z$,
into $\*X$ and out of $\*Z$,
and $\*Z(\*W)$ is the set of $\*Z$-clusters that are non-ancestors of any $\*W$-cluster in $G_{\*{C}_{\overline{\*X}}}$.
\end{restatable}

 We also show next that the do-calculus rules in C-DAGs are complete in the following sense:
%in the following sense:
%\footnote{\xcomment{EB: This seems a local version of completeness, or completeness of the rules. We usually say in the context of do-calc completeness for a certain task. Shouldn't this be contextualized? }} %Theorem \ref{thm:do-calc_completenes} shows the completeness of the do-calculus rules for C-DAGs. 
%\ref{thm:do-calc_soundness} ensures that
%whenever a rule of the do-calculus does not apply for a C-DAG, $G_{\*{C}}$, it also applies for every causal diagram $G$ compatible with $G_{\*{C}}$. Theorem 
%that whenever a rule of do-calculus does not apply for a C-DAG, $G_{\*{C}}$, there exists a causal diagram $G$ compatible with $G_{\*{C}}$ for which it also does not apply.

% Revisiting the example diagrams in Fig. ~\ref{fig:ignorabilitycluster} where we have a cluster $\*Z = \{Z_1, Z_2\}$

\begin{restatable}{theorem}{calccomplete}
\label{thm:do-calc_completenes}
(\textbf{Completeness of do-calculus}). %The rules of the do-calculus applied to C-DAGs are complete, in the sense that, 
If in a C-DAG $G_{\*{C}}$ a do-calculus rule does not apply, then there is a CBN $G$ compatible with $G_{\*{C}}$ for which it also does not apply.
\end{restatable}

\subsection*{Truncated Factorization %Product
\label{sec:truncate}}
An ADMG $G$ represents a CBN if the interventional distributions factorizes according to the graphical structure, known as the truncated factorization, i.e., for any $\*X \subseteq \*V$%. Specifically, 
\begin{align}
P(\*v\setminus \*x | do(\*x)) = \sum_{\*u} P(\*u) \!\!\! \prod_{k:V_k \in \*V \setminus \*X} \!\!\! P(v_k | pa_{v_k}, \*u_k),
\label{eq:g-tfactor}
\end{align}
where $Pa_{V_k}$ are the endogenous parents of $V_k$ in $G$ and $\*U_k \subseteq \*U$ are the latent parents of $V_k$. 

We show that the truncated factorization holds in C-DAGs as  if the  underlying ADMG is a CBN, in the following sense.
%Next theorem %\ref{thm:markovrelative} essentially 
%shows that C-DAGs can be treated as Causal Bayesian Networks (CBNs) \citep[Def.~16]{bar:etal2020} over macro-variables $\*C$

\begin{restatable}{theorem}{markovrelative}
\label{thm:markovrelative}
(\textbf{C-DAG as CBN})
Let $G_\*{C}$ %(\*{C}, \*{E}_\*{C})$ 
be a C-DAG compatible with an ADMG $G$. If $G$ satisfies the truncated factorization (\ref{eq:g-tfactor}) with respect to the interventional distributions,  
%If $G(\*V, \*E)$ is a causal diagram over $\*V$, then the C-DAG $G_\mathbf{C}(\*C, \*E_{\*C})$ constructed w.r.t. $G$ as Def.~\ref{def:cdag} is a CBN over $\*C$. Formally,
%, for any $\*X \subseteq \*V$, the interventional distribution $P(\*v\setminus \*x | do(\*x))$ factorizes according $G$, i.e.,
%\begin{align}
%P(\*v\setminus \*x | do(\*x)) = \sum_{\*u} P(\*u) \prod_{k:V_k \in \*V \setminus \*X} P(v_k | pa_{v_k}, \*u_k),\label{eq:g-tfactor}
%\end{align}
%where $Pa_{V_k}$ are the endogenous parents of $V_k$ in $G$ and $\*U_k \subseteq \*U$ are the exogenous parents of $V_k$, 
then, for any %admissible partition $\mathbf{C}$ of $\mathbf{V}$ and 
$\*X \subseteq \*C$, the interventional distribution $P(\*c\setminus \*x | do(\*x))$ factorizes according to $G_\mathbf{C}$, i.e., 
\begin{align}
P(\*c\setminus \*x| do(\*x)) = \sum_{\*u} P(\*u) \!\!\!  \prod_{k:\*C_k \in \*C \setminus \*X}\!\!\!  P(\*c_k | pa_{\*C_k}, \*u'_k), \label{eq:tfactor}
\end{align}
where $Pa_{\*C_k}$ are the parents of the cluster $\*C_k$,  %in $G_\*{C}$ 
and $\*U'_k \subseteq  \*U$ such that, for any $i, j$, $\*U'_i \cap \*U'_j \neq \emptyset$ if and only if there is a bidirected edge ($\*C_i \dashleftarrow \!\!\!\!\!\!\!\!\! \dashrightarrow \*C_j$) between $\*C_i$ and $\*C_j$ in $G_\*{C}$. 
\end{restatable}
Theorem \ref{thm:markovrelative} essentially 
shows that a C-DAG $G_\mathbf{C}$ can be treated as a CBN %\citep[Def.~16]{bar:etal2020} 
over the macro-variables $\*C$ if the underlying ADMG is a CBN.
%This powerful result is indeed critical to deriving causal inference rules and algorithms that are applicable to \textit{all} the causal diagrams compatible with a given C-DAG (Theorems \ref{thm:docalc-cdags}-\ref{thm:ids}) regardless of the unknown relationships within each cluster.

\subsection*{ID-Algorithm\label{sec:IDalg}}

Equipped with d-separation, do-calculus, and the truncated factorization in C-DAGs, causal inference algorithms developed for a variety of tasks that rely on a known causal diagram can be extended to C-DAGs \citep{bareinboim:pea16}. In this paper, we consider the problem of identifying causal effects from observational data using C-DAGs.

There exists a complete algorithm to determine whether $P(\*y | do(\*x))$ is identifiable from a causal diagram $G$ and the observational distribution $P(\*V)$ \citep{tian:02thesis, shpitser:pea06a,huang:val06-complete}. 
This algorithm, or ID-algorithm for short, is based on the truncated factorization,  %of the interventional distributions according to the graphical structure, known as the truncated factorization product, i.e.:%. Specifically, 
therefore, Theorem \ref{thm:markovrelative} allows us to prove that the ID-algorithm is sound and complete to systematically infer causal effects from the observational distribution $P(\*V)$ and partial domain knowledge encoded as a C-DAG $G_\*C$.
%\xst{that it is sound to apply the ID-algorithm in C-DAGs, by treating each cluster node as a single variable.} %, is sound and complete.

% Theorem \ref{thm:markovrelative} showed that the truncated factorization of the interventional distribution $P(\*y | do(\*x))$ holds in C-DAGs. This result along with the characterization and tools provided in the previous sections are crucial for proving that the ID-algorithm is sound and complete to systematically infer causal effects from the observational distribution $P(\*V)$ and partial domain knowledge encoded as a C-DAG $G_\*C$. %\xst{that it is sound to apply the ID-algorithm in C-DAGs, by treating each cluster node as a single variable.} %, is sound and complete.
\begin{restatable}{theorem}{idalgsoundcomplete}
\label{thm:ids}
(\textbf{Soundness and Completeness of ID-algorithm}).
The ID-algorithm is sound and complete when applied to a C-DAG $G_\*C$ for identifying causal effects of the form $P(\*y|do(\*x))$ from the observational distribution $P(\*V)$, where $\*X$ and $\*Y$ are sets of clusters in $G_\*C$ . 
\end{restatable}

The ID algorithm returns a formula for identifiable $P(\*y|do(\*x))$ that is valid in all causal diagrams compatible with the C-DAG $G_{\*C}$. The completeness result ensures that if the ID-algorithm fails to identify $P(\*y|do(\*x))$ from $G_{\*C}$, then there exists a causal diagram $G$ compatible with $G_{\*C}$ where the effect $P(\*y|do(\*x))$ is not identifiable. 
Appendix \ref{app:simulations} contains an experimental study evaluating the ability of C-DAGs to accurately assess the identifiability of  effects while requiring less domain knowledge for their construction.

%\hl{Since we are mentioning the most general algorithm for ctf id, should we say something about gID? -- lee:etal19}

%\xst{We have a valid algorithm for inferring causal effects from a combination of an observational distribution $P(\*V)$ and partial domain knowledge encoded as a C-DAG $G_\*C$.}

%\xst{Furthermore, we show in  Appendix \ref{app:calculus} (Theorem~\ref{thm:idalgcomplete}) that the ID-algorithm is not only sufficient but also necessary for identification from observational data (i.e., complete).}

%, and,  therefore, complete. % appendix \xadd{We also prove the completeness of the ID-algorithm in C-DAGs in the appendix.}

% %Given that C-DAGs are Causal Bayesian Networks, it follows from Lemmas  \ref{lem:ancestralsetGR} and \ref{lem:ccompGR} that 
% the ID-algorithm \citep{a} is sound and complete for identifying causal effects in C-DAGs. From a combination of observational distribution $P(\*V)$ and knowledge encoded as a C-DAG $G_\*C$, the ID-algorithm returns an estimand of the causal effect whenever the effect is identifiable in any causal diagrams $G$ compatible with $G_\*C$. If the ID-algorithm fails, then there exists at least one causal diagram compatible with the $G_\*C$ where the effect is not identifiable. 

% \begin{restatable}{theorem}{idalg}
% \label{lem:ccompGR}
% (\textbf{Soundness and completeness of the ID-algorithm})
% The ID-algorithm is sound and complete for the identification of causal effects when applied to C-DAGs. 
% \end{restatable}

\subsection*{Examples of Causal Identifiability in C-DAGs}

We show examples of identification in C-DAGs in practice. Besides illustrating identification of causal effects in the coarser graphical representation of a C-DAG, these examples demonstrate that clustering variables may lead to diagrams where effects are not identifiable. Therefore, care should be taken when clustering variables, to ensure not so much information is lost in a resulting C-DAG, such that identifiability is maintained when possible. 

%Note that, due to the coarsening of the diagram by clusters, it is possible an effect will be identifiable in some causal graph $G$, but will not be identifiable by the clustering yielding $G_{\*{C}}$. 
%This example along with Fig.~\ref{fig:a2_multipleXY} illustrate how any variable, whether a treatment, outcome or another variable in a graph, can be clustered with other variables. 

%By considering the clustering $\*{Z} =\{Z_1, Z_2\}$, any separation between $Z_1$ and $Z_2$ is lost. However, under the assumption of no unobserved confounders relative to $(Z_1, X)$ and $(Z_2, X)$, %as in diagrams (c) and (f), the effect of $X$ on $Y$ is identifiable regardless of the connections between $Z_1$ and $Z_2$. From the C-DAG $G_{\*C_1}$, the effect is given by $P(y| do(x)) = \sum_{\*z} P(y |x, \*z)P(\*z)$. If domain knowledge does not justify the assumption of no unobserved confounders relative to $(Z_1, X)$ or $(Z_2, X)$, then, from the C-DAG $G_{\*C_2}$ of diagrams (a), (b), (d), and (e), the effect is not identifiable. This results from the existence of the graphs compatible with $G_{\*C_2}$ (e.g. (d) and (e)), where the effect is not identifiable.
\vspace{0.3em}
\textbf{Identification in Fig.~\ref{fig:med-example}.} In diagram (a) the effect of $X$ on $Y$ is identifiable through backdoor adjustment \citep[pp.~79-80]{pearl:2k} over the set of variables $\{B, D\}$
%\xst{ because neither $B$ nor $D$ are descendants of $X$ and $\{B, D\}$ block all paths between $X$ and $Y$ with an arrow into $X$}.
In the C-DAG in Fig.~\ref{fig:med-example}(b), with cluster $\*{Z} = \{A, B, C, D\}$,  the effect of $X$ on $Y$ is identifiable through front-door adjustment \citep[p.~83]{pearl:2k} over $S$, given by $P(y | do(x)) = \sum_{s}P(s|x)\sum_{x^\prime}P(y|x^\prime, s)P(x^\prime)$.
%\xst{, since $S$ intercepts the directed path from $X$ to $Y$, there are no unblocked back-door paths from $X$ to $S$, and all back-door paths from $S$ to $Y$ are blocked by $X$}. 
Because this front-door adjustment holds for the C-DAG in Fig.~\ref{fig:med-example}(b) with which diagram (a) is compatible, this front-door adjustment identification formula is equivalent to the adjustment in the case of diagram (a) and gives the correct causal effect in any other compatible causal diagram. In the C-DAG in (c), the loss of separations from the creation of clusters $\*{Z} = \{A, B, C, D\}$ and $\*{W} = \{B, S\}$ render the effect no longer identifiable, indicating that there exists another graph compatible with (c) for which the effect cannot be identified. %Note that it is not possible to block all backdoor paths relative to $(X,Y)$, and $\*W$, a direct effect of $X$, is confounded. 

\begin{figure}[h!]
\begin{center}
\begin{minipage}[t]{.28\columnwidth}
\begin{subfigure}[b]{\linewidth}
\centering
        \begin{tikzpicture}[scale = .5]
			\node(X1) at (0,0) [label = below:$X_1$, point, blue];
			\node(X2) at (1.25,0)[label=below:$X_2$, point, blue];
		    \node(Y1) at (2.5, 0)[label=below:$Y_1$, point, red];
			\node(Y2) at (3.75, 0)[label = below:$Y_2$, point, red];
		    \node(Z1) at (1.25, 1.5)[label=above:$Z_1$, point];
			\node(Z2) at (2.5, 1.5)[label = above:$Z_2$, point];
			\path (X1) edge (X2);
			\path (X2) edge (Y1);
			\path (Y1) edge (Y2);
			\path (Z1) edge (Y1);
			\path (Z2) edge (X2);
			\path[bidirected] (Y1) edge[bend left =30] (Y2); 
			\path[bidirected] (Z1) edge[bend left =0] (X1); 
			\path[bidirected] (Z2) edge[bend left =0] (Y2); 
        \end{tikzpicture}
        \caption*{$(a)$ $G$}
\end{subfigure}
\end{minipage}
\begin{minipage}[t]{.22\columnwidth}
\begin{subfigure}[b]{\linewidth}
    \begin{minipage}[b][1.8cm][t]{\columnwidth}
    \centering
    \begin{tikzpicture}[scale = .5]
        \node[state, blue](X) at (0, 0){$\*X$}; 
 		\node[state, red](Y) at (1.8, 0) {$\*Y$};
		\node[state](Z) at (0.9, 2.2) {$\*Z$};
		\path (X) edge (Y);
		\path (Z) edge (X);
		\path (Z) edge (Y);
		\path[bidirected] (Z) edge[bend left =30] (Y); 
		\path[bidirected] (Z) edge[bend right =30] (X); 
    \end{tikzpicture}
    \end{minipage}
    \caption*{$(b)$ $G_{\*C_1}$}
\end{subfigure}
\end{minipage}
\begin{minipage}[t]{.22\columnwidth}
\begin{subfigure}[b]{\linewidth}
    \begin{minipage}[b][1.8cm][t]{\columnwidth}
    \centering
    \begin{tikzpicture}[scale = .5]
        \node[state, blue](X) at (0, 0){$\*X$}; 
 		\node[state, red](Y) at (2.5, 0) {$\*Y$};
	    \node(Z1) at (0.625, 1.5)[label=above:$Z_1$, point];
		\node(Z2) at (1.775, 1.5)[label = above:$Z_2$, point];
		\path (X) edge (Y);
		\path (Z1) edge (Y);
		\path (Z2) edge (X);
		\path[bidirected] (Z2) edge[bend left =0] (Y); 
		\path[bidirected] (Z1) edge[bend right =0] (X); 
    \end{tikzpicture}
    \end{minipage}
    \caption*{$(c)$ $G_{\*C_2}$}
\end{subfigure}
\end{minipage}
\begin{minipage}[t]{.22\columnwidth}
\centering
\begin{subfigure}[b]{\linewidth}
    \begin{minipage}[b][1.8cm][t]{\columnwidth}
    \centering
    \begin{tikzpicture}[scale = .5]
		\node(X1) at (0,0) [label = below:$X_1$, point, blue];
		\node(X2) at (1.25,0)[label=below:$X_2$, point, blue];
 		\node[state, red](Y) at (2.5, 0) {$\*Y$};
		\node[state](Z) at (1.25, 2.2) {$\*Z$};
		\path (X1) edge (X2);
		\path (X2) edge (Y);
		\path (Z) edge (X2);
		\path (Z) edge (Y);
		\path[bidirected] (Z) edge[bend left =30] (Y); 
		\path[bidirected] (Z) edge[bend right =30] (X1); 
    \end{tikzpicture}
    \end{minipage}
    \caption*{$(d)$ $G_{\*C_3}$}
\end{subfigure}
\end{minipage} 
		\caption{$(a)$: causal diagram $G$ where the effect $P(y_1, y_2 | do(x_1, x_2))$ is identifiable. $(b)$: C-DAG $G_{\*C_1}$ 
% 		resulting from 
		with clustering $\*X = \{X_1, X_2\}$, $\*Y = \{Y_1, Y_2\}$, and $\*Z = \{Z_1, Z_2\}$. $(c)$: C-DAG $G_{\*C_2}$ 
% 		resulting from 
		with clustering $\*X = \{X_1, X_2\}$ and $\*Y = \{Y_1, Y_2\}$. $(d)$: C-DAG $G_{\*C_3}$ 
% 		resulting from 
		with clustering $\*Y = \{Y_1, Y_2\}$ and $\*Z = \{Z_1, Z_2\}$. The effect $P(\*y | do(\*x))$ is not identifiable in $G_{\*C_1}$, but 
% 		$P(\*y | do(\*x))$ 
		is identifiable in $G_{\*C_2}$ and $P(\*y | do(x_1, x_2))$ is identifiable in $G_{\*C_3}$.}
		 \label{fig:a2_multipleXY}
\end{center}
\vspace{-0.5\intextsep}
\end{figure}
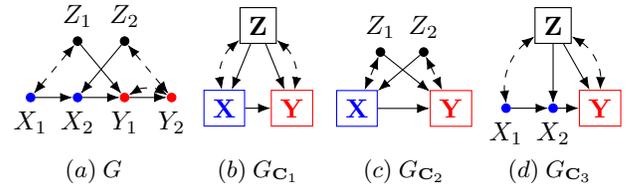

\textbf{Identification in Fig.~\ref{fig:a2_multipleXY}.} In causal diagram (a), the effect of %the joint intervention of 
$\{X_1, X_2\}$ on %both outcomes 
$\{Y_1, Y_2\}$ is identifiable by backdoor adjustment over $\{Z_1, Z_2\}$ as follows: $P(y_1, y_2|do(x_1, x_2)) = \sum_{z_1, z_2} P(y_1, y_2|x_1, x_2, z_1, z_2) P(z_1, z_2)$. Note, however, that the backdoor path cannot be blocked in the C-DAG $G_1$ (b) with clusters $\*X = \{X_1, X_2\}$, $\*Y = \{Y_1, Y_2\}$, and $\*Z = \{Z_1, Z_2\}$. In this case, the effect $P(\*y|do(\*x))$ is not identifiable. If the covariates $Z_1$ and $Z_2$ are not clustered together as shown in the C-DAG $G_{\*C_2}$ (c), the backdoor paths relative to $\*X$ and $\*Y$ can still be blocked despite the unobserved confounders between $Z_1$ and $\*X$ and between $Z_2$ and $\*Y$. So the effect $P(\*y|do(\*x))$ is identifiable by backdoor adjustment over $\{Z_1, Z_2\}$ as follows: $P(\*y|do(\*x)) = \sum_{z_1, z_2} P(\*y|\*x, z_1, z_2) P(z_1, z_2)$. If the treatments $X_1$ and $X_2$ are not clustered together as shown in the C-DAG $G_{\*C_3}$ (d), then the joint effect of $X_1$ and $X_2$ on the cluster $\*Y$ is identifiable and given by the following expression:  $P(\*y|do(x_1, x_2)) = \sum_{\*z, x'_1} P(\*y|x'_1, x_2, \*z) P(x'_1, \*z)$.

%\newpage
%\subsection{Causal Identification with Clustered Treatments and Outcomes}
%\label{app:idexample}

%In this section, we provide an additional example of identification of causal effects in C-DAGs where the treatment and outcome variables are clustered. 

\begin{figure}[h!]
\begin{minipage}[t]{.24\columnwidth}
\centering
\begin{subfigure}[b]{\linewidth}
\centering
        \begin{tikzpicture}[scale = 0.5]
			\node(X1) at (0,0) [label = below:$X_1$, point, blue];
			\node(X2) at (1.5, 0)[label=above:$X_2$, point, blue]; 
			\node(Y2) at (3, 0)[label = below:$Y_2$, point, red];
		    \node(Y1) at (1.5, -1)[label=below:$Y_1$, point, red]; 
			\path (X1) edge (X2);
			\path (X2) edge (Y1);
			\path (X2) edge (Y2);
			\path (X1) edge (Y1);
			\path (Y1) edge (Y2);
			\path[bidirected] (X1) edge[bend left =70] (Y2); 
        \end{tikzpicture}
        \caption*{$(a)$ $G_1$}
\end{subfigure}
\end{minipage}
\begin{minipage}[t]{.24\columnwidth}
\centering
\begin{subfigure}[b]{\linewidth}
\centering
        \begin{tikzpicture}[scale = 0.5]
			\node(X1) at (0,0) [label = below:$X_1$, point, blue];
			\node(X2) at (1.5, 0)[label=above:$X_2$, point, blue]; 
			\node(Y2) at (3, 0)[label = below:$Y_2$, point, red];
		    \node(Y1) at (1.5, -1)[label=below:$Y_1$, point, red]; 
			\path (X1) edge (X2);
			\path (X2) edge (Y1);
			\path (X2) edge (Y2);
			\path (X1) edge (Y1);
			\path (Y1) edge (Y2);
			\path[bidirected] (X1) edge[bend left =30] (X2); 
			\path[bidirected] (X1) edge[bend left =70] (Y2); 
        \end{tikzpicture}
        \caption*{$(b)$ $G_2$}
\end{subfigure}
\end{minipage}
\begin{minipage}[t]{.24\columnwidth}
\centering
\begin{subfigure}[b]{\linewidth}
\centering
        \begin{tikzpicture}[scale = 0.5]
			\node(X1) at (0,0) [label = below:$X_1$, point, blue];
			\node(X2) at (1.5, 0)[label=above:$X_2$, point, blue]; 
			\node(Y2) at (3, 0)[label = below:$Y_2$, point, red];
		    \node(Y1) at (1.5, -1)[label=below:$Y_1$, point, red]; 
			\path (X1) edge (X2);
			\path (X2) edge (Y1);
			\path (X2) edge (Y2);
			\path (X1) edge (Y1);
			%\path (Y1) edge (Y2);
			\path[bidirected] (X1) edge[bend left =70] (Y2); 
			\path[bidirected] (Y1) edge[bend right =0] (Y2); 
        \end{tikzpicture}
        \caption*{$(c)$ $G_3$}
\end{subfigure}
\end{minipage}
\begin{minipage}[t]{.24\columnwidth}
\centering
\begin{subfigure}[b]{\linewidth}
    \begin{minipage}[b][1.5cm][t]{\columnwidth}
    \centering
    \begin{tikzpicture}[scale = 0.5]
        \node[state, blue](X) at (0, 2){$\*X$}; 
 		\node[state, red](Y) at (2, 2) {$\*Y$};
		\path (X) edge (Y);
		\path[bidirected] (X) edge[bend left =60] (Y); 
    \end{tikzpicture}
    \end{minipage}
    \caption*{$(d)$ $G_{\*C}$}
\end{subfigure}
\end{minipage}
    % \captionsetup{labelformat=empty}
		\caption{$(a), (b),$ and $(c)$ are causal diagrams compatible with the C-DAG $G_\*C$ in $(d)$ where $\*X = \{X_1, X_2\}$ and $\*Y = \{Y_1, Y_2\}$. The causal effect $P(y_1, y_2 | do(x_1, x_2))$ is identifiable in $(a)$ but not in $(b)$ or $(c)$. Consequently, the effect $P(\*y | do(\*x))$ is not identifiable from the C-DAG $G_\*C$.}
		 \label{fig:a1_multipleXY}
\end{figure}
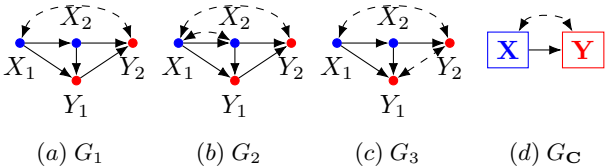

%\textbf{Identification in Fig.~\ref{fig:a1_multipleXY}.} 
\textbf{Identification in Fig.~\ref{fig:a1_multipleXY}}. In the causal diagram (a), the effect of the joint intervention to 
$\{X_1, X_2\}$ on both outcomes 
$\{Y_1, Y_2\}$ is identifiable as follows: $P(y_1, y_2|do(x_1, x_2)) = P(y_1|x_1, x_2)\sum_{x'_1}P(y_2|x'_1, x_2, y_1)P(x'_1)$. By clustering the two treatments as $\*X$ and the two outcomes as $\*Y$, we lose the information that $X_2$ is not a confounded effect of $X_1$ and that $Y_1$ and $Y_2$ are not confounded. If this is the case, as in causal diagrams $G_2$ (b) and $G_3$ (c), the effect would not be identifiable. Note that the C-DAG (d), representing causal diagrams (a), (b), and (c), is the bow graph, where the effect $P(\*y|do(\*x))$ is also not identifiable. 

%\input{sections/simulations}

%Also, Appendix \ref{app:simulations} contains an experimental study illustrating the ability of C-DAGs to accurately assess the identifiability of a causal effect while requiring less domain knowledge for their construction.
\section{C-DAGs for $\mathcal{L}_3$-Inferences} %: Counterfactuals}
\label{sec:ctf}

%In this section, 
Now, we study counterfactual ($\mathcal{L}_3$) inferences in C-DAGs. We assume that the underlying graph $G$ over $\*V$ is induced by an SCM $\mathcal{M}$, and our goal is to perform counterfactual reasoning about macro-variables with $G_{\*C}$ that are always valid in $\mathcal{M}$ (while both $G$ and the SCM $\mathcal{M}$ are unknown).

We show that for any SCM over $\mathbf{V}$ with causal diagram $G$, there is an equivalent SCM over macro-variables $C$ that induces C-DAG $G_C$ and makes the same predictions about counterfactual distributions over the macro-variables.

\begin{restatable}{theorem}{scmcdags}
\label{thm:ctf_cdags}
Let $G_\*{C}$ be a C-DAG compatible with an ADMG $G$. Assume $G$ is induced by an SCM $\mathcal{M}$, then 
%Let $\mathcal{M}$ be an SCM over $\mathbf{V}$ with causal diagram $G(\*V, \*E)$. For any valid C-DAG $G_{\*C}$ constructed w.r.t. $G$ as Def.~\ref{def:cdag}, 
there exists an SCM $\mathcal{M}_{\*C}$ over macro-variables $\*C$ such that its induced causal diagram is $G_{\*C}$ and, for any set of counterfactual variables $\*Y_{\*x}, \dots, \*Z_{\*w}$ where  $\*Y, \*X, \ldots, \*Z, \*W \subseteq \*C$, $P_{\mathcal{M}}(\*y_{\*x}, \ldots, \*z_{\*w}) = P_{\mathcal{M}_{\*C}}(\*y_{\*x}, \ldots, \*z_{\*w})$. %where the random variable $\*Y_\*x$ denotes the potential response of $\*Y$ to the action $do(\*X=\*x)$. 
\end{restatable}

Following this result, algorithms developed for a variety of counterfactual inference tasks that rely on a known causal diagram, such as the CTFID algorithm \citep{correa21nestedctf}, can %potentially 
be used in the context of C-DAGs. 
%For example, the CTFID algorithm \citep{correa21nestedctf} for the identification of nested counterfactuals from an arbitrary combination of observational and experimental distributions can be extended to C-DAGs.  %To prove some completeness of these result is somewhat more involved, since it requires finding one element of the EC in which the counterfactual distribution is not identifiable.  
% Formally,
% \begin{restatable}{corollary}{idstar}
% \label{cor:cifid}
% (\textbf{Soundness and Completeness of CTFID-algorithm}).
% The CTFID algorithm is sound and complete when applied to a C-DAG $G_\*C$ for identifying a counterfactual query $P(\*y_{\*x} \ldots, \*z_{\*w})$ from  an arbitrary combination of observational and experimental distributions , where $\*X$, $\*Y$, $\ldots, \*Z$, $\*W$ are sets of clusters in $G_\*C$ . 
% \end{restatable}

For example, consider the C-DAG $G_{C_1}$ in Fig. \ref{fig:C-EC}, where $X$ is a drug, $Y$ is a disease, and $\*Z$ is a cluster of factors potentially affecting $X$ and $Y$. Suppose that a patient who took the drug ($X = 1$) would like to know what his chances of being cured ($Y=1$) would have been had he not taken the drug ($X=0$). This quantity is defined by $P(Y_{X=0} =1 | X=1)$. The CTFID algorithm applied to $G_{C_1}$ will conclude that $P(Y_{X=0}=1 | X=1)$ is identifiable and given by $\sum_zP(Y=1|X=0, Z=z)P(Z=z|X=1)$. This formula is correct in all ADMGs compatible with $G_{C_1}$, regardless of the relationships within cluster $Z$.

%\textcolor{red}{
After all, we note that inferences in the lower layers assume less knowledge than the higher layers. On the one hand, some results about the lower layers are implied by  the higher layers. For instance, if $G$ is induced by an SCM, then $G$ represents a CBN and a BN, and Thm.~\ref{thm:ctf_cdags} implies that if the underlying $G$ is induced by an SCM, then  $G_{\*C}$ represents a CBN and a BN.   %Thms.~\ref{thm:markovrelative} and ~\ref{thm:cdag_bn}. 
If $G$ is a CBN, then it is necessarily a BN, therefore Thm.~\ref{thm:markovrelative} implies that if the underlying $G$ represents a CBN, then  $G_{\*C}$ represents  a BN. %Thm.~\ref{thm:cdag_bn}. 
On the other hand, if one does not want to commit to the SCM generative process, but can only ascertain that the truncated factorization holds (e.g., for $\mathcal{L}_2$), it's still possible to leverage the machinery developed without any loss of inferential power or making unnecessary assumptions about the upper layers.  

%Thm.~\ref{thm:ctf_cdags} implies that, for any $\*Y, \*X \subseteq \*C$, the potential response $\*Y_\*x$ can be equivalently computed as the solution of $\*Y$ in the submodel $\mathcal{M}_{\*x}$ or in the submodel $\mathcal{M}_{{\*C}_{\*x}}$. This allows us to construct a counterfactual C-DAG by applying to a C-DAG the \texttt{make-cg} algorithm introduced by  \citep{shpitser:pea07} and show that
%counterfactual tools such as the ID$^*$ algorithm \citep{shpitser:pea08-r336} is sound and complete for the task of identification of a counterfactual query from the observational distribution $P(\*V)$ and partial domain knowledge encoded as a C-DAG $G_\*C$. This is formalized in the next result:

%In this case, we can make use of the \texttt{make-cg} algorithm by \citep{shpitser:pea07} for the construction of a counterfactual C-DAGs given a C-DAG and a conjunction of counterfactual events. %, which extends the twin-network approach of \cite{balke:pea94a, balke:pea94b}.

%\hl{Should we introduce here the construction of a counterfactual C-DAG using make-cg, and prove the IC$^*$ for C-DAG?}

%\begin{restatable}{theorem}{idstar}
%\label{thm:ids}
%(\textbf{Soundness and Completeness of ID$^*$-algorithm}).The ID$^*$-algorithm is sound and complete when applied to a C-DAG $G_\*C$ for identifying a counterfactual query $P(\*y_{\*x} \ldots, \*z_{\*w})$ from the observational distribution $P(\*V)$, where $\*X$, $\*Y$, $\ldots, \*Z$, $\*W$ are sets of clusters in $G_\*C$ . 
%\end{restatable}
\section{Conclusions}

Causal diagrams provide an intuitive language for specifying the necessary assumptions for causal inferences. Despite all their power and successes, the substantive knowledge required to construct a causal diagram -- i.e., the causal and confounded relationships among all pairs of variables -- is unattainable in some critical settings found across society, including in the health and social sciences. This paper introduces a new class of graphical models that allow for a more relaxed encoding of knowledge. In practice, when a researcher does not fully know the relationships among certain variables, under some mild assumptions delineated by Def.~\ref{def:cdag}, these variables can be clustered together. (A causal diagram is an extreme case of a C-DAG where each cluster has exactly one variable. )
We prove fundamental results to allow causal inferences within C-DAG's equivalence class, which translate to statements about all diagrams compatible with the encoded constraints. 
We develop the formal machinery for probabilistic, interventional, and counterfactual reasoning in C-DAGs following Pearl's  hierarchy assuming the (unknown) underlying model over individual variables are BN ($\mathcal{L}_1$), CBN ($\mathcal{L}_2$), and SCM ($\mathcal{L}_3$), respectively. 
These results are critical for enabling C-DAGs use in ways comparable to causal diagrams. We hope these new tools  will allow researchers to represent complex systems in a simplified way, allowing for more relaxed causal inferences when substantive knowledge is largely unavailable and coarse.

\clearpage
\newpage
\section*{Acknowledgements}
This work was done in part while Jin Tian was visiting the Simons Institute for the Theory of Computing. Jin Tian was partially supported by NSF grant IIS-2231797. This research was supported by the NSF, ONR, AFOSR, DoE, Amazon, JP Morgan, The Alfred P. Sloan Foundation and the United States NLM T15LM007079.

\bibliography{aaai23}

\clearpage
\appendix
\section{Appendix: Supplemental material for ``Causal Effect Identification in Cluster DAGs''}

This appendix consists of four parts. In Section \ref{app:pathanalysis}, we describe how path analysis is extended to C-DAGs. In Section \ref{app:cdags_cbn} we provide a more detailed explanation on how a C-DAG can be characterized as a Causal Bayesian Network (CBN) over clustered variables, representing possibly multiple underlying CBNs over the original variables. In Section \ref{app:simulations}, we empirically illustrate the validity of C-DAGs as an abstraction of causal diagrams. 
Lastly, in Section \ref{appendix:proofs}, we present the proofs of all results in the paper, namely Propositions \ref{prop:cdags_connections} and \ref{prop:connection_directed_Gc}, Lemmas \ref{lem:noncollider_inactive}, \ref{lem:desc_collider_inactive}, and \ref{lem:graph_mutilation}, and Theorems \ref{thm:dsep_cdag_dag} to \ref{thm:ids}, along with some auxiliary results necessary for those proofs.

\subsection{Path Analysis in C-DAGs}
\label{app:pathanalysis}
Since a C-DAG represents a class of compatible causal diagrams, we must make sure that the status of a path in a C-DAG reflects the status of all corresponding paths in all compatible causal diagrams.

Let $\*{X}$, $\*{Y}$, $\*Z$ be clusters in the C-DAG $G_{\*{C}}$. Let the symbol $\ast$ represent either an arrow head or tail. Let $p$ be a path in $G_{\*{C}}$ between $\*{X}$ and $\*{Y}$ that goes only through $\*Z$.

We say that $p$ is \emph{inactive} in $G_{\*{C}}$ if in all causal diagrams $G$ compatible with $G_{\*{C}}$, all corresponding paths are inactive according to the d-separation rules. Note that if  $p$ is in the form $\*{X}\ast\!\! - \!\!\ast \*Z \rightarrow \*{Y}$ (i.e., $\*Z$ is a non-collider), then the corresponding paths in $G$ are those between $\*{X}$ and $\*{Y}$ going only through variables in $\*Z$ and containing the edge $Z \rightarrow Y$ for some $Z \in \*Z$ and some $Y \in \*Y$. Alternatively, if $p$ is in the form $\*{X}\ast\!\!\! \rightarrow \*Z \leftarrow \!\!\!\ast \*{Y}$ (i.e., $\*Z$ is a collider), then the corresponding paths in $G$ are those between $\*{X}$ and $\*{Y}$ going only through variables in $\*Z$ and containing the edge $X \ast\!\!\!\rightarrow Z$ for some $X \in \*X$ and some $Z \in \*Z$ and the edge $Z' \leftarrow\!\!\!\ast Y$ for some $Z'\in \*Z$ and some $Y \in \*Y$. 

We say $p$ is \emph{either active or inactive} if in some compatible causal diagram $G$ all the paths between $\*{X}$ and $\*{Y}$ going through only variables in $\*Z$ are inactive, while in some compatible causal diagram $G$ there exists at least one active path between $\*{X}$ and $\*{Y}$ going through only variables in $\*Z$. 

Next, we investigate the status of paths in a C-DAG under two possible scenarios for a cluster $\*Z$: 1) as a non-collider, or 2) as a collider or %one of its descendants
a descendant of a collider.

\subsection*{Cluster as a Non-Collider (Chains \& Forks)}

\begin{figure*}[t]
\begin{minipage}[t]{.24\linewidth}
\begin{subfigure}[b]{\linewidth}
\centering	
\begin{tikzpicture}[scale = 0.85]
			\node[state](Z) at (0,0) {$\*{Z}$};
			\node[state](X) at (-1.1, 0){$\*{X}$};%[label = left:X, point];
			\node[state](Y) at (1.1, 0){$\*{Y}$};%[label=right:Y, point]; 
			\path[{Rays[n=6]}-Latex] (X) edge (Z);
            % \path(X) edge (Z);
			\path (Z) edge (Y);
		\end{tikzpicture}
		\vspace{+0.15in}
		\caption*{$(a)$}
        \vspace{+0.1in}
\end{subfigure}
\begin{subfigure}[b]{\linewidth}
		\begin{tikzpicture}[scale = 0.85]
			\node(X) at (-1.7, 0)[label = below:X, point];
			\node(Y) at (1.7, 0)[label=below:Y, point]; 
			\node(Z2) at (0,0) {$Z_2$};
			\node(Z1) at (-.9,0) {$Z_1$};
			\node(Z3) at (.9,0) {$Z_3$};
			\path (X) edge (Z1);
			\path (Z3) edge (Y);
			\path (Z1) edge (Z2);
			\path (Z2) edge (Z3);
		\end{tikzpicture}
		\vspace{+0.01in}
		\caption*{$(b)$}
		\vspace{+0.05in}
\end{subfigure}
\begin{subfigure}[b]{\linewidth}
\centering		
		\begin{tikzpicture}[scale = 0.85]
			\node(X) at (-1.7, 0)[label = below:X, point];
			\node(Y) at (1.7, 0)[label=below:Y, point]; 
			\node(Z2) at (0,0) {$Z_2$};
			\node(Z1) at (-.9,0) {$Z_1$};
			\node(Z3) at (.9,0) {$Z_3$};
			\path (X) edge (Z1);
			\path (Z3) edge (Y);
			\path (Z1) edge (Z2);
			\path (Z3) edge (Z2);
		\end{tikzpicture}
		\vspace{+0.01in}
		\caption*{$(c)$}
\end{subfigure}
\end{minipage}
\hfill\vline\hfill
\begin{minipage}[t]{.24\linewidth}
\begin{subfigure}[b]{\linewidth}
\centering
\begin{tikzpicture}[scale = 0.85]
			\node[state](Z) at (0,0) {$\*{Z}$};
			\node[state](X) at (-1.1, 0){$\*{X}$};%[label = left:X, point];
			\node[state](Y) at (1.1, 0){$\*{Y}$};%[label=right:Y, point]; 
			\path (Z) edge (X);
			\path (Z) edge (Y);
		\end{tikzpicture}
		\vspace{+0.15in}
        \caption*{$(d)$}
        \vspace{+0.1in}
\end{subfigure}
\begin{subfigure}[b]{\linewidth}
\centering
		\begin{tikzpicture}[scale = 0.85]
			\node(X) at (-1.7, 0)[label = below:X, point];
			\node(Y) at (1.7, 0)[label=below:Y, point]; 
			\node(Z2) at (0,0) {$Z_2$};
			\node(Z1) at (-.9,0) {$Z_1$};
			\node(Z3) at (.9,0) {$Z_3$};
			\path (Z1) edge (X);
			\path (Z3) edge (Y);
			\path (Z1) edge (Z2);
			\path (Z2) edge (Z3);
		\end{tikzpicture}
			\vspace{+0.01in}
	        \caption*{$(e)$}
	        \vspace{+0.05in}
\end{subfigure}
\begin{subfigure}[b]{\linewidth}
\centering
		\begin{tikzpicture}[scale = 0.85]
			\node(X) at (-1.7, 0)[label = below:X, point];
			\node(Y) at (1.7, 0)[label=below:Y, point]; 
			\node(Z2) at (0,0) {$Z_2$};
			\node(Z1) at (-.9,0) {$Z_1$};
			\node(Z3) at (.9,0) {$Z_3$};
			\path (Z1) edge (X);
			\path (Z3) edge (Y);
			\path (Z1) edge (Z2);
			\path (Z3) edge (Z2);
		\end{tikzpicture}
		\vspace{+0.01in}
	 \caption*{$(f)$}
\end{subfigure}
\end{minipage}
\hfill\vline\hfill
\begin{minipage}[t]{.24\linewidth}
\begin{subfigure}[b]{\linewidth}
\centering
\begin{tikzpicture}[scale = 0.85]
			\node[state](Z) at (0,0) {$\*{Z}$};
			\node[state](X) at (-1.1, 0){$\*{X}$};%[label = below:X, point];
			\node[state](Y) at (1.1, 0){$\*{Y}$};%[label=below:Y, point]; 
			\path[{Rays[n=6]}-Latex] (X) edge (Z);
			\path[{Rays[n=6]}-Latex] (Y) edge (Z);
		\end{tikzpicture}
\vspace{+0.15in}
\caption*{$(g)$}
\vspace{+0.1in}
\end{subfigure}
\begin{subfigure}[b]{\linewidth}
\centering
		\begin{tikzpicture}[scale = 0.85]
			\node(X) at (-1.5, 0)[label = below:X, point];
			\node(Y) at (1.5, 0)[label=below:Y, point]; 
			\node(Z1) at (0,0) {$Z_1$};
			\node(Z2) at (0,-.9) {$Z_2$};
			\node(Z3) at (.9,-.9) {$Z_3$};
			\path (X) edge (Z1);
			\path (Y) edge (Z1);
			\path (Z1) edge (Z2);
			\path (Z3) edge (Z2);
		\end{tikzpicture}
\caption*{$(h)$}
\end{subfigure}
\begin{subfigure}[b]{\linewidth}
%\vspace{+0.1in}
\centering
		\begin{tikzpicture}[scale = 0.85]
			\node(X) at (-1.7, 0)[label = below:X, point];
			\node(Y) at (1.7, 0)[label=below:Y, point]; 
			\node(Z2) at (0,0) {$Z_2$};
			\node(Z1) at (-.9,0) {$Z_1$};
			\node(Z3) at (.9,0) {$Z_3$};
			\path (X) edge (Z1);
			\path (Y) edge (Z3);
			\path (Z1) edge (Z2);
			\path (Z2) edge (Z3);
		\end{tikzpicture}
		\vspace{+0.01in}
	\caption*{$(i)$}
\end{subfigure}
\end{minipage}
\hfill\vline\hfill
\begin{minipage}[t]{.24\linewidth}
\begin{subfigure}[b]{\linewidth}
\centering
 \begin{tikzpicture}[scale = 0.85]
			\node[state](Z) at (0,-1) {$\*{Z}$};
			\node[state](X) at (-1.2, 0){$\*{X}$};%[label = below:X, point];
			\node[state](Y) at (1.2, 0){$\*{Y}$};%[label=below:Y,
			\node(W)[state] at (0, 0) {$\*{W}$}; %[label=above:W, point]; 
			\path (X)[{Rays[n=6]}-Latex] edge (W);
			\path (Y)[{Rays[n=6]}-Latex] edge (W);
			\path (W) edge(Z);
		\end{tikzpicture} 
\caption*{$(j)$}
%\vspace{+0.1in}
\end{subfigure}
\begin{subfigure}[b]{\linewidth}
\centering
		\begin{tikzpicture}[scale = 0.85]
			\node(X) at (-1, 0)[label = below:$X$, point];
			\node(W1) at (0, 0) {$W_1$}; 
			\node(W2) at (1, 0) {$W_2$}; 
			\node(Y) at (2, 0)[label=below:$Y$, point]; 
			\node(Z1) at (0,-.9) {$Z1$};
			\node(Z2) at (1,-.9) {$Z2$};
			\path (X) edge (W1);
			\path (Y) edge (W2);
			\path (W2) edge (W1); 
			\path (W1) edge (Z1);
			\path (Z2) edge (Z1);
		\end{tikzpicture}
\caption*{$(k)$}
%\vspace{+0.1in}
\end{subfigure}
\begin{subfigure}[b]{\linewidth}
\centering
		\begin{tikzpicture}[scale = 0.85]
			\node(X) at (-1, 0)[label = below:$X$, point];
			\node(W1) at (0, 0) {$W_1$}; 
			\node(W2) at (1, 0) {$W_2$}; 
			\node(Y) at (2, 0)[label=below:$Y$, point]; 
			\node(Z1) at (0,-.9) {$Z1$};
			\node(Z2) at (1,-.9) {$Z2$};
			\path (X) edge (W1);
			\path (Y) edge (W2);
			\path (W1) edge (W2); 
			\path (W1) edge (Z1);
			\path (Z2) edge (Z1);
		\end{tikzpicture}
\caption*{$(l)$}
\end{subfigure}
\end{minipage}
		\caption{Causal diagrams $(b), (c)$ are compatible with the C-DAG $(a)$. Causal diagrams $(e), (f)$ are compatible with the C-DAG $(d)$. In C-DAGs $(a), (d)$, $\*{Z}$ is a non-collider between $\*{X}$ and $\*{Y}$. 
		The path is active in $(b), (e)$, but inactive in $(c)$ and $(f)$.
		Causal diagrams $(h),(i)$ are compatible with the C-DAG $(g)$, where $\*{Z}$ is a collider between $\*{X}$ and $\*{Y}$. Causal diagrams $(k),(l)$ are compatible with the C-DAG $(j)$, where $\*{W}$ is a collider between $\*{X}$ and $\*{Y}$, and $\*{Z}$ is a descendant of $\*{W}$. When $\*{Z}$ is conditioned on, the path is active in $(h), (k)$, but inactive in $(i), (l)$.}
	\label{fig:clusterPathAnalysis}
%\vspace{-1\intextsep}
\end{figure*}
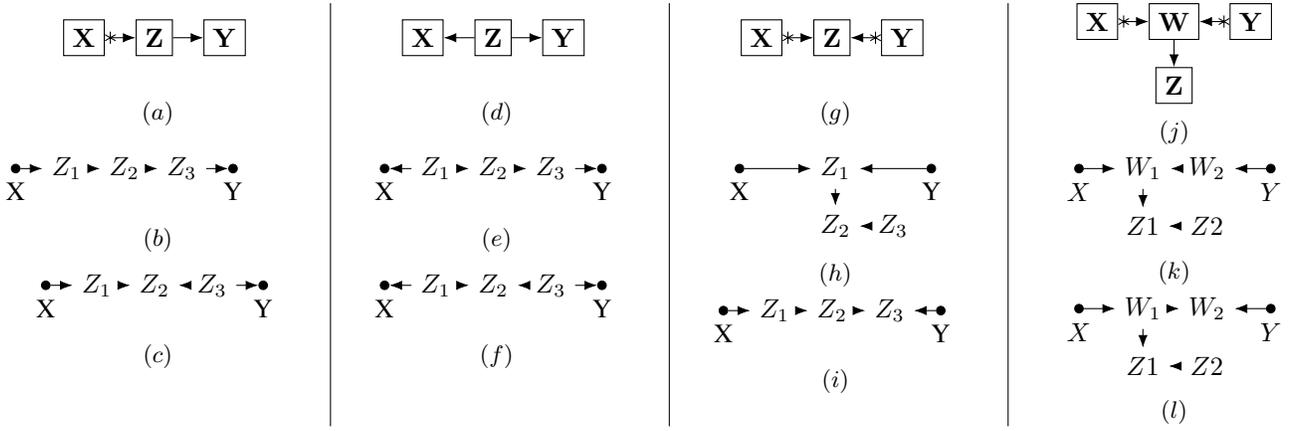

Consider a C-DAG where $\*{Z}$ is a non-collider between $\*{X}$ and $\*{Y}$ as shown in Fig. \ref{fig:clusterPathAnalysis}(a) and (d). The causal diagrams (b) and (c) in Fig.~\ref{fig:clusterPathAnalysis}, over $\*{V} = \{X, Z_1, Z_2, Z_3, Y\}$, are compatible with the C-DAG in Fig.~\ref{fig:clusterPathAnalysis}(a), where $\*{X} = \{X\}$, $\*{Z} = \{Z_1, Z_2, Z_3\}$, and $\*{Y} = \{Y\}$. Note that in diagram (b) the path between $X$ and $Y$ is active when $\*Z$ is not conditioned on. However, in diagram (c), the path between $X$ and $Y$ is inactive, even though the abstracted cluster $\*{Z}$ is a non-collider.
Similarly, in Fig.~\ref{fig:clusterPathAnalysis}, the causal diagrams (e) and (f) over $\*{V} = \{X, Z_1, Z_2, Z_3, Y\}$ are compatible with the C-DAG $G_{\*{C}}$ in %Fig.~\ref{fig:clusterPathAnalysis}
(d), where $\*{X} = \{X\}$, $\*{Z} = \{Z_1, Z_2, Z_3\}$, and $\*{Y} = \{Y\}$. In diagram (e), the path is active while in diagram (f) the path is inactive when $\*{Z}$ is not conditioned on. From these examples, it is clear that when a cluster $\*{Z}$ acts as a non-collider between two other clusters $\*{X}$ and $\*{Y}$, the path between $\*{X}$ and $\*{Y}$ through $\*{Z}$ may be either active or inactive. 

\begin{remark}
In a C-DAG, the path $\*{X} \ast\!\!\! - \!\!\!\ast \*{Z} \rightarrow \*{Y}$ may be either active or inactive  when $\*{Z}$ is not conditioned on.
\end{remark}

Conditioning on a cluster will be considered as equivalent to conditioning on all the variables in it. In all diagrams (b), (c), (e) and (f), the path between $X$ and $Y$ is inactive when conditioning on $\{Z_1, Z_2, Z_3\}$ and we show that this property will hold regardless of the connections within a cluster %non-collider 
as stated in the following lemma. 

\begin{restatable}{lemma}{noncollider}
\label{lem:noncollider_inactive}
In a C-DAG, the path $\*{X} \ast\!\!\! - \!\!\!\ast \*{Z} \rightarrow \*{Y}$ is inactive when non-collider $\*{Z}$ is conditioned on.
\end{restatable}
%\noncollider*
 \begin{proof}
 Let $G$ be any causal diagram compatible with the C-DAG $G_C$: $\*X \ast\!\! - \!\!\ast \*{Z} \rightarrow \*{Y}$. We show $\*X$ is blocked from $\*Y$ by $\*Z$ in $G$. 
 We show the contrapositive.

Assume there is an active path between $\*X$ and $\*Y$ not blocked by $\*Z$ in $G$ . Then there is path $p$  between a variable $X \in \*{X}$ and a variable $Y \in \*{Y}$ such that all the non-terminal nodes on $p$ are in $\*{Z}$ (otherwise $\*X$ will be adjacent to  $\*Y$ in $G_C$) and are colliders (otherwise $p$ will be blocked by $\*Z$). Let $V \in \*{Z}$ be the collider adjacent

to $Y$. For $V$ to be a collider there must either be an edge from $Y$ into $V$ or a bidirected edge between $Y$ and $V$. In both cases, the edge is into $V$. This contradicts the hypothesis of $\*{Z}$ being connected to $\*{Y}$ by the edge $\*{Z} \rightarrow \*{Y}$, which does not allow a variable in $\*{Y}$ to be either an ancestor of a variable in $\*{Z}$ or connected to a variable in $\*{Z}$ by a bidirected edge.
\end{proof}

%\begingroup
%\begin{wrapfigure}{r}{0.35\textwidth}
%\vspace{-2\intextsep}
%\endgroup	

%\newpage	
\subsection*{Cluster as a Collider or %its Descendant
a Descendant of a Collider}

Consider a C-DAG where $\*{Z}$ is a collider between $\*{X}$ and $\*{Y}$ as shown in Fig.~\ref{fig:clusterPathAnalysis}(g), or a descendant of a collider $\*W$ between $\*{X}$ and $\*{Y}$ as shown in Fig.~\ref{fig:clusterPathAnalysis}(j). 
The causal diagrams (h) and (i) in Fig.~\ref{fig:clusterPathAnalysis} over $\*{V} = \{X, Z_1, Z_2, Z_3, Y\}$ are compatible with the C-DAG in Fig.~\ref{fig:clusterPathAnalysis}(g), where $\*{X} = \{X\}$, $\*{Z} = \{Z_1, Z_2, Z_3\}$, and $\*{Y} = \{Y\}$. The path between $X$ and $Y$ is active when conditioning on $\{Z_1, Z_2, Z_3\}$ in diagram (h), but it is inactive in diagram (i). 
Similarly, the causal diagrams (k) and (l) in Fig.~\ref{fig:clusterPathAnalysis} over $\*{V} = \{X, W_1, W_2, Z_1, Z_2, Y\}$ are compatible with the C-DAG in Fig.~\ref{fig:clusterPathAnalysis}(j), where $\*{X} = \{X\}$, $\*{Z} = \{Z_1, Z_2\}$, $\*{W} = \{W_1, W_2\}$, and $\*{Y} = \{Y\}$. In diagram (k), the path from $X$ and $Y$ is active when conditioning on $\{Z_1, Z_2\}$, while in diagram (l) the path is inactive when conditioning on $\{Z_1, Z_2\}$. 
These examples illustrate that when a cluster $\*{Z}$ acts as a collider or a descendant of a collider between two other clusters $\*{X}$ and $\*{Y}$, the path between $\*{X}$ and $\*{Y}$ through $\*{Z}$ may be either active or inactive when conditioning on $\*{Z}$.

\begin{remark}
In a C-DAG, the path $\*{X} \ast\!\!\!\rightarrow \*{Z} \leftarrow\!\!\!\ast \*{Y}$ may be either active or inactive when $\*{Z}$ or some descendant of $\*{Z}$ is conditioned on.
\end{remark}

Note that the path between $X$ and $Y$ is inactive in all diagrams (h), (i), (k), (l). We show this property holds in general regardless of the connections within a cluster %(descendant of a) collider 
as stated in the following lemma. 
\begin{restatable}{lemma}{desccollider}
\label{lem:desc_collider_inactive}
In a C-DAG, the path $\*{X} \ast\!\!\!\rightarrow \*{Z} \leftarrow\!\!\!\ast \*{Y}$ is inactive when none of the  descendants of $\*{Z}$ (nor $\*Z$) are conditioned on. 
\end{restatable}

%\desccollider*

\begin{proof}
Let $G$ be any causal diagram  compatible with the C-DAG $\*X \ast\!\!\!\rightarrow \*{Z} \leftarrow\!\!\!\ast \*{Y}$. We show $\*X$ is blocked from $\*Y$ in $G$ when conditioning neither on $\*Z$ nor on a descendant of $\*Z$. 
Consider any path $p$ in $G$ between a variable $X \in \*{X}$ and a variable $Y \in \*{Y}$ going through some variables in $\*{Z}$. Then there exist $X' \in \*X$ and $V \in \*{Z}$, both in $p$, such that $X' \ast\!\!\!\rightarrow V$. Further, there exist $V' \in \*{Z}$ and $Y' \in \*Y$, both in $p$, such that $Y' \ast\!\!\!\rightarrow V'$. Then $p$ is of the form $X \ast\!\!-\!\!\ast \ldots  \ast\!\!-\!\!\ast X' \ast\!\!\rightarrow V \ast\!\!-\!\!\ast \ldots  \ast\!\!-\!\!\ast  V' \leftarrow\!\!\ast Y'\ast\!\!-\!\!\ast \ldots  \ast\!\!-\!\!\ast Y$ and must contain a collider between $V$ and $V'$ that is in $\*{Z}$. Therefore, $p$ must be inactive when none of the descendants of such a collider (nor the collider) is conditioned on.
\end{proof}

%\newpage 
\subsection{Characterization of C-DAGs as Causal Bayesian Networks}
\label{app:cdags_cbn}

In Section \ref{sec:cdags}, Remark 3, we noted that it does not follow directly from Def.~\ref{def:cdag} that a C-DAG $G_{\*C}$ has the semantics and properties of a causal diagram or Causal Bayesian Network (CBN) over the cluster nodes $\*C_i$, for $i = 1, \ldots, k$. Here we clarify how the results proved in this work allow us to characterize a C-DAG as a CBN over clustered variables, representing possibly multiple underlying CBNs over the original variables.

\begin{figure}[t]
%\begin{wrapfigure}{r}{.53\textwidth}
\vspace{-1.5\intextsep}
\begin{minipage}[t]{.32\linewidth}
\begin{subfigure}[b]{\linewidth}
\centering	
\begin{tikzpicture}[scale = .6]
			\node(X) at (0,0) [label = below:$X$, point];
			\node(Y) at (2, 0)[label = below:$Y$, point];
			\node(Z1) at (0, 2)[label=left:{$Z_1$}, point]; 
			\node(Z3) at (2, 2)[label=right:{$Z_3$}, point]; 
			\node(Z2) at (1, 1) {$Z_2$};     %[label=below:{$Z_2$}, point]; 
			\path (X) edge (Y); 
			\path[bend right =30] (Z1) edge (Y);
			\path (Z2) edge (Y);
			\path (Z3) edge (Y);
			\path (Z1) edge (Z2);
			\path (Z2) edge (Z3);
			\path[bidirected] (Z1) edge[bend right =30] (X);
			\path[bidirected] (Z1) edge[bend left =30] (Z2);
			\path[bidirected] (Z2) edge[bend left =30] (Z3);
			\path[bidirected] (Z3) edge[bend left =30] (Y);
	\end{tikzpicture}
		\caption*{$G_1$}
		\vspace{0.0em}
\end{subfigure}
\end{minipage}%	
\begin{minipage}[t]{.32\linewidth}
\begin{subfigure}[b]{\linewidth}
\centering	
\begin{tikzpicture}[scale = .6]
			\node(X) at (0,0) [label = below:$X$, point];
			\node(Y) at (2, 0)[label = below:$Y$, point];
			\node(Z1) at (0, 2)[label=left:{$Z_1$}, point]; 
			\node(Z3) at (2, 2)[label=right:{$Z_3$}, point]; 
			\node(Z2) at (1,1) {$Z_2$};     %[label=below:{$Z_2$}, point]; 
			\path (X) edge (Y);
			\path (Z2) edge (Z1);
			\path (Z2) edge (Z3);
			\path[bidirected] (Z1) edge[bend right =30] (X);
			\path[bidirected] (Z2) edge[bend right =30] (X);
			\path[bidirected] (Z2) edge[bend left =30] (Y);
			\path[bidirected] (Z1) edge[bend left =40] (Z2);
			\path[bidirected] (Z2) edge[bend left =40] (Z3);
			\path[bidirected] (Z3) edge[bend left =30] (Y);
	\end{tikzpicture}
		\caption*{$G_3$}
		\vspace{0.0em}
\end{subfigure}
\end{minipage}%
\begin{minipage}[t]{.32\linewidth}
\begin{subfigure}[b]{\linewidth}
\centering	
\begin{tikzpicture}[scale = .6]
			\node(X) at (0,0) [label = below:$X$, point];
			\node(Y) at (2, 0)[label = below:$Y$, point];
			\node(Z1) at (0, 2)[label=left:{$Z_1$}, point]; 
			\node(Z3) at (2, 2)[label=right:{$Z_3$}, point]; 
			\node(Z2) at (1,1) {$Z_2$};     %[label=below:{$Z_2$}, point]; 
			\path (Y) edge (X);
			\path (Z2) edge (Z1);
			\path (Z2) edge (Z3);
			\path[bend right =30] (Z1) edge (Y); 
			\path (Z2) edge (Y); 
            \path (Z3) edge (Y); 
			\path[bidirected] (Z1) edge[bend right =30] (X);
			\path[bidirected] (Z2) edge[bend right =30] (X);
			\path[bidirected] (Z1) edge[bend left =40] (Z2);
			\path[bidirected] (Z2) edge[bend left =40] (Z3);
	\end{tikzpicture}
		\caption*{$G_5$}
		\vspace{0.0em}
\end{subfigure}
\end{minipage}
% \begin{minipage}[t]{.24\linewidth}
% \begin{subfigure}[b]{\linewidth}
% \centering	
% \begin{tikzpicture}[scale = .6]
% 			\node(X) at (0,0) [label = below:$X$, point];
% 			\node(Y) at (2, 0)[label = below:$Y$, point];
% 			\node(Z1) at (0, 2)[label=left:{$Z_1$}, point]; 
% 			\node(Z3) at (2, 2)[label=right:{$Z_3$}, point]; 
% 			\node(Z2) at (1, 1) {$Z_2$};     
% 			\path[bend right =30] (Z1) edge (Y);
% 			\path (Z2) edge (Y);
% 			\path (Z3) edge (Y);

% 			\path[bidirected] (Z1) edge[bend right =30] (X);
% 			\path[bidirected] (Z1) edge[bend left =30] (Z2);
% 			\path[bidirected] (Z2) edge[bend left =30] (Z3);
% 			\path[bidirected] (Z3) edge[bend left =30] (Y);
% 	\end{tikzpicture}
% 		\caption*{$G_3$}
% 		\vspace{0.0em}
% \end{subfigure}
% \end{minipage}

\begin{minipage}[t]{.32\linewidth}
\begin{subfigure}[b]{\linewidth}
\centering	
\begin{tikzpicture}[scale = .6]
			\node(X) at (0,0) [label = below:$X$, point];
			\node(Y) at (2, 0)[label = below:$Y$, point];
			\node(Z1) at (0, 2)[label=left:{$Z_1$}, point]; 
			\node(Z3) at (2, 2)[label=right:{$Z_3$}, point]; 
			\node(Z2) at (1, 1)   [label=below:{$Z_2$}, point]; % {$Z_2$}; 
			\path (X) edge (Y);
			%\path (Z1) edge (Y);
			\path (Z2) edge (Y);
			\path (Z3) edge (Y);
			\path[bidirected] (Z1) edge[bend right =30] (X);
			\path[bidirected] (Z1) edge[bend left =30] (Z2);
			\path[bidirected] (Z2) edge[bend left =30] (Z3);
			\path[bidirected] (Z3) edge[bend left =30] (Y);
	\end{tikzpicture}
		\caption*{$G_2$}
		\vspace{0.0em}
\end{subfigure}
\end{minipage}	
\begin{minipage}[t]{.32\linewidth}
\begin{subfigure}[b]{\linewidth}
\centering	
\begin{tikzpicture}[scale = .6]
			\node(X) at (0,0) [label = below:$X$, point];
			\node(Y) at (2, 0)[label = below:$Y$, point];
			\node(Z1) at (0, 2)[label=left:{$Z_1$}, point]; 
			\node(Z3) at (2, 2)[label=right:{$Z_3$}, point]; 
			\node(Z2) at (1,1) {$Z_2$};     %[label=below:{$Z_2$}, point]; 
			\path (X) edge (Y);
			\path (Z2) edge (Z1);
			\path (Z2) edge (Z3);
			\path[bidirected] (Z1) edge[bend right =30] (X);
			\path[bidirected] (Z1) edge[bend left =20] (Z3);
			\path[bidirected] (Z3) edge[bend left =30] (Y);
	\end{tikzpicture}
		\caption*{$G_4$}
		\vspace{0.0em}
\end{subfigure}
\end{minipage}
\begin{minipage}[t]{.32\linewidth}
\begin{subfigure}[b]{\linewidth}
\centering	
\begin{tikzpicture}[scale = .6]
			\node(X) at (0,0) [label = below:$X$, point];
			\node(Y) at (2, 0)[label = below:$Y$, point];
			\node(Z1) at (0, 2)[label=left:{$Z_1$}, point]; 
			\node(Z3) at (2, 2)[label=right:{$Z_3$}, point]; 
			\node(Z2) at (1,1) {$Z_2$}; 
			\path (Y) edge (X);
			\path (Z1) edge (Z2);
			\path (Z2) edge (Z3);
			\path (Z3) edge (Y); 
			\path[bidirected] (Z1) edge[bend right =30] (X);
	\end{tikzpicture}
		\caption*{$G_6$}
		\vspace{0.0em}
\end{subfigure}
\end{minipage}

\begin{minipage}[t]{.32\linewidth}
\begin{subfigure}[b]{\linewidth}
\centering	
		\begin{tikzpicture}[scale = .6]
			\node(X) at (0,0) [label = below:X, point, blue];
			\node(Y) at (2, 0)[label = below:Y, point, red];
			\node[state](Z) at (1, 2) {$\*Z$}; 
			\path (X) edge (Y);
			%\path (Z) edge (X);
			\path (Z) edge (Y);
			\path[bidirected] (Z) edge[bend right = 60] (X);
			\path[bidirected] (Z) edge[bend left = 60] (Y);
		\end{tikzpicture} 
		\caption*{$G_{\*C_1}$}
\end{subfigure}
\end{minipage}
% \begin{minipage}[t]{.24\linewidth}
% \begin{subfigure}[b]{\linewidth}
% \centering	
% 		\begin{tikzpicture}[scale = .6]
% 			\node(X) at (0,0) [label = below:X, point, blue];
% 			\node(Y) at (2, 0)[label = below:Y, point, red];
% 			\node[state](Z) at (1, 2) {$\*Z$}; 
% 			\path (Z) edge (Y);
% 			\path[bidirected] (Z) edge[bend right = 60] (X);
% 			\path[bidirected] (Z) edge[bend left = 60] (Y);
% 		\end{tikzpicture} 
% 		\caption*{$G_{\*C_2}$}
% \end{subfigure}
% \end{minipage}
\begin{minipage}[t]{.32\linewidth}
\begin{subfigure}[b]{\linewidth}
\centering	
		\begin{tikzpicture}[scale = .6]
			\node(X) at (0,0) [label = below:X, point, blue];
			\node(Y) at (2, 0)[label = below:Y, point, red];
			\node[state](Z) at (1, 2) {$\*Z$}; 
			\path (X) edge (Y);
			\path[bidirected] (Z) edge[bend right = 60] (X);
			\path[bidirected] (Z) edge[bend left = 60] (Y);
		\end{tikzpicture} 
		\caption*{$G_{\*C_2}$}
\end{subfigure}
\end{minipage}
\begin{minipage}[t]{.32\linewidth}
\begin{subfigure}[b]{\linewidth}
\centering	
		\begin{tikzpicture}[scale = .6]
			\node(X) at (0,0) [label = below:X, point, blue];
			\node(Y) at (2, 0)[label = below:Y, point, red];
			\node[state](Z) at (1, 2) {$\*Z$}; 
			\path (Y) edge (X);
			\path (Z) edge (Y);
			\path[bidirected] (Z) edge[bend right = 60] (X);
		\end{tikzpicture} 
		\caption*{$G_{\*C_3}$}
\end{subfigure}
\end{minipage}

        \caption{With cluster $\*Z = \{Z_1, Z_2, Z_3\}$, $G_{\*C_1}$ is the C-DAG for $G_1$ and $G_2$; $G_{\*C_2}$ is the C-DAG for $G_3$ and $G_4$; and 
        $G_{\*C_3}$ is the C-DAG for $G_5$ and $G_6$. } 
	\label{fig:C-EC2}
\vspace{-1\intextsep}
\end{figure}
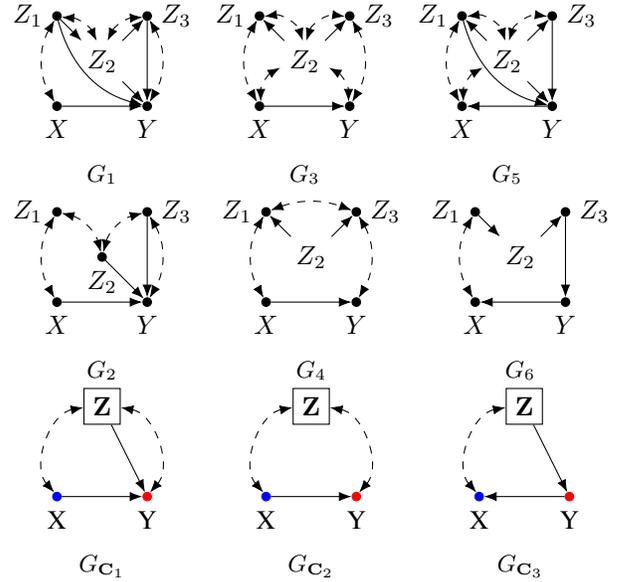

The semantics of a C-DAG is defined in terms of the class of all compatible causal diagrams and corresponding SCMs over $\*V$. In this way, a C-DAG can be seen as an equivalence class of causal diagrams sharing the relationships among the clusters while allowing for any possible relationships among the variables within each cluster. For instance, in Fig.~\ref{fig:C-EC2}, the DAGs $G_1$, $G_2$ induce different collections of interventional distributions over the set of variables $\*V$ (i.e, $\*P^1_{*}(\*v) \neq \*P^2_{*}(\*v)$). However, when considering the cluster $\*Z = \{Z_1, Z_2, Z_3\}$, both $G_1$ and $G_2$ are represented by the C-DAG $G_{\*{C}_1}$, which induces the same collection of interventional distributions over the clusters $\*C$ (i.e., $\*P^1_{*}(\*c) = \*P^2_{*}(\*c) = \*P^{\*{C}_1}_{*}(\*c)$). In this sense, $G_1$ and $G_2$ can be thought of as being members of the equivalence class represented by $G_{\*{C}_1}$. The same can be concluded for $G_3$, $G_4$, represented by $G_{\*{C}_2}$, and for $G_5$, $G_6$, represented by $G_{\*{C}_3}$.

%\end{wrapfigure} 

%
% \begin{minipage}[t]{.24\linewidth}
% \begin{subfigure}[b]{\linewidth}
% \centering	
% \begin{tikzpicture}[scale = .6]
% 			\node(X) at (0,0) [label = below:$X$, point];
% 			\node(Y) at (2, 0)[label = below:$Y$, point];
% 			\node(Z1) at (0, 2)[label=left:{$Z_1$}, point]; 
% 			\node(Z3) at (2, 2)[label=right:{$Z_3$}, point]; 
% 			\node(Z2) at (1, 1) {$Z_2$};     %[label=below:{$Z_2$}, point]; 
% 		%	\path (X) edge (Y);
% 			\path (Z1) edge (Z2);
% 			\path (Z3) edge (Z2);
% 			\path (Z3) edge (Y);

% 			\path[bidirected] (Z1) edge[bend right =30] (X);
% 			\path[bidirected] (Z3) edge[bend left =30] (Y);
% 	\end{tikzpicture}
% 		\caption*{$G_4$}
% 		\vspace{0.0em}
% \end{subfigure}
% \end{minipage}
 
In Section \ref{sec:dsep}, Theorem \ref{thm:dsep_cdag_dag}, we showed that a d-separation in a C-DAG $G_\*C$ implies the same d-separation in all compatible causal diagrams. Given this pleasant result, we can characterize a C-DAG as a graphical encoder of conditional independencies over the set of clusters $\*C$, but this is still insufficient to give the causal interpretation necessary for the application of causal inferences tools. In fact, the same set of conditional independencies can be encoded by different DAGs, where vertices are not necessarily arranged following a causal order. This means that different DAGs can induce the same observational distribution, while differing with respect to the induced collection of interventional distributions. For example, in Fig.~\ref{fig:C-EC2}, for DAGs $G_i$, with $i = 1, 3, 5$, it holds that $P^i(\*v) = P(\*v)$, even though $P^i(\*v|do(\*x)) \neq P^j(\*v|do(\*x))$ for any $i \neq j$ and some $\*X \subseteq \*V$. Now, at the cluster level, for all DAGs $G_i$, with $i = 1, \ldots, 6$, $P^i(\*c) = P(\*c)$), while only DAGs represented by the same C-DAG induce the same collection of interventional distributions over $\*C$.
%in Fig.~\ref{fig:C-EC}, for DAGs $G_i$, with $i = 1, 3, 5$, it holds that $P^i(\*v) = P(\*v)$, even though $P^i(\*v|do(\*x)) \neq P^j(\*v|do(\*x))$ for any $i \neq j$ and some $\*X \subseteq \*V$. 

To formalize a C-DAG $G_{\*C}$ as a CBN over the cluster variables $\*C_i \in \*C$, it is necessary to show that any distribution $P_{\*x}(\*v)$ resulting from the intervention $do(\*X = \*x)$, for any $\*X \subseteq \*C$, including the observational distribution $P(\*v)$, where $\*X = \emptyset$, factorizes according to the graphical structure of $G_{\*C}$. This result, proved in Theorem \ref{thm:markovrelative}, allows us to characterize a single C-DAG $G_\*{C}$ as an encoder of the space of all interventional distributions $\*P_{*}$ at the cluster level, i.e., over the clusters of variables $\*C_i \in \*C$. More formally, a C-DAG $G_\*{C}$, compatible with a causal diagram $G$, %associated with an SCM $\mathcal{M}=\langle \*U, \*V, \mathcal{F}, P(\*U)\rangle$, 
is a CBN for $\*P_{*}^\mathcal{M}(\*c)$, the collection of interventional distributions induced by $\mathcal{M}$, over the clusters $\*C$.

%Note that, given a partition $\*C$ of $\*V$, a larger set of DAGs can induce the same observational distribution over the set of clusters $\*C$. For each cluster $\*C_i$ of size $m_i$, the number of possible internal structures is $N_i \times 2^{m_i (m_i -1)/2}$, where $N_i$ is the number of possible DAGs without bidirected edges over $m_i$ variables, given by a super-exponential recurrence formula (see  \url{http://oeis.org/A003024}). In Fig.~\ref{fig:C-EC}, for all DAGs $G_i$, with $i = 1, \ldots, 6$, $P^i(\*c) = P(\*c)$), while only DAGs represented by the same C-DAG induce the same collection of interventional distributions over $\*C$.

This characterization is crucial for the correct abstraction of the possible underlying causal models and validity of causal tools applied to causal C-DAGs. Identifying an effect $P(y | do(x))$ in a causal C-DAG  means identifying such an effect for the entire class of causal diagrams represented. In $G_{\*{C}_2}$, the identification formula $P(y | do(x)) = P(y|x)$ is valid for $G_3$ and $G_4$, as well as all other diagrams in the class. This effect is different in the causal diagrams represented by $G_{\*{C}_3}$ (e.g., $G_5$ and $G_6$) where $P(y|do(x)) = P(y)$. On the other hand, no expression can be derived for $P(y | do(x))$ from $G_{\*{C}_1}$ since the encoded partial knowledge is compatible with some causal diagram in which the effect is not identifiable such as $G_1$.

%\newpage
\subsection{Simulations}
\label{app:simulations}

We evaluate C-DAGs in their ability to determine an equivalent causal effect from an observational dataset to the effect determined from any compatible causal diagram, by their respective identification formulas. Given a causal diagram, representing true and complete causal structural knowledge, an identifiable effect should be computable by an identification formula from a compatible C-DAG, representing partial structural knowledge, in which the effect is also identifiable. This means that given the limited knowledge of a C-DAG, an effect can be determined that would be equivalent to whatever the underlying (unknown) causal diagram is. 

\begin{figure}[t]
%\begin{wrapfigure}{r}{0.505\textwidth}
\vspace{-0.8\intextsep}
\begin{minipage}[t]{.32\linewidth}
\begin{subfigure}[b]{\linewidth}
\centering	
\begin{tikzpicture}[scale = .6]
			\node(X) at (0,0) [label = below:X, point, blue];
			\node(Y) at (2, 0)[label = below:Y, point, red];
			\node[state](Z) at (1, 2) {$\*Z$}; 
			\path (X) edge (Y);
			\path (Z) edge (X);
			\path (Z) edge (Y);
		\end{tikzpicture} 
		\caption*{$G_{\*C_1}$}
		\vspace{1em}
\end{subfigure}
\end{minipage}
\begin{minipage}[t]{.32\linewidth}
\begin{subfigure}[b]{\linewidth}
\centering	
    \begin{tikzpicture}[scale = .6]
			\node(X) at (0,0) [label = below:X, point, blue];
			\node(Y) at (2, 0)[label = below:Y, point, red];
			\node(Z1) at (0, 2)[label=left:{$Z_1$}, point]; 
			\node(Z2) at (2, 2)[label=right:{$Z_2$}, point];
			\path (X) edge (Y);
			\path (Z1) edge (X);
			\path (Z1) edge (Y);
			\path (Z2) edge (Y);
			\path (Z2) edge (X);
			\draw[gray, fill=gray!30, opacity =.5] (-.1,1.7) rectangle (2.1, 2.3);
		\end{tikzpicture} 
		\caption*{$(a)$}
		\vspace{0.7em}
\end{subfigure}
\end{minipage}
% \begin{minipage}[t]{.32\linewidth}
% \begin{subfigure}[b]{\linewidth}
% \centering	
%     \begin{tikzpicture}[scale = .6]
% 			\node(X) at (0,0) [label = below:X, point, blue];
% 			\node(Y) at (2, 0)[label = below:Y, point, red];
% 			\node(Z1) at (0, 2)[label=left:{$Z_1$}, point]; 
% 			\node(Z2) at (2, 2)[label=right:{$Z_2$}, point];
% 			\path (X) edge (Y);
% 			\path (Z1) edge (X);
% 			\path (Z2) edge (Y);
% 			\draw[gray, fill=gray!30, opacity =.5] (-.1,1.7) rectangle (2.1, 2.3);
% 		\end{tikzpicture} 
% 		\caption*{$(b): Y_x \indep X \mid \*Z$}
% \end{subfigure}
% \end{minipage}
\begin{minipage}[t]{.32\linewidth}
\begin{subfigure}[b]{\linewidth}
\centering	
\begin{tikzpicture}[scale = .6]
			\node(X) at (0,0) [label = below:X, point, blue];
			\node(Y) at (2, 0)[label = below:Y, point, red];
			\node(Z1) at (0, 2)[label=left:{$Z_1$}, point]; 
			\node(Z2) at (2, 2)[label=right:{$Z_2$}, point];
			\path (X) edge (Y);
			\path (Z1) edge (X);
			\path (Z1) edge (Y);
			\path (Z2) edge (Y);
			\path (Z2) edge (X);
			\path[bidirected] (Z1) edge[bend left =10] (Z2);
			\draw[gray, fill=gray!30, opacity =.5] (-.1,1.7) rectangle (2.1, 2.3);
		\end{tikzpicture} 
		\caption*{$(b)$}
		\vspace{0.7em}
\end{subfigure}
\end{minipage}	
% \begin{minipage}[t]{.32\linewidth}
% \begin{subfigure}[b]{\linewidth}
% \centering	
% \begin{tikzpicture}[scale = .6]
% 			\node(X) at (0,0) [label = below:X, point, blue];
% 			\node(Y) at (2, 0)[label = below:Y, point, red];
% 			\node(Z1) at (0, 2)[label=left:{$Z_1$}, point]; 
% 			\node(Z2) at (2, 2)[label=right:{$Z_2$}, point];
% 			\path (X) edge (Y);
% 			\path (Z1) edge (X);
% 			\path (Z2) edge (Y);
% 			\path[bidirected] (Z1) edge[bend left =10] (Z2);
% 			\draw[gray, fill=gray!30, opacity =.5] (-.1,1.7) rectangle (2.1, 2.3);
% 		\end{tikzpicture} 
% 		\caption*{$(d): Y_x \indep X \mid \*Z$}
% \end{subfigure}
% \end{minipage}
%\vspace{\intextsep}

\begin{minipage}[t]{.32\linewidth}
\begin{subfigure}[b]{\linewidth}
\centering	
		\begin{tikzpicture}[scale = .6]
			\node(X) at (0,0) [label = below:X, point, blue];
			\node(Y) at (2, 0)[label = below:Y, point, red];
			\node[state](Z) at (1, 2) {$\*Z$}; 
			\path (X) edge (Y);
			\path (Z) edge (X);
			\path (Z) edge (Y);
			\path[bidirected] (Z) edge[bend right = 60] (X);
			\path[bidirected] (Z) edge[bend left = 60] (Y);
		\end{tikzpicture} 
		\caption*{$G_{\*C_2}$}
\end{subfigure}
\end{minipage}
\begin{minipage}[t]{.32\linewidth}
\begin{subfigure}[b]{\linewidth}
\centering	
\begin{tikzpicture}[scale = .6]
			\node(X) at (0,0) [label = below:X, point, blue];
			\node(Y) at (2, 0)[label = below:Y, point, red];
			\node(Z1) at (0, 2)[label=left:{$Z_1$}, point]; 
			\node(Z2) at (2, 2)[label=right:{$Z_2$}, point];
			\path (X) edge (Y);
			\path (Z1) edge (X);
			\path (Z1) edge (Y);
			\path (Z2) edge (Y);
			\path (Z2) edge (X);
			\path[bidirected] (Z1) edge[bend right = 60] (X);
			\path[bidirected] (Z2) edge[bend left = 60] (Y);
			\draw[gray, fill=gray!30, opacity =.5] (-.1,1.7) rectangle (2.1, 2.3);
		\end{tikzpicture} 
		\caption*{$(c)$}
\end{subfigure}
\end{minipage}
% \begin{minipage}[t]{.32\linewidth}
% \begin{subfigure}[b]{\linewidth}
% \centering	
% \begin{tikzpicture}[scale = .6]
% 			\node(X) at (0,0) [label = below:X, point, blue];
% 			\node(Y) at (2, 0)[label = below:Y, point, red];
% 			\node(Z1) at (0, 2)[label=left:{$Z_1$}, point]; 
% 			\node(Z2) at (2, 2)[label=right:{$Z_2$}, point];
% 			\path (X) edge (Y);
% 			\path (Z1) edge (X);
% 			\path (Z2) edge (Y);
% 			\path[bidirected] (Z1) edge[bend right = 60] (X);
% 			\path[bidirected] (Z2) edge[bend left = 60] (Y);
% 			\draw[gray, fill=gray!30, opacity =.5] (-.1,1.7) rectangle (2.1, 2.3);
% 		\end{tikzpicture} 
% 		\caption*{$(f): Y_x \indep X \mid \*Z$}
% \end{subfigure}
% \end{minipage}
\begin{minipage}[t]{.32\linewidth}
\begin{subfigure}[b]{\linewidth}
\centering	
\begin{tikzpicture}[scale = .6]
			\node(X) at (0,0) [label = below:X, point, blue];
			\node(Y) at (2, 0)[label = below:Y, point, red];
			\node(Z1) at (0, 2)[label=left:{$Z_1$}, point]; 
			\node(Z2) at (2, 2)[label=right:{$Z_2$}, point];
			\path (X) edge (Y);
			\path (Z1) edge (X);
			\path (Z1) edge (Y);
			\path (Z2) edge (Y);
			\path (Z2) edge (X);
			\path[bidirected] (Z1) edge[bend right = 60] (X);
			\path[bidirected] (Z2) edge[bend left = 60] (Y);
			\path[bidirected] (Z1) edge[bend left =10] (Z2);
			\draw[gray, fill=gray!30, opacity =.5] (-.1,1.7) rectangle (2.1, 2.3);
		\end{tikzpicture} 
		\captionsetup{labelformat=empty}
		\caption*{$(d)$}
\end{subfigure}
\end{minipage}
%\begin{minipage}[t]{.32\linewidth}
% \begin{subfigure}[b]{\linewidth}
% \centering	
% 		\begin{tikzpicture}[scale = .6]
% 			\node(X) at (0,0) [label = below:X, point, blue];
% 			\node(Y) at (2, 0)[label = below:Y, point, red];
% 			\node(Z1) at (0, 2)[label=left:{$Z_1$}, point]; 
% 			\node(Z2) at (2, 2)[label=right:{$Z_2$}, point];
% 			\path (X) edge (Y);
% 			\path (Z1) edge (X);
% % 			\path (Z1) edge (Y);
% 			\path (Z2) edge (Y);
% % 			\path (Z2) edge (X);
% 			\path[bidirected] (Z1) edge[bend right = 60] (X);
% 			\path[bidirected] (Z2) edge[bend left = 60] (Y);
% 			\path[bidirected] (Z1) edge[bend left =10] (Z2);
% 			\draw[gray, fill=gray!30, opacity =.5] (-.1,1.7) rectangle (2.1, 2.3);
% 		\end{tikzpicture} 
% 		\caption*{$(h): Y_x \nindep X \mid \*Z$}
% \end{subfigure}
% \end{minipage}
        \caption{$G_{\*C_1}$ is the C-DAG for diagrams (a) and (b). $G_{\*C_2}$ is the C-DAG for diagrams (c) and (d). $P(y | do(x))$ is identifiable by backdoor adjustment over the cluster $\*Z = \{Z_1, Z_2\}$ in $G_{\*C_1}$ and, therefore, in \emph{all} compatible causal diagrams (e.g., diagrams (a)-(b)).
        However, $P(y | do(x))$ is not identifiable in $G_{\*C_2}$. This means that there is \emph{some} compatible causal diagram for which $P(y | do(x))$ is not identifiable (e.g., diagram (d)).} 
	\label{fig:ignorabilitycluster}
%\vspace{-1.75\intextsep}
\end{figure}
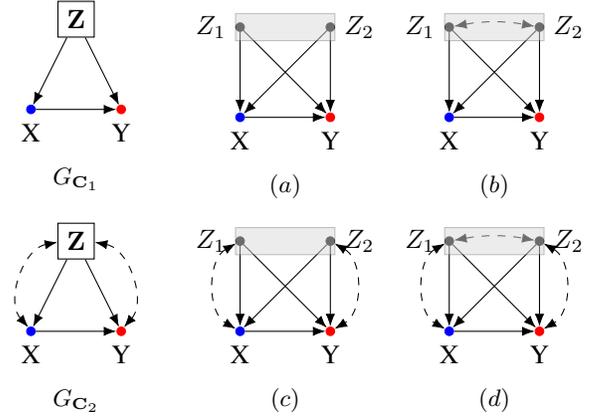 
%\end{wrapfigure} 

In Fig.~\ref{fig:ignorabilitycluster}, $P(y| do(x))$ is identifiable by backdoor adjustment in $G_{\*C_1}$ (i.e., $P(y| do(x)) = \sum_{\*z} P(y |x, \*z)P(\*z)$), but it not identifiable in $G_{\*C_2}$. Diagrams (a)-(b) are  compatible with $G_{\*C_1}$ and diagrams (c)-(d) are compatible with $G_{\*C_2}$. Those are examples of compatible causal diagrams in which two variables are clustered to create $\*Z$. 

For our simulations, we consider more complex causal diagrams involving 10 variables clustered to create $\*Z$. 

For C-DAG $G_{\*C_1}$, we randomly generated 100 causal diagrams that are compatible with this C-DAG. Each generated compatible causal diagram includes randomly generated relationships among the 10 variables in cluster $\*Z$, and to $\*X$ and $\*Y$, while adhering to the constraints imposed by the C-DAG.
% \newpage

\begin{figure*}[t]
\centering
% \begin{subfigure}[t]{.02\linewidth}
% A)
% \end{subfigure}
\begin{subfigure}[b]{.49\linewidth}
\includegraphics[width=\textwidth]{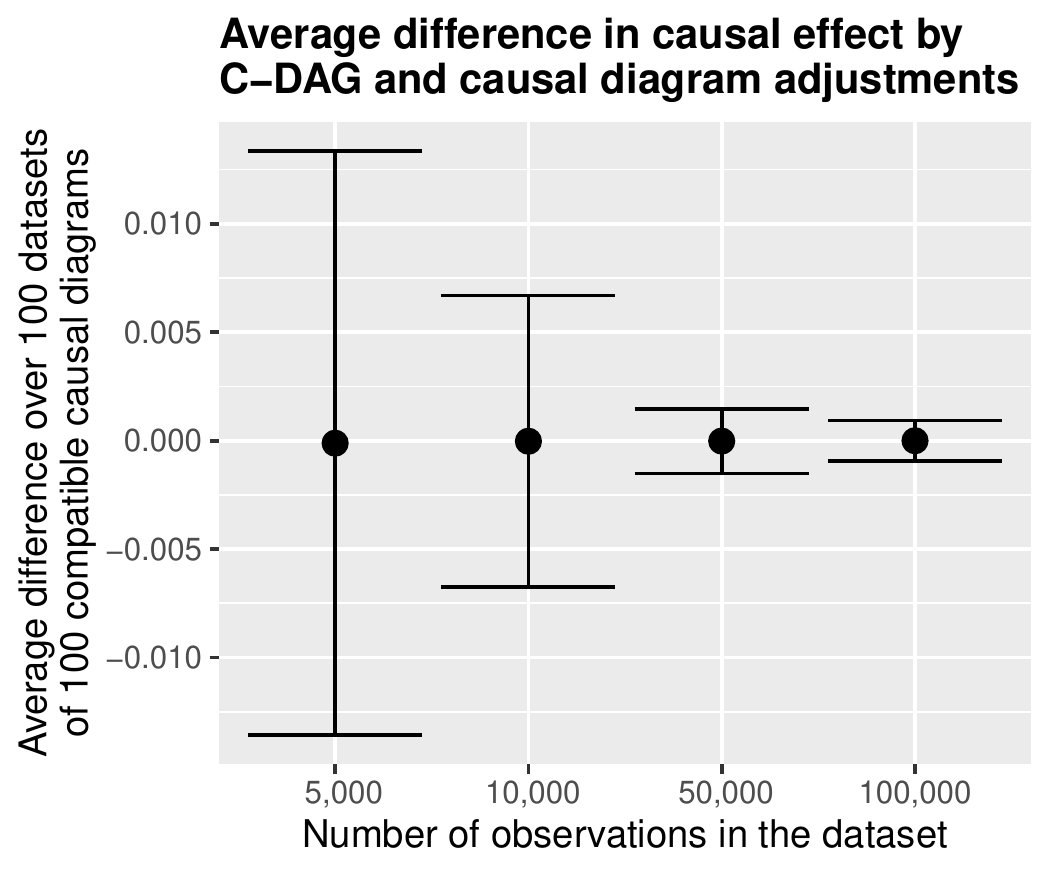}
\caption*{(A)}
\end{subfigure}
% \begin{subfigure}[t]{.02\linewidth}
% B)
% \end{subfigure}
\begin{subfigure}[b]{.49\linewidth}
\includegraphics[width=\textwidth]{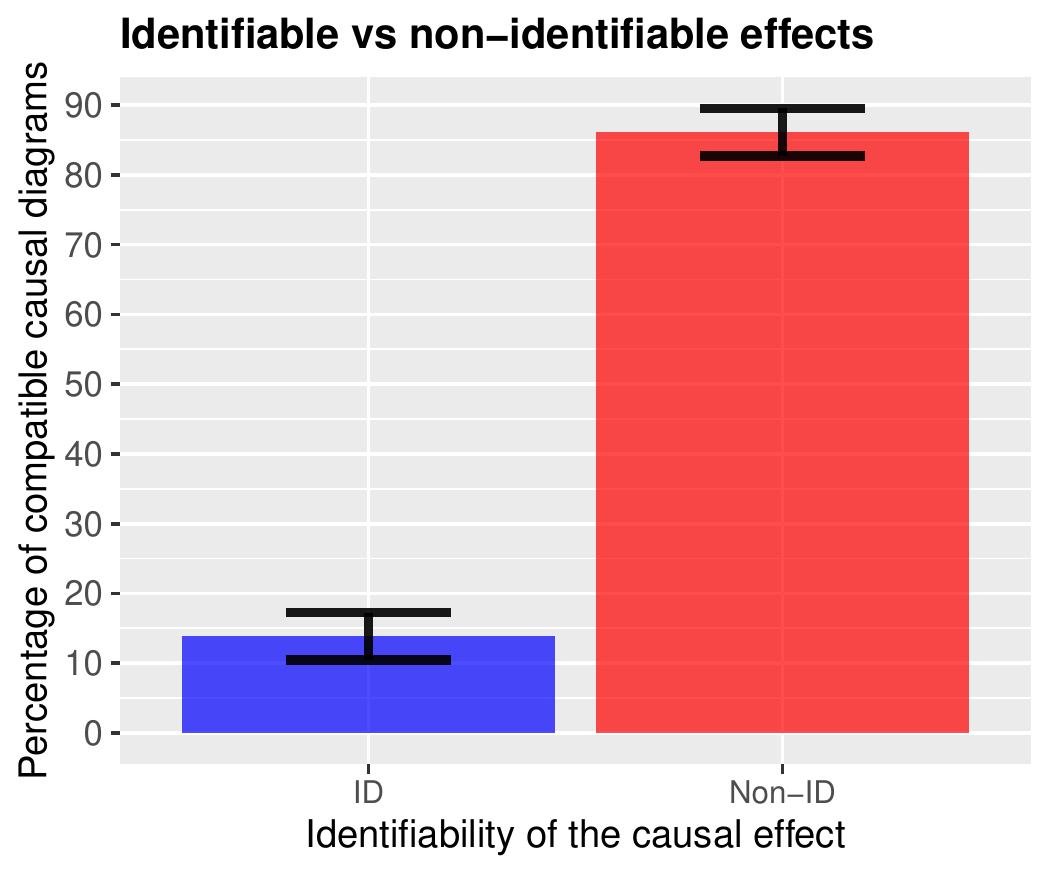}
\caption*{(B)}
\end{subfigure}
\caption{(A) Average and standard error of the difference between the effects computed using the identification formula returned by the ID algorithm when the C-DAG $G_{\*C_1}$ in Fig.~\ref{fig:ignorabilitycluster} is given and when a random compatible causal diagram is given. The average is over 100 causal diagrams with 100 data sets consisting of $N$ observations, with $N = \{5000, 10000, 50000, 100000\}$. (B) Percentage of causal diagrams compatible with the C-DAG $G_{\*C_2}$ in Fig.~\ref{fig:ignorabilitycluster} in which the causal effect was identifiable (ID) and non-identifiable (Non-ID), in  a sample of 100 randomly generated causal diagrams.}
\label{fig:simulationsBD1Cluster}
\end{figure*}

Identification expressions were determined by the ID-algorithm given the C-DAG as well as the compatible causal diagrams. According to each causal diagram, we generated 100 data sets consisting of $N$ observations. For each data set, we computed effect sizes using the identification formula derived from the causal diagram used to generate the data set and the one derived from the C-DAG. Then, we obtained the average difference between these two calculated effect sizes over the samples for all causal diagrams. The simulation was repeated with increasing values of $N = \{5000, 10000, 50000, 100000\}$. The simulations were implemented in R v4.1.2 \citep{rcore2021}, using the packages \textit{dagitty} v0.3.1 \citep{dagitty2016}, \textit{igraph} v1.2.8.9014 \citep{igraph2006}, \textit{pcalg} v2.7.3 \citep{pcalg2012}, and \textit{causaleffect} v1.3.13 \citep{causaleffect2017}. The results are shown in Fig.~\ref{fig:simulationsBD1Cluster}-A).

These empirical results illustrate a negligible average difference between the effect sizes computed by the identification formula for a causal diagram, and by the identification formula for the compatible C-DAG. Moreover, the standard error greatly decreases as the number of observations in the dataset increases. This illustrates that C-DAGs can be used to determine equivalent, accurate effect sizes, while necessitating significantly less domain knowledge for their construction.

For the C-DAG $G_{\*C_2}$ in Fig.~\ref{fig:ignorabilitycluster},
we evaluate the non-identifiability of the causal effect in a subset of the causal diagrams compatible with the C-DAG. Since the interventional distribution $P(y|do(x))$ is not identifiable given $G_{\*C_2}$, it's expected that there is at least one causal diagram compatible with $G_{\*C_2}$ for which the effect is not identifiable. For these experiments, we generated 100 random causal diagrams compatible with the C-DAG  $G_{\*C_2}$ and used the ID-algorithm to determine if the effect was identifiable or not in the true causal diagram. This simulation was repeated 100 times. As shown in Fig.~\ref{fig:simulationsBD1Cluster}-B), the simulation shows that the effect was not identifiable in 86.13\% $\pm$ 3.41 (mean $\pm$ standard error)  %$\approx$90\% 
of the random causal diagrams generated and compatible with C-DAG $G_{\*C_2}$. 

\subsection{Proofs}\label{appendix:proofs}

We start by showing in Proposition \ref{prop:cdags_connections} that any adjacency between variables is preserved in a compatible C-DAG.

\begin{restatable}{proposition}{adjacencies}
\label{prop:cdags_connections}
(\textbf{Preservation of adjacencies}) Let $G_{\*{C}}(\*{C}, \*{E}_\*{C})$ be a C-DAG compatible with an ADMG $G(\*{V}, \*{E})$. Consider distinct clusters $\*C_i, \*C_j \in \*C$. If $V_i, V_j \in \*V$ are adjacent in $G$ and belong to $\*C_i,\*C_j$ respectively, then $\*C_i$ and $\*C_j$ are adjacent in $G_{\*C}$. Further, if $\*C_i$ and $\*C_j$ are adjacent in $G_{\*C}$, then there exists $V_i\in \*C_i$ and $V_j\in \*C_j$ such that $V_i$ and $V_j$ are adjacent in $G$.
\end{restatable}

\begin{proof}

%Let $G_{\*{C}}$ be a C-DAG. 
Assume that %, in a compatible causal diagram $G$, 
two variables $V_i, V_j \in \*V$ are connected by an edge in $G$. %If $V_i$ and $V_j$ are in the same cluster in $G_{\*{C}}$, then they will never be separated from each other. So, $V_i$ and $V_j$ are always connected. On the other hand, 
If $V_i \in \*C_i$ and $V_j \in \*C_j$, where $\*C_i$ and $\*C_j$ are two distinct clusters in $G_{\*{C}}$, then by the definition of cluster DAGs (Definition \ref{def:cdag}), if $V_i \rightarrow V_j$, then $\*C_i \rightarrow \*C_j$. Further, if $V_i \dashleftarrow \!\!\!\!\!\!\!\!\! \dashrightarrow V_j$, then $\*C_i \dashleftarrow \!\!\!\!\!\!\!\!\! \dashrightarrow \*C_j$. So, $\*C_i$ and $\*C_j$ are also connected by an edge in $G_{\*{C}}$. 

Now, assume that $\*C_i$ and $\*C_j$ are connected by an edge in $G_{\*{C}}$. If $\*C_i \rightarrow \*C_j$, then, by Definition \ref{def:cdag}, there exists $V_i \in \*C_i$ and $V_j \in \*C_j$ such that $V_i \in Pa(V_j)$ in $G$. Therefore, $V_i$ and $V_j$ are connected by the edge $V_i \rightarrow V_j$ in $G$. Similarly, if $\*C_i  \dashleftarrow \!\!\!\!\!\!\!\!\! \dashrightarrow \*C_j$, then, by Definition \ref{def:cdag}, there exists $V_i \in \*C_i$ and $V_j \in \*C_j$ such that $V_i \dashleftarrow \!\!\!\!\!\!\!\!\! \dashrightarrow V_j$, that is,  $V_i$ and $V_j$ are connected by an edge in $G$. 
\end{proof}

We show in Proposition \ref{prop:connection_directed_Gc} that connections via directed paths are also preserved. This implies that both order and ancestral relationships are preserved in C-DAGs. To prove, we need first the following lemma:

% TODO lem
\begin{lemma}
\label{lem:connectionGc}
(\textbf{Preservation of paths})
Let $G_{\*{C}}(\*{C}, \*{E}_\*{C})$ be a C-DAG compatible with an ADMG $G(\*{V}, \*{E})$ and $\*C_i, \*C_j \in \*C$ be two distinct clusters. If two variables $V_i \in \*C_i$ and $V_j \in \*C_j$ are connected in $G$ by a path (of any sorts), then the clusters $\*C_i$ and $\*C_j$ are connected  by a path in $G_{\*C}$. 
\end{lemma}
\begin{proof}
First %note that, if $\*C_i = \*C_j$, then $\*C_i$ and $\*C_j$ are trivially connected in $G_{\*C}$. Also,
by Proposition \ref{prop:cdags_connections}, if %$\*C_i \neq \*C_j$ and 
$V_i$ and $V_j$ are adjacent, then $\*C_i$ and $\*C_j$ are connected in $G_{\*C}$ by an edge. %by the same  edges(s) between $V_i$ and $V_j$ in $G$. 
Now, assume that $V_i$ and $V_j$ are not adjacent and let $p$ be a path connecting $V_i$ and $V_j$ in $G$. Let $p'$ be any subpath of $p$ that is split into two different clusters of the form $V_1 \ast\!\!-\!\!\ast \ldots \ast\!\!-\!\!\ast V_k \ast\!\!-\!\!\ast V_{k+1} \ast\!\!-\!\!\ast \ldots \ast\!\!-\!\!\ast V_n$, where $V_1, \ldots, V_k \in C_k$, $V_{k+1}, \dots, V_n \in C_{k+1}$, and $\ast$ represents either an edge head or tail. By Definition \ref{def:cdag}, $\*C_k$ and $\*C_{k+1}$ will be connected by the same type of edge between $V_k$ and $V_{k+1}$ in $G_{\*C}$. By using the same argument for every subpath of $p$ that is split into two different clusters, we conclude that all clusters containing variables in $p$ are connected by a path and, thus, $\*C_i$ and $\*C_j$ are connected by a path in $G_{\*C}$.
\end{proof}

\begin{restatable}{proposition}{directedpaths}
\label{prop:connection_directed_Gc}
(\textbf{Preservation of directed paths})
Let $G_{\*{C}}(\*{C}, \*{E}_\*{C})$ be a C-DAG compatible with an ADMG $G(\*{V}, \*{E})$ and $\*C_i, \*C_j \in \*C$ be two distinct clusters. If two variables $V_i \in \*C_i$ and $V_j \in \*C_j$ are connected in $G$ by a directed path from $V_i$ to $V_j$, then the clusters $\*C_i$ and $\*C_j$ are connected  by a directed path from $\*C_i$ to $\*C_j$ in $G_{\*C}$. 
\end{restatable}

\begin{proof}
The proof is analogous to that of Lemma \ref{lem:connectionGc}, noticing that the edges forming the path connecting $\*C_i$ to $\*C_j$ in $G_{\*C}$ are always pointing towards $\*C_j$.
If %$\*C_i \neq \*C_j$ and 
$V_i$ and $V_j$ are adjacent, then $V_i \in Pa(V_j)$ and, by definition of C-DAGs, $\*C_i \rightarrow \*C_j$. Now, assume that $V_i$ and $V_j$ are connected, but not adjacent, and let $p$ be a directed path from $V_i$ to $V_j$. %Then $p$ is of the form $V_i \rightarrow \ldots \rightarrow V_k \rightarrow V_{k+1} \rightarrow \ldots \rightarrow V_j$. 
Let $p'$ be any subpath of $p$ that is split into two different clusters %. Then $p'$ is 
of the form $V_1 \rightarrow \ldots \rightarrow V_k \rightarrow V_{k+1} \rightarrow \ldots \rightarrow V_n$, where $V_1, \ldots, V_k \in C_k$, $V_{k+1}, \dots, V_n \in C_{k+1}$. Since $V_k \rightarrow V_{k+1}$ in $G$, we have, by Definition \ref{def:cdag}, that $\*C_k \rightarrow \*C_{k+1}$ in $G_{\*C}$. By using the same argument for every subpath of $p$ that is split into two different clusters, we conclude that all clusters containing variables in $p$ are connected by a directed path pointing away from $V_i$ and towards $V_j$, thus forming a directed path from $\*C_i$ to $\*C_j$ in $G_{\*C}$.
\end{proof}

Next, we show the proofs for the main results in this work, including soundness and completeness of d-separation, do-calculus, and ID-algorithm in C-DAGs.

\dsepcdags*

\begin{proof}
We first prove the soundness of d-separation in C-DAGs by showing that if $\*X$ and $\*Y$ are d-separated by $\*Z$ in $G_{\*C}$, then, in any ADMG $G$ compatible with $G_{\*C}$, $\*X$ and $\*Y$ are d-separated by $\*Z$ in $G$. 

We show the contrapositive. Assume that $\*X$ and $\*Y$ are d-separated by $\*Z$ in $G_{\*C}$, but in a compatible ADMG $G$ there exists a path $p'$ between a variable $X \in \*X$ and $Y \in \*Y$ that is active when the set of variables contained in clusters in $\*Z$ are conditioned on.  By Proposition \ref{prop:cdags_connections} and Lemma \ref{lem:connectionGc}, no connection is destroyed through clustering, so $p'$ is contained in a path $p$ of $G_{\*C}$ between clusters $\*X$ and $\*Y$. Since $\*X$ and $\*Y$ are d-separated by $\*Z$ in $G_{\*C}$, $p$ is blocked and clusters $\*X$ and $\*Y$ are not adjacent. Therefore, by Definition \ref{def:dsepcdag}, there is at least one triplet of clusters in $p$ that is blocked. Let this triplet be $\langle \*C_i, \*C_m, \*C_j \rangle$, where $\*C_m$ is distinct from $\*X$ and $\*Y$. 
Consider the subpath $p'_{ij}$ of $p'$ contained in the triplet $\langle \*C_i, \*C_m, \*C_j \rangle$ in $p$. Since $p'$ is active by assumption, every subpath of $p'$ is active, including $p'_{ij}$. The triplet $\langle \*C_i, \*C_m, \*C_j \rangle$ is blocked either if $\*C_m$ is a mediator or common cause such that $\*C_m \in \*Z$ or if $\*C_m$ is a collider such that $\*C_m \not \in \*Z$ and no descendant of $\*C_m$ is in $\*Z$. In the former case, Lemma \ref{lem:noncollider_inactive} shows that any path between $\*C_i$ and $\*C_j$ in a compatible ADMG $G$ is inactive when $\*C_m$ is conditioned on. In the latter case, Lemma \ref{lem:desc_collider_inactive} shows that any path between $\*C_i$ and $\*C_j$ in a compatible ADMG $G$ is inactive when $\*C_m$ is not conditioned on. So, the subpath $p'_{ij}$ of $p'$ is inactive, and, thus, $p'$ is also active.  

We now prove the completeness of d-separation for C-DAGs by showing that if $\*X$ and $\*Y$ are not d-separated by $\*Z$ in $G_{\*C}$, then there exists an ADMG $G$ compatible with $G_\*C$ where $\*X$ and $\*Y$ are not d-separated by $\*Z$ in $G$. 

%Let $p$ be the path between $\*X$ and $\*Y$ in $G_\*C$. 
If $\*X$ and $\*Y$ are adjacent, then, by Definition \ref{def:cdag}, in any ADMG $G$ compatible with $G_{\*C}$, there always exists a pair of variables $X \in \*X$ and $Y \in \*Y$ that are adjacent. Now, if $\*X$ and $\*Y$ are not adjacent and are not d-separated by $\*Z$ in $G_{\*C}$, then by Definition \ref{def:dsepcdag}, there exists a path $p$ between $\*X$ and $\*Y$ in $G_\*C$ such that every triplet $\langle \*C_i, \*C_m, \*C_j \rangle$ in it is active. This means that for all mediators and common causes $\*C_m$ contained in $p$, $\*C_m$ must not be in $\*Z$,  and for all colliders $\*C_n$ contained in $p$, either $\*C_n \in \*Z$ or some descendant of $\*C_n$ is in $\*Z$. By remarks 1, 2, 3, and 4, these triplets may be either active or inactive. Therefore, there exists some ADMG $G$ compatible with $G_\*C$ in which a path $p'$ between $\*X$ and $\*Y$ in $G$ goes through all (and only) clusters in $p$ and all of its subpaths contained in a triplet of clusters $\langle \*C_i, \*C_m, \*C_j \rangle$ in $p$ are active. In this case, $p'$ is active and $\*X$ and $\*Y$ are not d-separated by $\*Z$ in $G$.
\end{proof} 

%\subsection{Proofs for Section \ref{sec:do-calculus}}

% Proof of BN, Theorem 2
\cdagsbn*

\begin{proof}
Consider the following topological order over all clusters in $\*C$:
$\*C_1 \prec \*C_2 \prec  \ldots \prec \*C_K$. 
By %Proposition \ref{prop:cdags_ancestral}, 
Proposition~\ref{prop:connection_directed_Gc}, the ancestral relationships are preserved relative to any compatible ADMG. Therefore, such a topological order is consistent with a  topological order of the compatible graph $G$, regardless of the internal order of the variables in each cluster. 

Given the topological order over clusters, we construct a consistent topological order for the compatible ADMG $G$ over $\*V$ as follows:
$(V_{11} \prec \ldots \prec V_{1n_1}) \prec (V_{21} \prec \ldots \prec V_{2n_2})\prec \ldots \prec (V_{K1} \prec \ldots \prec V_{Kn_K})$. 

By assumption, the observational distribution $P(\*v) = P(\*c) $ factorizes according to $G$ by the following:
\begin{align}
& P(\*c) \nonumber \\ 
& \quad \quad 
= \sum_{\*u} P(\*u) \prod_{k:  \*C_k \in \*C} \prod_{i: V_{ki} \in \*C_k } P(v_{ki} | pa(v_{ki}), \*u_{ki}).\label{eq:ag-tfactor1} 
\end{align}
% $P(\*V)$ is semi-Markov relative to $G$ and factorizes as follows:
% \begin{align}
% P(\*c) = P(\*v) = \sum_{\*u} P(\*u) \prod_{k, i: V_{ki} \in \*C_k \in \*C} P(v_{ki} | pa(v_{ki}), \*u_{ki}),
% \end{align}
where $Pa(V_{ki})$ are the endogenous parents of $V_{ki}$ and $\*U_{ki}$ are the exogenous parents of $V_{ki}$, including those that are shared with some other variable $V_{k'j}$ (represented by $V_{k'j} \dashleftarrow \!\!\!\!\!\!\!\!\! \dashrightarrow V_{ki}$ in $G$).

Consider $V_{ki} \in \*C_k$. The set $Pa(V_{ki})$ of endogenous parents of $V_{ki}$ consists of the union of the set of the endogenous parents of $V_{ki}$ within the cluster $\*C_k$, denoted as $Pa^{(k)}_{ki}$, and the sets of the endogenous parents of $V_{ki}$ in all of the clusters that come before $\*C_k$ in the topological order, denoted as $Pa^{(1)}_{ki}, \ldots, Pa^{(k-1)}_{ki}$.  %Let $\*U_{k'j} \in \*U$ be the set of exogenous parents of variable $V_{k'j}$ in the cluster $\*C_{k'}$, for $k' = 1, \ldots K$. 
Define $\*U'^{(k')}_{ki} = \*U_{ki} \cap (\bigcup_j \*U_{k'j})$ for $k'\neq k$. Also let any exogenous parent of $V_{ki}$ that is not shared with other variables outside of $\*C_k$ be in $\*U'^{(k)}_{ki}$. Note that $\*U_{ki} = \bigcup_{k'} \*U'^{(k')}_{ki}$.
In words, $\*U_{ki}$ is the union of $\*U'^{(1)}_{ki}, \ldots, \*U'^{(K)}_{ki}$, which are the sets of exogenous parents of $V_{ki}$ that are also exogenous parents of some variables in clusters $\*C_1, \ldots, \*C_K$, respectively. Note that each variable $U \in \*U'^{(k')}_{ki}$ indicates a connection $V_{k'j} \dashleftarrow \!\!\!\!\!\!\!\!\! \dashrightarrow V_{ki}$ in $G$ and, therefore, a connection  $\*C_{k'} \dashleftarrow \!\!\!\!\!\!\!\!\! \dashrightarrow \*C_{k}$ in $G_{\*C}$. Also note that $\*U'^{(k)}_{ki}$ is the set of exogenous parents of $V_{ki}$ that may also be  exogenous parents of some other variable $V_{kj}$ in $\*C_k$. The connection $V_{kj} \dashleftarrow \!\!\!\!\!\!\!\!\! \dashrightarrow V_{ki}$ in $G$ is part of the internal structure of the cluster $\*C_k$, which is abstracted away in $G_\*C$. 

Using this notation, we can rewrite the conditional distribution for $V_{ki}$ as follows:
$$
\begin{aligned}
&P(v_{ki} | pa(v_{ki}), \*u_{ki}) = \\
& \quad \quad 
P(v_{ki} | pa^{(1)}_{ki}, \ldots, pa^{(k-1)}_{ki}, pa^{(k)}_{ki}, \*u'^{(1)}_{ki}, \ldots, \*u'^{(K)}_{ki})
\end{aligned}
$$

Note that all variables $V_{k_1}, \ldots, V_{k_{i-1}} \in \*C_k$ precede $V_{ki}$ in the topological order and, therefore, are non-descendants of $V_{ki}$. %Denote the set of non-descendants of $V_{ki}$ that are not parents of $V_{ki}$ as  $Nd^{(k)}_{ki} = \*C_k \setminus (Pa^{(k)}_{ki} \cup V_{ki})$.  
Then, we have 
\begin{align}
\begin{split}
    & (V_{ki} \indep V_{k_1}, \ldots, V_{k_{i-1}} | \\
    & \quad \quad 
    Pa^{(1)}_{ki}, \ldots, Pa^{(k-1)}_{ki}, Pa^{(k)}_{ki}, \*U'^{(1)}_{ki}, \ldots, \*U'^{(K)}_{ki}).
\end{split}
\label{eq:nd1}
\end{align}

% Therefore, we can rewrite the factor as follows:
% $$P(v_{ki}| pa(v_{ki}), \*u_{ki}) = P(v_{ki} | v_{k_1}, \ldots, v_{k_{i-1}}, pa^{(1)}_{ki}, \ldots, pa^{(k-1)}_{ki}, \*u^{(1)}_{ki}, \ldots, \*u^{(K)}_{ki})$$

Now, consider the set of endogenous parents of all other variables in the cluster, excluding those that are in $\*C_k$ or are also endogenous parents of $V_{ki}$:
$$
\begin{aligned}
& Pa^*(\{ V_{k1}, \ldots, V_{k_{n_k}} \}) =  Pa(\{ V_{k1}, \ldots, V_{k_{n_k}} \}) \setminus \\
& \quad \quad
(Pa( V_{ki} )  \cup \*C_k) \\
&= 
\left( Pa^{(1)}_{k1} \cup \ldots \cup Pa^{(1)}_{k_{n_k}} \cup \ldots \cup Pa^{(k-1)}_{k1} \cup \ldots \cup Pa^{(k-1)}_{k_{n_k}} \right) \\
& \quad \quad 
\setminus  (Pa( V_{ki} )  \cup \*C_k)
\end{aligned}
$$

Note that all variables in $Pa^*(\{ V_{k1}, \ldots, V_{k_{n_k}} \})$ are non-descendants of $V_{ki}$. Then, we have:
\begin{align}
\begin{split}
& (V_{ki} \indep Pa^*(\{ V_{k1}, \ldots, V_{k_{i-1}} \}) | \\
& \quad \quad 
Pa^{(1)}_{ki}, \ldots, Pa^{(k-1)}_{ki}, Pa^{(k)}_{ki}, \*U'^{(1)}_{ki}, \ldots, \*U'^{(K)}_{ki}). 
\end{split}
\label{eq:nd2}
\end{align}
Further, consider the set of variables that are in the same cluster as some parents of $\{V_{k1}, \ldots, V_{k_{nk}}\}$, excluding the parents themselves, denoted as 
$\*C^*(\{ Pa_{k1}, \ldots, Pa_{k_{nk}}\})$. Denote by $\*C^{*(k')}_{kj}$, for $k' = 1, \ldots, k-1$, the variables in cluster $k'$ that are together with a parent of $V_{kj}$, but are not parents of any variable $V_{kj'} \in \*C_k$, for $j' = 1, \ldots, n_k$.
$$
\begin{aligned}
& \*C^*(\{Pa_{k1}, \ldots, Pa_{kn_k} \}) = \\ 
& \quad \quad 
\*C^{*(1)}_{k1} \cup \ldots \cup \*C^{*(1)}_{kn_k}  \cup \ldots \cup \*C^{*(k-1)}_{k1}  \cup \ldots \cup \*C^{*(k-1)}_{kn_k}.
\end{aligned}
$$

Since all these variables are also non-descendants of $V_{ki}$, we have:
\begin{align}
\begin{split}
    & (V_{ki} \indep \*C^*(\{Pa_{k1}, \ldots, Pa_{kn_k} \}) | \\
    & \quad \quad 
    Pa^{(1)}_{ki}, \ldots, Pa^{(k-1)}_{ki}, Pa^{(k)}_{ki}, \*U'^{(1)}_{ki}, \ldots, \*U'^{(K)}_{ki}).
\end{split}
\label{eq:nd3}
\end{align}

Lastly, consider the set of exogenous parents of $V_{k1}, \ldots, V_{k_{n_k}}$ that are not exogenous parents of $V_{ki}$:
$$
\begin{aligned}
& \*U^*_{\{{k1}, \ldots, {kn_k} \}} = \\
& \quad \quad 
\left( \*U'^{(1)}_{k1} \cup \ldots \cup \*U'^{(1)}_{kn_k} \cup \ldots \cup \*U'^{(K)}_{k1} \cup \ldots \cup \*U'^{(K)}_{kn_k} \right) \setminus \\
& \quad \quad 
\left( \*U'^{(1)}_{ki}, \ldots, \*U'^{(K)}_{ki} \right).
\end{aligned}
$$

We also have 
\begin{align}
\begin{split}
    &(V_{ki} \indep \*U^*_{\{{k1}, \ldots, {kn_k} \}}  | \\ 
    & \quad \quad 
    Pa^{(1)}_{ki}, \ldots, Pa^{(k-1)}_{ki}, Pa^{(k)}_{ki}, \*U'^{(1)}_{ki}, \ldots, \*U'^{(K)}_{ki}).
\end{split}
\label{eq:nd4}
\end{align}

Given the conditional independence relations in \eqref{eq:nd1}, \eqref{eq:nd2}, \eqref{eq:nd3}, and \eqref{eq:nd4}, we can rewrite the conditional distribution of $V_{ki}$ to also condition on such non-descendants of $V_{ki}$ as follows:
$$
\begin{aligned}
&P(v_{ki}| pa(v_{ki}), \*u_{ki}) =  P(v_{ki} | v_{k_1}, \ldots, v_{k_{i-1}}, \\ 
& \quad \quad 
pa^{(1)}_{k1}, \ldots, pa^{(1)}_{kn_k}, \ldots, pa^{(k-1)}_{k1},  \ldots, pa^{(k-1)}_{kn_k},  \\
& \quad \quad 
\*c^{*(1)}_{k1},  \ldots, \*c^{*(1)}_{kn_k}, \ldots, \*c^{*(k-1)}_{k1},  \ldots, \*c^{*(k-1)}_{kn_k}, \\
& \quad \quad 
\*u'^{(1)}_{k1}, \ldots, \*u'^{(1)}_{kn_k}, \ldots, \*u'^{(K)}_{k1}, \ldots, \*u'^{(K)}_{kn_k}).
\end{aligned}
$$

Note that the parents of $V_{ki}$ that are in the cluster $\*C_k$ are contained in $\{V_1, \ldots, V_{k_{i-1}}\}$. %i.e. $\left( Pa(V_{ki}) \cap\*C_k \right)  \subseteq \{V_1, \ldots, V_{k_{i-1}}\}$. 
Also note that $\left( Pa(\{ V_{k1}, \ldots, V_{kn_k} \}) \setminus \*C_k \right)  \cup \*C^*(\{Pa_{k1}, \ldots, Pa_{kn_k} \}) = \bigcup Pa(\*C_k)_{G_\*C} \setminus \*C_k$. Denote as $\*U'_k  = \bigcup_i \*U_{ki} = \bigcup_{j,k'} \*U^{'(k')}_{kj}$, which is the set of the exogenous parents of all variables $V_{kj} \in \*C_k$.

Given that, we can rewrite the conditional distribution of each $V_{ki} \in \*C_i$ as follows:
%$$
\begin{align}
&P(v_{ki}| pa(v_{ki}), \*u_{ki}) = P(v_{ki} | v_{k_1}, \ldots, v_{k_{i-1}}, pa(\*c_k), \*u'_{k}).
\end{align}
%$$

By considering the factors for every variable $V_{ki} \in \*C_k$, for $i = 1 \ldots n_k$, we can write the conditional distribution of the set of variables in the cluster $\*C_k$ as follows:
%$$
\begin{align}
\begin{split}
\label{eq:factck}
&P(\*c_k | pa(\*c_k), \*u'_k) = %\\
%& \quad \quad 
\prod_{i: V_{ki} \in \*C_k} P(v_{ki} | v_{k_1}, \ldots, v_{k_{i-1}}, pa(\*c_k), \*u'_{k}).
\end{split}
\end{align}
%$$

Note that, for any two clusters $\*C_i, \*C_j \in \*C$, $\*U'_i \cap \*U'_j \neq \emptyset$ if, and only if, there exist a variable $V_{il} \in \*C_i$ and a variable $V_{jl'} \in \*C_j$ with a common exogenous parent. This follows from the construction of the sets $\*U'_i$ and $\*U'_j$. Note that $\*U'_i$ includes all and only the exogenous parents of $V_{il}$, for $l=1, \ldots, n_i$, that are also exogenous parents of any other variable in the other clusters, including $V_{jl'}$. Similarly, $\*U'_j$, includes all and only the exogenous parents of $V_{jl'}$, for $l'=1, \ldots, n_j$, that are also exogenous parents of any other variable in the other clusters, including $V_{il}$. Since, by Definition \ref{def:cdag}, a bidirected edge between $\*C_i$ and $\*C_j$ exists if $\exists V_{il} \in \*C_i$ and $V_{jl'} \in \*C_j$ such that $V_{il} \dashleftarrow \!\!\!\!\!\!\!\!\! \dashrightarrow V_{jl'}$ (i.e., $V_{il}$ and $V_{jl'}$ have a common exogenous parent), we have  $\*U'_i \cap \*U'_j \neq \emptyset$ if and only if $\*C_i \dashleftarrow \!\!\!\!\!\!\!\!\! \dashrightarrow \*C_j$ in $G_\*{C}$.

Therefore we can rewrite the factorization over the variables $V_{ki} \in \*V$ in $G$, shown in Equation \ref{eq:ag-tfactor1}, as a factorization over the clusters $\*C_k \in \*C$ in $G_\*C$ as follows:
\begin{align}
\begin{split}
& P(\*c) = %\\
%& \quad \quad 
\sum_{\*u} P(\*u) \prod_{k: \*C_k \in \*C} P(\*c_k | pa(\*c_k), \*u'_{k}). 
\end{split}
\label{eq:agc-tfactor1}
\end{align}

\end{proof}

\mutilation*

\begin{proof}
Note that we abuse the notation by letting $\*X, \*Z \subset \*C$ denote sets of clusters in $G_{\*C}$ and also letting $\*X, \*Z \subset \*V$ denote, in $G$, the set of variables contained in the clusters in $\*X, \*Z \subset \*C$.

We prove that the mutilated graph $G_{\overline{\*X}}$ is compatible with $G_{\*C_{\overline{\*{X}}}}$ by showing that the C-DAG $G_{\*C_{\overline{\*{X}}}}$ constructed by removing from $G_{\*C}$ the edges into $\*X$ is the same as the one constructed from $G_{\overline{\*X}}$ using Definition \ref{def:cdag}.

An edge in $G$ is cut in the construction of $G_{\overline{\*X}}$ if and only if it is into some variable $X \in \*X_i \in \*X$. Note that, if every such an edge is between two variables that are in the same cluster $\*X_i$, then the C-DAG constructed from $G_{\overline{\*X}}$ will be the same as the one constructed from $G$. In other words, $G_{\*C_{\overline{\*{X}}}} = G_{\*C}$. Further, the C-DAG $G_{\*C}$ constructed from $G$ contains no edge into any cluster $\*X_i \in \*X$. Therefore, no edge will be cut in the construction of $G_{\*C_{\overline{\*{X}}}}$ from $G_{\*C}$, leading to the same conclusion that $G_{\*C_{\overline{\*{X}}}} = G_{\*C}$.

Now, consider any edge of the type $V \ast \!\!\rightarrow X$ in $G$, where $X \in \*{X}_i \in \*X$  and $V \in \*C_i \in \*C$ such that $\*C_i \neq \*X_i$ (note that $\*C_i$ can be a cluster in $\*X$). Every of such an edge is cut in the construction of $G_{\overline{\*X}}$ from $G$. Therefore, the C-DAG constructed from $G_{\overline{\*X}}$ has no edge into any cluster $\*{X}_i \in \*X$. Further, by Definition \ref{def:cdag}, the existence of $V \ast \!\!\rightarrow X$ in $G$ guarantees the existence of the edge $\*C_i \ast \!\!\rightarrow \*X_i$ in $G_\*C$ and every such an edge in $G_{\*C}$ will be cut in the construction of $G_{\*C_{\overline{\*{X}}}}$ from $G_{\*C}$. Note that all other edges that are cut from $G$ to construct $G_{\overline{\*X}}$ are those between two variables that are in the same cluster $\*X_i \in \*X$. As previously discussed, such cuts do not change the C-DAG constructed from $G_{\overline{\*X}}$. Therefore, the C-DAG constructed from $G_{\overline{\*X}}$ is exactly the $G_{\*C}$ after cutting the edges into every cluster $\*X_i \in \*X$.

Similar to the proof that for the compatibility between $G_{\overline{\*X}}$ and $G_{\*C_{\overline{\*{X}}}}$, we prove that the mutilated graph $G_{\underline{\*Z}}$ is compatible with $G_{\*C_{\underline{\*{Z}}}}$ by showing that the C-DAG $G_{\*C_{\underline{\*{Z}}}}$ constructed by removing from $G_{\*C}$ the edges out of $\*Z$ is the same as the one constructed from $G_{\underline{\*Z}}$ using Definition \ref{def:cdag}.

An edge in $G$ is cut in the construction of $G_{\underline{\*Z}}$ if and only if it is out of some variable $Z \in \*Z_i \in \*Z$. If every such an edge is between two variables that are in the same cluster $\*Z_i$, then the C-DAG constructed from $G_{\underline{\*Z}}$ will be the same as the one constructed from $G$. In other words, $G_{\*C_{\underline{\*{Z}}}} = G_{\*C}$. Further, the C-DAG $G_{\*C}$ constructed from $G$ contains no edge out of any cluster $\*Z_i \in \*Z$. Therefore, no edge will be cut in the construction of $G_{\*C_{\underline{\*{Z}}}}$ from $G_{\*C}$, leading to the same conclusion that $G_{\*C_{\underline{\*{Z}}}} = G_{\*C}$.

Now, consider any edge of the type $Z \rightarrow V$ in $G$, where $Z \in \*{Z}_i \in \*Z$  and $V \in \*C_i \in \*C$ such that $\*C_i \neq \*Z_i$ (note that $\*C_i$ can be a cluster in $\*Z$). Every of such an edge is cut in the construction of $G_{\underline{\*Z}}$ from $G$. Therefore, the C-DAG constructed from $G_{\underline{\*Z}}$ has no edge out of any cluster $\*{Z}_i \in \*Z$. Further, by Definition \ref{def:cdag}, the existence of $Z \rightarrow V$ in $G$ guarantees the existence of the edge $\*Z_i \rightarrow \*C_i$ in $G_\*C$ and every such an edge in $G_{\*C}$ will be cut in the construction of $G_{\*C_{\underline{\*{Z}}}}$ from $G_{\*C}$. Note that all other edges that are cut from $G$ to construct $G_{\underline{\*Z}}$ are those between two variables that are in the same cluster $\*Z_i \in \*Z$. As previously discussed, such cuts do not change the C-DAG constructed from $G_{\underline{\*Z}}$. Therefore, the C-DAG constructed from $G_{\underline{\*Z}}$ is exactly the $G_{\*C}$ after cutting the edges into every cluster $\*Z_i \in \*Z$. 

Since the processes of cutting edges for constructing $G_{\overline{\*X}}$ and $G_{\underline{\*Z}}$ are independent of each other, we conclude that $G_{\*C_{{\overline{\*X}\underline{\*Z}}}}$ is compatible with $G_{\overline{\*X}\underline{\*Z}}$, i.e., the C-DAG $G_{\*C_{{\overline{\*X}\underline{\*Z}}}}$ constructed by cutting from $G_{\*C}$ the edges into $\*X$ and out of $\*Z$ is the same as the one constructed from $G_{\overline{\*X}\underline{\*Z}}$ using Def.~\ref{def:cdag}.

\end{proof}

\docalc*

\begin{proof}%[Proof of Soundness of Do-calculus in C-DAGs]
%Let $G_{\*{C}}$ be a C-DAG over a set of clusters $\*C$ and $G$ be a compatible causal diagram over a set of variables $\*V$. Also, let $\*W, \*X, \*Y$, and $\*Z$ be disjoint subsets of clusters in $\*C$. 
Note that we abuse the notation by letting $\*W, \*X, \*Y, \*Z \subset \*V$ also denote the set of variables contained in the clusters in $\*W, \*X, \*Y, \*Z \subset \*C$, respectively.

We first show that Rule 1 is sound. By Lemma \ref{lem:graph_mutilation}, $G_{\overline{\*X}}$ is compatible with $G_{\*{C}_{\overline{\*X}}}$. Then by Theorem \ref{thm:dsep_cdag_dag}, if $(\*{Y} \indep \*{Z} | \*{X}, \*{W})_{G_{\*{C}_{\overline{\*X}}}}$ then $(\*Y \indep \*Z | \*X, \*W)_{G_{\overline{\*X}}}$. Therefore Rule 1 holds by the Rule 1 of do-calculus in causal diagrams.

We now show that Rule 2 is sound. 
By Lemma \ref{lem:graph_mutilation}, $G_{\*{C}_{\overline{\*X} \underline{\*Z}}}$ is compatible with $G_{\overline{\*X}\underline{\*Z}}$. Then by Theorem \ref{thm:dsep_cdag_dag}, if $(\*{Y} \indep \*{Z} | \*{X}, \*{W})_{G_{\*{C}_{\overline{\*X}\underline{\*Z}}}}$ then $(\*{Y} \indep \*{Z} | \*{X}, \*W)_{G_{\overline{\*X}\underline{\*Z}}}$. Therefore Rule 2 holds by the Rule 2 of do-calculus in causal diagrams.

Lastly, we show that Rule 3 is sound. By Lemma \ref{lem:graph_mutilation}, $G_{\overline{\*X}}$ is compatible with $G_{\*{C}_{\overline{\*X}}}$. Let $\*Z_i \in \*Z$ be a cluster in $G_\*C$. If $\exists W \in \*W$ and $Z \in \*Z_i$ such that $Z \in An(W)$ in $G_{\overline{\*X}}$, then, by Proposition \ref{prop:connection_directed_Gc}, %and Proposition \ref{prop:cdags_ancestral}
there exists in $G_{\*{C}_{\overline{\*X}}}$ a directed path from $\*Z_i$ to some cluster $\*W_i \in \*W$ and, thus, $\*Z_i$ is an ancestor of $\*W$ in $G_{\*{C}_{\overline{\*X}}}$. In this case, any existing edge $\*C_i \ast\!\!\rightarrow \*Z_i$, where $\*C_i \in \*C$, in $G_{\*{C}_{\overline{\*X}}}$ also exists in the graph $G_{\*{C}_{\overline{\*X}\overline{\*Z(\*W)}}}$, which is obtained by cutting from $G_{\*{C}_{\overline{\*X}}}$ the edges into the $\*Z$-clusters that are not ancestors of any $\*W$-clusters. On the other hand, an edge $\*C_i \ast\!\!\rightarrow \*Z_i$ in $G_{\*{C}_{\overline{\*X}}}$ also exists in a C-DAG $G'_{\*C_{\overline{\*X}\overline{\*Z(\*W)}}}$ compatible with $G_{\overline{\*X}\overline{\*Z(\*W)}}$ if $\exists W \in \*W$, $Z \in \*Z_i$, and $V \in \*C_i$ such that $Z \in An(W)$ and $V \in Pa(Z)$ in $G_{\overline{\*X}}$. Therefore, $G_{\*C_{\overline{\*X}\overline{\*Z(\*W)}}}$ has the same or more edges than $G'_{\*{C}_{\overline{\*X}\overline{\*Z(\*W)}}}$ and, thus, a separation in $G_{\*{C}_{\overline{\*X}\overline{\*Z(\*W)}}}$ implies the same separation in $G'_{\*C_{\overline{\*X}\overline{\*Z(\*W)}}}$ which, by Theorem \ref{thm:dsep_cdag_dag}, implies the same separation in any compatible graph $G_{\overline{\*X}\overline{\*Z(\*W)}}$. In particular, if $(\*{Y} \indep \*{Z} | \*{X}, \*{W})_{G_{\*{C}_{\overline{\*X}\overline{\*Z(\*W)}}}}$, then 
$(\*{Y} \indep \*{Z} | \*{X}, \*W)_{G_{\overline{\*X}\overline{\*Z(\*W)}}}$. Therefore Rule 3 holds by the Rule 3 of do-calculus in causal diagrams.

\end{proof}

\calccomplete*

\begin{proof}%[Proof of Completeness of Do-calculus in C-DAGs]
Let $G_{\*{C}}$ be a C-DAG over a set of clusters $\*C$ and $G$ be a compatible causal diagram over a set of variables $\*V$. Also, let $\*W, \*X, \*Y$, and $\*Z$ be disjoint subsets of clusters in $\*C$. Note that we abuse the notation by letting $\*W, \*X, \*Y, \*Z \subset \*V$ also denote the set of variables contained in the clusters in $\*W, \*X, \*Y, \*Z \subset \*C$, respectively.

We first show that if $(\*{Y} \nindep \*{Z} | \*{X}, \*{W})_{G_{\*{C}_{\overline{\*X_\*C}}}}$ then $(\*Y \nindep \*Z | \*X, \*W)_{G_{\overline{\*X}}}$ for at least one causal diagram $G$ that is compatible with the C-DAG $G_{\*{C}}$. Lemma \ref{lem:graph_mutilation} ensures that the C-DAG  $G_{\*{C}_{\overline{\*X}}}$ is compatible with $G_{\overline{\*X}}$. This, along with the completeness part of Theorem \ref{thm:dsep_cdag_dag}, implies that if $(\*{Y} \nindep \*{Z} | \*{X}, \*{W})_{G_{\*{C}_{\overline{\*X}}}}$, then there exists a causal diagram $G$ compatible with $G_{\*{C}}$ for which $(\*Y \nindep \*Z | \*X, \*W)_{G_{\overline{\*X}}}$ and, therefore, Rule 1 does not hold by the Rule 1 of do-calculus in causal diagrams.

We now show that if $(\*{Y} \nindep \*{Z} | \*{X}, \*{W})_{G_{\*{C}_{\overline{\*X}\underline{\*Z}}}}$ then $(\*Y \nindep \*Z | \*X, \*W)_{G_{\overline{\*X}\underline{\*Z}}}$ for at least one causal diagram $G$ that is compatible with the C-DAG $G_{\*{C}}$. Lemma \ref{lem:graph_mutilation} ensures that the C-DAG $G_{\*{C}_{\overline{\*X} \underline{\*Z}}}$ is compatible with $G_{\overline{\*X}\underline{\*Z}}$. 
This, along with the completeness part of Theorem \ref{thm:dsep_cdag_dag}, implies that if $(\*{Y} \nindep \*{Z} | \*{X}, \*{W})_{G_{\*{C}_{\overline{\*X}\underline{\*Z}}}}$,  then there exists a causal diagram $G$ compatible with $G_{\*{C}}$ for which $(\*Y \nindep \*Z | \*X, \*W)_{G_{\overline{\*X}\underline{\*Z}}}$ and, therefore, Rule 2 does not hold by the Rule 2 of do-calculus in causal diagrams.

Lastly, we show that if
$(\*{Y} \nindep \*{Z} | \*{X}, \*{W})_{G_{\*{C}_{\overline{\*X}\overline{\*Z(\*W)}}}}$ then $(\*Y \nindep \*Z | \*X, \*W)_{G_{{\overline{\*X}\overline{\*Z(\*W)}}}}$ for at least one causal diagram $G$ that is compatible with the C-DAG $G_{\*{C}}$. Consider the causal diagram $G$ over $\*V$ compatible with $G_{\*{C}}$ where, for every $\*C_i, \*C_j \in \*C$, any two variables $V_i, V_j \in \*V$ such that $V_i \in \*C_i$ and $V_j \in \*C_j$, there is a connection $V_i \rightarrow V_j$ if $\*C_i \rightarrow \*C_j$ and $V_i \dashleftarrow \!\!\!\!\!\!\!\!\! \dashrightarrow V_j$ if $\*C_i \dashleftarrow \!\!\!\!\!\!\!\!\! \dashrightarrow \*C_j$. In this case, if for some cluster $\*Z_i \in \*Z$ there exist $Z \in \*Z_i$ and $W \in \*W$, such that $Z \in An(W)$, then, $\forall Z' \in \*Z_i$, we have $Z' \in An(W)$. As previously discussed in the proof of the soundness of the Rule 3 (Theorem \ref{thm:docalc-cdags}), in this case $G_{\*C_{\overline{\*X}\overline{\*Z(\*W)}}}$ is compatible with $G_{{\overline{\*X}\overline{\*Z(\*W)}}}$. Further, in this case, for every path $p$ in $G_\*C$, there exists a path $p'$ in $G$ of the same form of $p$ and going through only one variable in each cluster in $p$. Therefore, every d-connection in $G_{\*C}$ corresponds to a d-connection in $G$. %Is this clear or we need a proof?
In particular, if $(\*{Y} \nindep \*{Z} | \*{X}, \*{W})_{G_{\*C_{\overline{\*X}\overline{\*Z(\*W)}}}}$, then $(\*{Y} \nindep \*{Z} | \*{X}, \*{W})_{G_{{\overline{\*X}\overline{\*Z(\*W)}}}}$ and Rule 3 does not hold in $G$ by the Rule 3 of do-calculus in causal diagrams.

\end{proof}

%\subsection{Proofs for Section \ref{sec:IDalg} \label{app:calculus}}

% Proof of Factorization
\markovrelative*

\begin{proof}

By assumption, the post-interventional distribution $P(\*v\setminus \*x | do(\*x)) = P(\*c\setminus \*x | do(\*x)) $ factorizes according to $G$ by the following truncated factorization:
\begin{align}
& P(\*c\setminus \*x | do(\*x)) \nonumber \\
& \quad \quad 
= \sum_{\*u} P(\*u) \prod_{k:  \*C_k \in \*C \setminus \*X} \prod_{i: V_{ki} \in \*C_k } P(v_{ki} | pa(v_{ki}), \*u_{ki}). \label{eq:ag-tfactor}
\end{align}
% $P(\*V)$ is semi-Markov relative to $G$ and factorizes as follows:
% \begin{align}
% P(\*c) = P(\*v) = \sum_{\*u} P(\*u) \prod_{k, i: V_{ki} \in \*C_k \in \*C} P(v_{ki} | pa(v_{ki}), \*u_{ki}),
% \end{align}
where $Pa(V_{ki})$ are the endogenous parents of $V_{ki}$ and $\*U_{ki}$ are the exogenous parents of $V_{ki}$, including those that are shared with some other variable $V_{k'j}$ (represented by $V_{k'j} \dashleftarrow \!\!\!\!\!\!\!\!\! \dashrightarrow V_{ki}$ in $G$).

We can use the same procedure in the proof of Theorem \ref{thm:cdag_bn} to show that, for any cluster $\*C_k \in \*C$, the conditional distribution $P(\*c_k | pa(\*c_k), \*u'_k)$ factorizes as a product of conditional distributions over the variables in $\*C_k$ as shown in Equation \ref{eq:factck}. Further, from the discussion regarding the construction of the sets $\*U'_i \cap \*U'_j$ in the proof of Theorem \ref{thm:cdag_bn}, we also conclude that  $\*U'_i \cap \*U'_j \neq \emptyset$ iff $\*C_i \dashleftarrow \!\!\!\!\!\!\!\!\! \dashrightarrow \*C_j$ in $G_\*{C}$.

Therefore we can rewrite the truncated factorization over the variables $V_{ki} \in \*V$ in $G$, shown in Equation \ref{eq:ag-tfactor}, as a truncated factorization over the clusters $\*C_k \in \*C$ in $G_\*C$ as follows:
\begin{align}
\begin{split}
& P(\*c\setminus \*x | do(\*x)) = %\\
%& \quad \quad 
\sum_{\*u} P(\*u) \prod_{k: \*C_k \in \*C \setminus \*X} P(\*c_k | pa(\*c_k), \*u'_{k}). 
\label{eq:agc-tfactor}
\end{split}
\end{align}
\end{proof}

%TODO lem
\begin{restatable}{lemma}{ancestral}
\label{prop:cdags_ancestral}
(\textbf{Preservation of ancestral relationships}) Let $G_{\*{C}}(\*{C}, \*{E}_\*{C})$ be a C-DAG compatible with causal diagram $G(\*{V}, \*{E})$. 
For any $\*C_i \in \*C$, $\forall V \in \*{C}_i,  (An(V)_G \cup \{V\}) \subseteq  (An(\*{C}_i)_{G_{\*{C}}} \cup \*{C}_i)$ and $(De(V)_G \cup \{V\}) \subseteq (De(\*{C}_i)_{G_{\*{C}}} \cup \*{C}_i)$.
\end{restatable}

\begin{proof}
Suppose that exists $V_j \in \*{C}_j$ and $V \in \*{C}_i$ such that $V_j$ is an ancestor of $V$. If $\*{C}_j = \*{C}_i$, then $V_j \in \*C_i$. Otherwise, by Proposition \ref{prop:connection_directed_Gc}, the directed path $p$ from $V_j$ to $V$ in $G$ is preserved and represented by a directed path from $\*C_j$ to $\*C_i$ in $G_\*C$. Therefore $C_j \in An(\*C_i)_{G_\*C} \cup \*C_i$. Then $V_j \in An(\*C_i)_{G_\*C} \cup \*C_i$. This holds for any $V_j \in An(V)_G \cup \{V\}$, hence we obtain $(An(V)_G \cup \{V\}) \subseteq (An(\*C_i)_{G_\*C} \cup \*C_i)$.

Analogously, suppose that there exists $V_j \in \*{C}_j$ and $V \in \*{C}_i$ such that $V_j$ is a descendant of $V$. If $\*{C}_j = \*{C}_i$, then $V_j \in \*C_i$. Otherwise, by Proposition \ref{prop:connection_directed_Gc}, the directed path $p$ from $V$ to $V_j$ in $G$ is preserved and represented by a directed path from $\*C_i$ to $\*C_j$ in $G_\*C$. Therefore, $C_j \in De(\*C_i)_{G_\*C} \cup \*C_i$. Then $V_j \in De(\*C_i)_{G_\*C} \cup \*C_i$ and we obtain $(De(V)_G \cup \{V\}) \subseteq (De(\*C_i)_{G_\*C} \cup \*C_i)$.

\end{proof} 

%lem
\begin{restatable}{lemma}{ancestralsetGR}
(\textbf{Reduction to an ancestral set})
\label{lem:ancestralsetGR}
Given a C-DAG $G_\*C$, let $\*A \subseteq \*C$. If $\*{A}$ is an ancestral set in $G_{\*{C}}$ then, $Q[\*A] = \sum_{\*C \setminus \*A} Q[\*C]$ in any compatible causal diagram $G$, where $\*A, \*C$ represent the sets of variables contained in the clusters in $\*A, \*C$, respectively.
\end{restatable}

\begin{proof}
    
    For any C-DAG $G_{\*{C}}$, let $\*A \subseteq \*C$. 
    %Proof of Soundness
    We first show that if $\*{A}$ is an ancestral set in $G_{\*{C}}$ then, for any compatible causal diagram $G$, the set of all variables contained in $\*{A}$ is ancestral in $G$. 
    
    Again we will use the same notation $\*A$ to also indicate  the set of variables in $\*V$ in $G$ contained in the cluster $\*A \subset \*C$ in $G_\*C$.
    
    Assume that $\*A$ is ancestral in $G_{\*C}$. This means $\*{A}$ contains all $An(\*{A})_{G_{\*C}}$. Now, consider any graph $G$ compatible with the C-DAG, $G_{\*{C}}$. By Lemma \ref{prop:cdags_ancestral}, $An(\*{A})_{G} \subseteq An(\*{A})_{G_\*C}$, so if $\*{A}$ is an ancestral set in $G_{\*{C}}$, then $\*{A}$ is ancestral in $G$. This means that the distribution $Q[\*{A}]$ induced by any compatible graph $G$ can be calculated directly from $Q[\*{C}]$ by marginalizing the clusters in $\*{C} \setminus \*{A}$. 

\end{proof}
%%%%% PROOF %%%%%%%

% lem
\begin{restatable}{lemma}{ccompGR}
\label{lem:ccompGR}
(\textbf{Preservation of c-components})
Given a C-DAG $G_\*C$, for any causal diagram $G$ compatible with $G_\*C$, if any two variables $V_i, V_j \in \*V$ are in the same c-component in $G$, then the clusters $\*C_i,\*C_j \in \*C$ containing $V_i$ and $V_j$, respectively, are in the same c-component in $G_\*C$. 
\end{restatable}

%\ccompGR*
\begin{proof}
We show that if two variables $V_i, V_j \in \*V$ belong to a c-component in $G$, then they will belong to the same c-component in $G_\*C$. If $V_i$ and $V_j$ are in the same cluster in $G_{\*C}$, then they are always together in a c-component of $G_\*C$. If $V_i \in \*C_i$ and $V_j \in \*C_j$, where $\*C_i, \*C_j \in \*C$ are distinct clusters, then, since $V_i$ and $V_j$ are in the same c-component in $G$, there exists a path of bidirected edges of the form $V_i \dashleftarrow \!\!\!\!\!\!\!\!\! \dashrightarrow \ldots \dashleftarrow \!\!\!\!\!\!\!\!\! \dashrightarrow V'_i \dashleftarrow \!\!\!\!\!\!\!\!\! \dashrightarrow V'_j \dashleftarrow \!\!\!\!\!\!\!\!\! \dashrightarrow \ldots  \dashleftarrow \!\!\!\!\!\!\!\!\! \dashrightarrow V_j$, where $V'_i \in \*C_i$ and $V'_j \in \*C_j$. Then, by Definition \ref{def:cdag}, $\*C_i \dashleftarrow \!\!\!\!\!\!\!\!\! \dashrightarrow \*C_j$ in $G_\*C$. This follows for every 
pair of variables $V_i, V_j$ that are in the same c-component. Therefore, 
any path of bidirected edges between variables in $G$ is represented by a path of bidirected edges between clusters in $G_\*C$. Consequently, any c-component in $G$ implies a c-component in $G_\*C$.
\end{proof}

\idalgsoundcomplete*
\begin{proof}
\textbf{(Soundness):} For any set of clusters $\*S \subseteq \*C$, let the quantity $Q[\*S]$ denote the post-intervention distribution of $\*S$ under the intervention to all other clusters in $\*C$, i.e.:
$Q[\*S] = P(\*s|do(\*c \setminus \*s))$. 
By the truncated factorization in C-DAGs shown in Theorem \ref{thm:markovrelative}, we have:
$$Q[\*S] = P(\*s|do(\*c \setminus \*s)) =  \sum_{\*u} P(\*u) \prod_{k:\*C_k \in \*S} P(\*c_k | pa(\*c_k), \*u_k).$$

Let $\*X, \*Y \subseteq \*C$ be distinct sets of clusters in a C-DAG $G_\*C$. Also, let $\*D = An(\*Y)$ in $G_{\*C}[\*C \setminus \*X]$, where $G_{\*C}[\*C \setminus \*X]$ is the subgraph of $G_C$ containing only the clusters in $\*C$ not in $\*X$. %$G_{\*C_{\overline{\*X}}}$ and 
Since $\*D$ contains all ancestors of $\*Y$ in $G_{\*C}[\*C \setminus \*X]$ and $Q[\*C \setminus \*X] = P(\*c \setminus \*x|do(\*x))$, we have:
\begin{align}
P(\*y|do(\*x)) = \sum_{\*c \setminus (\*x \cup \*y)} Q[\*C \setminus \*X] = \sum_{\*d \setminus \*y} Q[\*D].
\label{eq:ancred}
\end{align}
Note that, by Lemma \ref{prop:cdags_ancestral}, $\*D$ contains all ancestors of $\*Y$ in $G[\*V\setminus \*X]$, for any causal diagram $G$ compatible with $G_\*C$. Therefore, by Lemma \ref{lem:ancestralsetGR}, the reduction shown in Equation \eqref{eq:ancred} is valid in any $G$ compatible with $G_\*C$.

Now, let $\*D_1, \ldots, \*D_l$ be the c-components of the subgraph $G_{{\*C}}[\*D]$. Also, let $\*S_1, \ldots, \*S_k$ be the c-components of $G_{{\*C}}$. 

Consider the quantity $Q[\*S_j] = P(\*s_j | do(\*c \setminus \*s_j)$. Given a topological order $\*C_1 \prec \ldots  \prec \*C_n$ over all clusters in $G_\*C$, $Q[\*S_j]$ is computable from $P(\*c)$ as follows \citep{tian:02thesis, tian:pea02-general-id}:
\begin{align}
Q[\*S_j] = \prod_{\{i|\*C_i \in \*S_i\}} P(\*C_i | \*C_1, \ldots, \*C_{i-1}).
\label{eq:qsj}
\end{align}
%By using the Q-decomposition of $P(\*c)$, %TODO prove that?

By Lemma \ref{lem:ccompGR}, any c-component of a causal diagram $G$ compatible with $G_{{\*C}}$ is contained in some c-component $\*S_j$ of $G_{{\*C}}$. Further, Theorem \ref{thm:markovrelative} shows that a factorization of any intervention distribution %(including the observational distribution, where the intervention set is empty) 
over the clusters in $\*C$ according to a C-DAG $G_\*C$ is valid %a valid factorization of the same intervention distribution
over the variables in $\*V$ in any compatible causal diagram $G$. Thus, in any $G$ compatible with $G_\*C$, the quantity $Q[\*S_j]$ shown in Equation \eqref{eq:qsj} corresponds to the distribution $P(\*s_j | do(\*v \setminus \*s_j)$, where $\*S_j$ represents the set of variables contained in all clusters in $\*S_j$. 

By considering the Q-decomposition of $Q[\*D]$ (i.e., $Q[\*D] = \prod_i Q[\*D_i]$), the ID-algorithm decomposes the problem of identifying $P(\*y|do(\*x))$ into the smaller sub-problems of identifying the quantities $Q[\*D_i]$, for $i = 1, \ldots, l$, from some $Q[\*S_j]$, where $\*D_i \subseteq \*S_j$:
$$P(\*y|do(\*x)) = \sum_{\*d \setminus \*y} \prod_i Q[\*D_i].$$

The ID-algorithm repeatedly applies the reduction to an ancestral set and the factorization into c-components until it fails or outputs an expression for $Q[\*D_i]$ in terms of $Q[\*S_j]$. As previously indicated, such manipulations of the probability distribution over clusters correspond to valid manipulations in the probability distribution over variables. Therefore, if all $Q[\*D_i]$'s are identifiable in $G_{{\*C}}$, then, in any $G$ compatible with $G_{{\*C}}$, all $Q[\*D_i]$'s are also identifiable and thus, $P(\*y|do(\*x))$ is identifiable.

\textbf{(Completeness):} Let $G_{\*C}$ be a C-DAG and $\*X, \*Y \subset \*C$ be two sets of clusters. We will show that if the ID-algorithm fails to identify $P(\*y|do(\*x))$ in $G_{\*C}$, then there exists a causal diagram $G$ compatible with $G_{\*C}$ where the effect $P(\*y|do(\*x))$ is not identifiable and, therefore, the ID-algorithm would fail to identify the effect $P(\*y|do(\*x))$ in $G$.

Assume that the ID-algorithm fails in $G_{\*C}$. Then, for some $\*X' \subseteq \*X$ and $\*Y' \subseteq \*Y$, there exists a hedge  $\langle \mathcal{F}_{\*C}, \mathcal{F}'_{\*C} \rangle$ for $P(\*y'|do(\*x'))$ in $G_{\*C}$. Note that $\mathcal{F}_{\*C}, \mathcal{F}'_{\*C}$ are $\*R$-rooted C-forests, where $\*R \subset An(\*Y)_{G_{{\*C}_{\overline{\*X}}}}$, such that $\mathcal{F}'_{\*C} \subseteq \mathcal{F}_{\*C}$, $\mathcal{F}'_{\*C} \cap \*X = \emptyset$, and $\mathcal{F}_{\*C} \cap \*X \neq \emptyset$. We will show that we can construct a causal diagram $G$, compatible with $G_{\*C}$, with a hedge $\langle \mathcal{F}, \mathcal{F}' \rangle$ for $P(\*y'|do(\*x'))$. 

Consider the causal diagram $G$ over $\*V$ where, for every $\*C_i, \*C_j \in \*C$, any two variables $V_i, V_j \in \*V$ such that $V_i \in \*C_i$ and $V_j \in \*C_j$, there is a connection $V_i \rightarrow V_j$ if $\*C_i \rightarrow \*C_j$ and $V_i \dashleftarrow \!\!\!\!\!\!\!\!\! \dashrightarrow V_j$ if $\*C_i \dashleftarrow \!\!\!\!\!\!\!\!\! \dashrightarrow \*C_j$. Note that, by definition of cluster causal diagrams, such edges do not violate the compatibility of $G$ with $G_{\*C}$. Now, for each $\*C_i \in \*C$, let $V_{i1} \prec \ldots \prec V_{iN_i}$ be a topological order over the set of variables that constitute $\*C_i$. Let the internal structure of $\*C_i$ be a chain with an edge $V_{ik} \rightarrow V_{i(k+1)}$ and a bidirected edge $V_{ik} \dashleftarrow \!\!\!\!\!\!\!\!\! \dashrightarrow V_{i(k+1)}$, for $k = 1, \ldots, N_i$. Note that such an internal structure is consistent with the topological order and does not violate the compatibility with $G_{\*C}$. 

Construct $\mathcal{F}'$ and $\mathcal{F}$, edge subgraphs of $G$, as follows: consider the subgraphs over the variables that are contained in the clusters in $\mathcal{F}_\*C$ and $\mathcal{F}'_\*C$, respectively; remove all outgoing edges from any variable in $\*R$; then remove a set of directed edges so that every observable variable has at most one child, while preserving $\*R$ as the maximal root set. This construction process ensures that the resulting edge subgraphs $\mathcal{F}'$ and $\mathcal{F}$ satisfy the properties of $\*R$-rooted C-forests in $G_{\*C}$. Note that, since the variables within the clusters form a chain from a cluster to another, the ancestral relations in $\mathcal{F}_\*C$ and $\mathcal{F}'_\*C$ are preserved so that  $\mathcal{F}'$ and $\mathcal{F}$ have $\*R$ as a root set. Also, since no bidirected edge is removed, both $\mathcal{F}'$ and $\mathcal{F}$ have only one c-component.  Lastly, since $\mathcal{F}$ and $\mathcal{F}'$ are constructed over the variables in $\mathcal{F}_\*C$ and $\mathcal{F}'_\*C$, respectively,
%where $\mathcal{F}'_{\*C} \subseteq \mathcal{F}_{\*C}$, $\mathcal{F}'_{\*C} \cap \*X = \emptyset$, and $\mathcal{F}_{\*C} \cap \*X \neq \emptyset$.
we have $\mathcal{F}' \subseteq \mathcal{F}$, $\mathcal{F}' \cap \*X = \emptyset$, and $\mathcal{F}\cap \*X \neq \emptyset$.
Therefore, for some $\*X' \subseteq \*X$ and $\*Y' \subseteq \*Y$, there exist $\*R$-rooted C-forests $\mathcal{F}, \mathcal{F}'$ that form a hedge for $P(\*y'|do(\*x'))$ in $G$, which, in turn, implies that the ID-algorithm will fail to identify $P(\*y|do(\*x))$ in $G$. 
\end{proof}

\scmcdags*

\begin{proof}
Given a partition $\*C = \{\*C_1, \ldots \*C_K\}$ of $\*V$, denote $V_{ij}$ the $j$th variable in cluster $\*C_i \in \*C$. We assume that $G$ over $\*V$ is induced by an SCM $\mathcal{M} = \langle \*V, \*U, \mathcal{F}, P(\*U) \rangle$, where 
each variable $V_{ij} \in \*C_i$ is given by a function $f_{ij} \in \mathcal{F}$ as follows:
$$
\begin{aligned}
    V_{ij} \leftarrow f_{ij}(\*{Pa}_{ij}, \*{U}_{ij}),
\end{aligned}
$$
where $\*{Pa}_{ij} \subseteq \*V$ is the set of endogenous variables representing parent nodes of $V_{ij}$ in $G$ and $\*{U}_{ij} \subseteq \*U$ is the set of exogenous variable  representing the latent parents of $V_{ij}$.

Now, let $\tilde{\*C} = \{\tilde{C}_1, \ldots, \tilde{C}_K\}$ be a set of ``macro-variables'' where, for $i = 1 \ldots K$, letting $N_i$ be the number of variables in cluster $\*C_i$, $\tilde{C}_i = (V_{i1}, \ldots, V_{iN_i})$ is a $N_i$-dimensional random vector. We construct a function for $\tilde{C}_i$  as follows:  
\begin{align}\label{eq:vscm}
\tilde{C}_i = 
\begin{pmatrix} 
V_{i1} \\
\vdots \\
V_{iN_i}
\end{pmatrix} = 
\begin{pmatrix} 
f'_{i1}(\*{Pa}'_{i1}, \*{U}'_{i1}) \\
\vdots \\
f'_{iN}(\*{Pa}'_{iN_i}, \*{U}'_{iN_i}) \\
\end{pmatrix}, 
\end{align}
where, for $j = 1, \ldots N_i$, 
%\xst{ $\*{Pa}'_{ij} = \bigcup \{ \*{Pa}_{ik} : V_{ik} \in An(V_{ij}) \cap \*C_i \}  \setminus \*C_i$, i.e., the set of endogenous variables that are not in $\*C_i$ representing parent nodes of ancestors of $V_{ij}$ in $G$ that are in $\*C_i$;} 
$\*{Pa}'_{ij} = Pa(An(V_{ij}) \cap \*C_i) \setminus \*C_i$, i.e., the set of endogenous variables that are not in $\*C_i$ but are parents of nodes in $G$ that are both ancestors of $V_{ij}$ and in $\*C_i$; and 
%\xst{  $\*{U}'_{ij} =\bigcup \{ \*{U}_{ik} : V_{ik} \in An(V_{ij}) \cap \*C_i \}$, i.e., the set of exogenous variables representing latent parent nodes of ancestors of $V_{ij}$ in $G$ that are in $\*C_i$} 
$\*{U}'_{ij} = \{ \*{U}_{ik} : V_{ik} \in An(V_{ij}) \cap \*C_i \}$, i.e., the set of exogenous variables representing latent parents of nodes that are both ancestors of $V_{ij}$ and in $\*C_i$.
%\xadd{
$f'_{ij}$ results from $f_{ij}(\*{Pa}_{ij}, \*{U}_{ij})$ in which each endogenous variable $V_{ik}$ in $\*{Pa}_{ij}$ that is in $\*C_i$ is recursively replaced by its own function $f_{ik}$ until none of the arguments are in $\*C_i$ (this is feasible due to the acyclicity of SCMs and can be done following the topological order over the variables in $\*C_i$ induced by $G$). Therefore, $f'_{ij}$ is just an expanded form of $f_{ij}$ in terms of endogenous variables that are not in $\*C_i$, and  the assignment $V_{ij} \leftarrow f'_{ij}(\*{Pa}'_{ij}, \*{U}'_{ij})$ does not change the original assignment of $V_{ij}$ given a fixed $\*U$ value.%}

Now, define $\*{Pa}'_{\*C_i} = \*{Pa}'_{i1} \cup \ldots \cup \*{Pa}'_{iN_i}$ and  $\*{U}_{\tilde{C}_i} = \*{U}'_{i1} \cup \ldots \cup \*{U}'_{iN_i}$. %Then, $\tilde{C}_i \leftarrow f'_i(\*{Pa}'_{\*C_i}, \*{U}'_{\*C_i})$, where $f'_i$ is a function of variables in $\*V \cup \*U$. 
Take the smallest subset $\*C'$ of $\*C$ such that $\*{Pa}'_{\*C_i} \subseteq \*C'$ and denote $\*{Pa}_{\tilde{C}_i}$ as the set of macro-variables in $\tilde{C}$ corresponding to the clusters in $\*C'$. We rewrite the function for $\tilde{C}_i$ in Eq.~(\ref{eq:vscm}) as
\begin{align} \label{eq:macroscm}
    \tilde{C}_i \leftarrow f_i(\*{Pa}_{\tilde{C}_i}, \*{U}_{\tilde{C}_i}) \equiv \begin{pmatrix} 
f'_{i1}(\*{Pa}'_{i1}, \*{U}'_{i1}) \\
\vdots \\
f'_{iN}(\*{Pa}'_{iN_i}, \*{U}'_{iN_i}) \\
\end{pmatrix} 
\end{align}
We construct an SCM $\mathcal{M}_{\tilde{\*C}} = \langle \tilde{\*C}, \*U, \mathcal{F}_{\tilde{\*C}}, P(\*U) \rangle$, where $\mathcal{F}_{\tilde{\*C}}$ consists of the set of functions $f_i$ specified in Eq.~(\ref{eq:macroscm}), for $i  = 1, \ldots, K$. 

This SCM $\mathcal{M}_{\tilde{\*C}}$ induces a causal diagram $G_{\tilde{\*C}}$ over macro-variables $\tilde{\*C}$ where $\tilde{C}_i \leftarrow \tilde{C}_j$ iff $\tilde{C}_j \in \*{Pa}_{\tilde{C}_i}$; and $\tilde{C}_i \dashleftarrow \!\!\!\!\!\!\!\!\! \dashrightarrow \tilde{C}_j$ iff $\*U_{\tilde{C}_i} \cap \*U_{\tilde{C}_j} \neq \emptyset$. We have that
\begin{enumerate}
    \item $\tilde{C}_j \in \*{Pa}_{\tilde{C}_i}$ iff ${\*C}_j \cap \*{Pa}'_{\*C_i} \neq \emptyset$ iff $\exists k, {\*C}_j \cap \*{Pa}'_{ik} \neq \emptyset$ iff a node in ${\*C}_j$ is a parent of some node in $\*C_i$ since every node in $\*{Pa}'_{ik}$ is a parent of some node in $\*C_i$.
    \item $\*U_{\tilde{C}_i} \cap \*U_{\tilde{C}_j} \neq \emptyset$ iff $\exists k, k', \*{U}'_{ik} \cap \*{U}'_{jk'} \neq \emptyset$ iff  there exists a latent node in $\*{U}'_{ik} \cap \*{U}'_{jk'}$ that is a shared parent of a node in  $\*C_i$ and a node in  $\*C_j$ since every latent node in $\*{U}'_{ik}$ is a parent of some node in $\*C_i$ and every latent node in $\*{U}'_{jk'}$ is a parent of some node in $\*C_j$. %a node $V_{k'j}$ in ${\*C}_j$ has a shared latent parent with some node that is both in $\*C_i$ and ancestor of $V_{ik}$.
\end{enumerate}
Therefore, the causal diagram $G_{\tilde{\*C}}$ is exactly the C-DAG $G_{\*C}$ of $G$ given by Definition \ref{def:cdag}.

By the construction of the set of functions $f_i$ in Eq.~(\ref{eq:macroscm}), we have $\tilde{\*C}_i(\*u) = (V_{i1}(\*u), \ldots, V_{iN_i}(\*u))= {\*C}_i(\*u)$ for any fixed $\*U=\*u$ value. Performing an intervention  $do(\tilde{\*C}_i =\*c_i)$ in $\mathcal{M}_{\tilde{\*C}}$ for any $i$ replaces the function $f_i$ in Eq.~(\ref{eq:macroscm}) with $\tilde{\*C}_i \leftarrow \*c_i$ which is the same operation as performing an intervention $do(\*C_i =\*c_i)$ in $\mathcal{M}$. Therefore,  for any  $\*Y, \*X \subseteq \*C$ and corresponding macro-variables $\tilde{\*Y}, \tilde{\*X}$, we have $\*Y_{\*x}(\*u) = \tilde{\*Y}_{\tilde{\*x}}(\*u) $. We obtain that for any set of counterfactual variables $\*Y_{\*x} \ldots \*Z_{\*w}$ where $\*Y, \*X, \ldots, \*Z, \*W \subseteq \*C$,
\begin{align}
P_{\mathcal{M}} (\*y_{\*x} \ldots \*z_{\*w}) &= \sum_{\*u | \*Y_{\*x}(\*u)=\*y, \ldots, \*Z_{\*w}(\*u)=\*z } P(\*u) \nonumber\\
& = \sum_{\*u | \tilde{\*Y}_{\tilde{\*x}}(\*u)=\tilde{\*y}, \ldots, \tilde{\*Z}_{\tilde{\*w}}(\*u)=\tilde{\*z} } P(\*u) \nonumber\\
 &= P_{\mathcal{M}_{\tilde{\*C}}} (\tilde{\*y}_{\tilde{\*x}} \ldots \tilde{\*z}_{\tilde{\*w}}).
\end{align}

\end{proof}

\end{document}